\documentclass[twocolumn]{aastex631}

\usepackage{apjfonts}
\usepackage{graphics,graphicx,float}
\usepackage{xfrac,amsmath} 
\usepackage{hyperref,enumitem}
\usepackage{color,multirow}

\newcommand\W{{$\lambda$}}
\newcommand{\CH}[1]{\colhead{#1}}

\newcommand\CM{{$\checkmark$}}

\graphicspath{{./figures/}}

\begin{document}
\accepted{January 8th, 2024}

\shortauthors{Clark et al.}
\shorttitle{Aperture Effects}

\title{Aperture and Resolution Effects on Ultraviolet Star-Forming Properties: Insights from Local Galaxies and Implications for High-Redshift Observations}\footnote{
Based on observations made with the NASA/ESA Hubble Space Telescope,
obtained from the Data Archive at the Space Telescope Science Institute, which
is operated by the Association of Universities for Research in Astronomy, Inc.,
under NASA contract NAS 5-26555.}

\author[0000-0003-3334-4267]{Ilyse Clark}
\affiliation{Department of Astronomy, The University of Texas at Austin, 2515 Speedway, Stop C1400, Austin, TX 78712, USA}
\affiliation{Center for Astrophysics \& Space Sciences, University of California San Diego, 9500 Gilman Dr, La Jolla, CA 92093, USA}

\author[0000-0002-4153-053X]{Danielle A. Berg}
\affiliation{Department of Astronomy, The University of Texas at Austin, 2515 Speedway, Stop C1400, Austin, TX 78712, USA}

\author[0000-0003-2685-4488]{Claus Leitherer}
\affiliation{Space Telescope Science Institute, 3700 San Martin Drive, Baltimore, MD 21218, USA}

\author[0000-0002-2644-3518]{Karla Z. Arellano-C\'{o}rdova}
\affiliation{Department of Astronomy, The University of Texas at Austin, 2515 Speedway, Stop C1400, Austin, TX 78712, USA}

\author[0000-0002-2090-9751]{Andreas A. C. Sander}
\affiliation{Astronomisches Rechen-Institut, Zentrum f{\"u}r Astronomie der Universit{\"a}t Heidelberg, M{\"o}nchhofstr. 12-14, 69120 Heidelberg, Germany}

\correspondingauthor{Ilyse Clark} 
\email{ilyseclark@gmail.com}


\begin{abstract}
We present an analysis of the effects of spectral resolution and aperture scales on derived galaxy properties using far-ultraviolet (FUV) spectra of local star-forming galaxies from the International Ultraviolet Explorer ($R\sim250$, FOV$\sim10\arcsec\times20\arcsec$) and Cosmic Origins Spectrograph on the Hubble Space Telescope ($R\sim15,000$, FOV$\sim$2\farcs5). Using these spectra, we measured FUV luminosities, spectral slopes, dust attenuation, and equivalent widths. We find that galaxies with one dominant stellar cluster have FUV properties that are independent of aperture size, while galaxies with multiple bright clusters are sensitive to the total light fraction captured by the aperture. Additionally, we find significant correlations between the strength of stellar and interstellar absorption-lines and metallicity, indicating metallicity-dependent line-driven stellar winds and interstellar macroscopic gas flows shape the stellar and interstellar spectral lines, respectively. The observed line-strength versus metallicity relation of stellar-wind lines agrees with the prediction of population synthesis models for young starbursts. In particular, measurements of the strong stellar \ion{C}{4} \W\W1548,1550 line provide an opportunity to determine stellar abundances as a complement to gas-phase abundances. We provide a relation between the equivalent width of the \ion{C}{4} line and the oxygen abundance of the galaxy. We discuss this relation in terms of the stellar-wind properties of massive stars. 
As the driving lines in stellar winds are mostly ionized iron species, the \ion{C}{4} line 
may eventually offer a method to probe $\alpha$-element-to-iron ratios in star-forming galaxies 
once consistent models with non-solar abundance ratios are available. 
These results have important implications for the galaxy-scale, low-resolution observations of 
high-redshift galaxies from JWST ($R\sim100-3,500$). 
\end{abstract}  


\section{Introduction}\label{sec:1}

The advent of space-based observations permitted the study of star-forming galaxies in the ultraviolet (UV)
where  newly formed stars reach the peak of their spectral energy distributions (SED). 
Pioneering studies of nearby star-forming galaxies were done with the International Ultraviolet Explorer (IUE), which operated between 1978 and 
1998 \citep{Macchetto76,Boggess78}. IUE employed a 45-cm telescope whose spectroscopic modes 
provided resolutions of 0.2 and 6~\AA, although only the latter proved useful for observations of galaxies 
due to the limited light-gathering power of the telescope. 
The IUE spectral range of 1150 to 3200~\AA\ was recorded along two separate optical paths below and above 1950~\AA. 
Both optical paths shared an obround entrance aperture of dimensions $10'' \times 20''$ 
(the small $3''$ diameter aperture was rarely used). 
After the mission concluded, the final processed science products were delivered to the Mikulski Archive for Space 
Telescopes (MAST), where they are available to the community for analysis \citep{Nichols96,Garhart97}.

Since the end of the IUE mission, the Hubble Space Telescope (HST) has been the dominant observatory for 
UV observations.
HST's current spectrographs, the Space Telescope Imaging Spectrograph (STIS) and Cosmic Origins Spectrograph 
(COS), have greatly expanded and improved upon the legacy of the IUE in the 1150 -- 3200~\AA\ wavelength range. 
The vast array of available entrance apertures, gratings, and improved sensitivity of these instruments have enabled 
transformative science not previously possible with IUE. 
Nevertheless, there is still scientific value in the IUE data set that has not yet been unlocked, even if the IUE's data 
quality cannot compete with that of HST in most aspects. 
Therefore, the goal of the present study is to compare IUE and HST spectroscopic data of nearby star-forming 
galaxies to gain insight into how the instrumental properties affect the interpretation of the data, as well 
as understanding the physical processes operating in these galaxies. 
This is of particular interest because of the limited lifetime of the HST. Exploring the capabilities of the IUE data will help identify the best use of future HST UV science in its limited lifetime, and inform interpretations of low-resolution rest-frame UV spectra of high-redshift galaxies taken with the James Webb Space Telescope (JWST).

\citet{Kinney93} published a comprehensive atlas of spectra of star-forming and active galaxies obtained with IUE 
while IUE was still accumulating new data and before the final data products were released in 1998.
\citet{heckman98} analyzed the data set of Kinney et al. to establish trends of various measurements with galaxy 
parameters. 
The present work builds on and extends these past works.
In particular, our study makes several improvements:
(i) New data are included that were not available to Heckman et al. 
(ii) The final data release permits a homogeneous analysis of the data. 
(iii) Available COS high-resolution spectra allow us to resolve and remove Galactic foreground absorption,
which was not feasible for Heckman et al. 
(iv) Synthetic galaxy spectra for quantitative analysis of the data have only become available after the work of 
Heckman et al., allowing a comparison with synthetic spectra in the present work.  

Almost all galaxies originally targeted by IUE have also been observed with HST's spectrographs in much 
greater detail and with superior quality. 
In this study, we will focus on data collected with COS whose light-collecting power is 
particularly well-suited for z~0 extragalactic spectroscopy.   
The present work provides three major scientific advancements in the UV spectral analysis of nearby galaxies:
(i) We constructed an atlas of nearby star-forming galaxies with both IUE and HST/COS spectra.
\citet{Leitherer11} previously published an atlas presenting spectra obtained with the HST first-generation 
UV spectrographs, the Faint Object Spectrograph (FOS), and Goddard High Resolution Spectrograph (GHRS), but no such 
atlas exists for the latest generation of spectrographs. 
(ii) We perform a comparative analysis between the IUE and COS spectra that allows us to differentiate galaxy 
properties on pc and kpc scales.
IUE's large $10'' \times 20''$ entrance aperture provides the large scale (kpc-scale),
integrated galaxy properties, while HST/COS' 2\farcs5 aperture probes smaller, pc-sized scales.  
(iii) The data quality of the IUE spectra in terms of signal-to-noise (S/N) and spectral resolution is inferior to 
that of HST spectra of local galaxies \citep[e.g.,][]{berg22} but is comparable to that of spectra in the high-redshift universe obtained with JWST. 
Recent JWST spectra of $z>7$ galaxies have already revealed the potential for deep galaxy evolution studies,
but their interpretation requires comparison to detailed analysis that is only possible in local galaxies
\citep[e.g.,][]{arellanocordova22,trump22}.
The results of this paper, therefore, provide guidance for the planning and interpretation of observations 
of galaxies close to the era of reionization. \looseness=-2

The remainder of the paper is organized as follows: 
We describe the archival UV and optical spectral observations used in this work
in Section~\ref{sec:2}. The sample selection and its properties are in Section~\ref{sec:2.1}. The processing of the IUE and HST/COS data is described in Sections~\ref{sec:2.2} and \ref{sec:2.3}, respectively. A general comparison of the two data sets is performed in Section~\ref{sec:2.4}. The ground-based optical spectra and the determination of metallicities are discussed in Section~\ref{sec:2.5}. Section~\ref{sec:3} covers our measurements. The determination of the $\beta-$slopes is in Section~\ref{sec:3.1}, followed by the reddening determinations in Section~\ref{sec:3.2}. Equivalent widths are derived in Section~\ref{sec:3.3}, with a comparison between IUE and COS in Section~\ref{sec:3.4}. In Section~\ref{sec:4} we study the relation of the derived properties with oxygen abundances. This is done for both the COS and the IUE samples in Section~\ref{sec:4.1} and \ref{sec:4.2}, respectively. In Section~\ref{sec:4.3} we interpret our results.
Finally, our conclusions are presented in Section~\ref{sec:5}.


\begin{deluxetable*}{lccccccclc}
    \setlength{\tabcolsep}{2pt}
     \tabletypesize{\normalsize}
     \tablewidth{0pt}  
     \tablecaption{UV Galaxy Sample Properties \label{tbl1}}
\tablehead{ 
\CH{}       & \CH{R.A., Decl.} & \CH{Morph.} & \CH{$E(B-V)_{\rm MW}$} & \CH{$v_{\rm rad}$} & \CH{$D$}& \CH{Ref.} & \CH{Scale} & \CH{12$+$} & \CH{COS}\\ 
[-2ex]
\CH{Galaxy} & \CH{(J2000)}   & \CH{Type} & \CH{(mag)}           & \CH{(km s$^{-1}$)} & \CH{(Mpc)} & \CH{}     
& \CH{$(\frac{\rm{pc}}{\rm{arcsec}})$} & \CH{log(O/H)} & \CH{Spec.}}     
\startdata
\ \ 1. SBS 0335-052   & 03:37:43.96, $+$05:02:39.6 & BCG    & 0.041 & 4053 & 53.90 & HF (1)     & 261 & 7.25$\pm$0.05 (1)  & \CM \\
\ \ 2. NGC 1705       & 04:54:13.41, $-$53:21:38.9 & SA0pec & 0.007 &  633 & 5.22  & TRGB (2)   & 25  & 8.01$\pm$0.03 (2)  & \CM \\
\ \ 3. NGC 1741       & 05:01:38.30, $-$04:15:25.0 & Pec    & 0.045 & 4039 & 54.60  & HF (1)     & 265 &  8.25$\pm$0.10 (3)  &     \\ 
\ \ 4. He 2-10	    & 08:36:15.96, $-$26:24:34.9 & I0     & 0.099 &  873 & 8.23  & TRGB (3)   & 40  & 8.55$\pm$0.02 (4)  & \CM \\
\ \ 5. IRAS 08339     & 08:38:23.15, $+$65:07:15.4 & Pec    & 0.083 & 5730 & 81.50 & HF (1)     & 395 & 8.42$\pm$0.07 (3)  & \CM \\
\ \ 6. I Zw 18	    & 09:34:02.00, $+$55:14:28.0 & cI     & 0.029 &  751 & 18.20 & TRGB (4)   & 88  & 7.19$\pm$0.06 (1)  & \CM \\
\ \ 7. NGC 3049       & 09:54:49.50, $+$09:16:16.0 &SB(rs)ab& 0.034 & 1455 & 24.1  & HF (1)   & 117   & 8.95$\pm$0.10 (5*)     &     \\
\ \ 8. NGC 3125	    & 10:06:33.37, $-$29:56:08.6 & S      & 0.068 & 1113 & 13.80 & HF (1)     & 67  & 8.30$\pm$0.02 (3)  & \CM \\
\ \ 9. NGC 3256	    & 10:27:51.30, $-$43:54:13.0 & Pec    & 0.108 & 2804 & 37.00 & HF (1)     & 179 & 8.77$\pm$0.09 (6*)  & \CM \\
10. Haro 2	        & 10:32:32.00, $+$54:24:02.0 & Im pec & 0.011 & 1430 & 25.50 & HF (1)     & 124 & 8.38$\pm$0.00 (7)  & \CM \\
11. NGC 3310        & 10:38:45.80, $+$53:30:12.0 &SAB(r)bc& 0.020 &  993 & 19.20 & HF (1)   & 93    & 8.83 $\pm$0.01 (8*)    &     \\
12. NGC 3351        & 10:43:57.70, $+$11:42:14.0 & SB(r)b & 0.025 &  778 & 9.64  & TRGB (5) & 47  &  8.82$\pm$0.01  (9*)     &     \\
13. NGC 3353	    & 10:45:22.40, $+$55:57:37.0 &Sb Pec & 0.006 &  944 & 18.50 & HF (1)     & 90  & 8.30$\pm$0.01 (10)  & \CM \\
14. UGCA 219	    & 10:49:05.00, $+$52:20:08.0 & Scp    & 0.011 & 2389 & 38.60 & HF (1)     & 187 & 7.86$\pm$0.04 (7)  & \CM \\
15. NGC 3690	    & 11:28:32.30, $+$58:33:43.0 & S pec  & 0.015 & 3121 & 48.50 & HF (1)     & 235 & 8.25$\pm$0.32  (11)  & \CM \\ 
16. NGC 3991	    & 11:57:31.10, $+$32:20:16.0 & Im pec & 0.019 & 3192 & 50.90 & HF (1)     & 247 & 8.50$\pm$0.15 (11)  & \CM \\
17. NGC 4194        & 12:14:09.50, $+$54:31:37.0 & Ibm pec& 0.014 & 2501 & 40.80 & HF (1)   & 198 &    8.82$\pm$0.02 (12)    &     \\  
18. NGC 4214	    & 12:15:39.20, $+$36:19:37.0 & IAB(s)m& 0.019 &  291 & 2.70  & TRGB (5)   & 13  & 8.36$\pm$0.10 (1)  & \CM \\
19. UGCA 281        & 12:26:15.90, $+$48:29:37.0  & Sm pec & 0.013 &  281 & 5.19  & TRGB (2) & 25  &   7.74$\pm$0.01 (13)    &     \\
20. NGC 4449	    & 12:28:11.10, $+$44:05:37.0 & Ibm    & 0.017 &  207 & 4.01  & TRGB (2)   & 19  & 8.31$\pm$0.07 (14) & \CM \\
21. NGC 4670	    & 12:45:17.10, $+$27:07:31.0 &SB(s)0/a& 0.013 & 1069 & 23.10 & HF (1)     & 112 & 8.38$\pm$0.10 (1)  & \CM \\
22. NGC 4861	    & 12:59:02.30, $+$34:51:34.0 & SB(s)m & 0.009 &  835 & 9.95  & TRGB (3)   & 48  & 8.01$\pm$0.05 (1)  & \CM \\
23. NGC 5236	    & 13:37:00.90, $-$29:51:56.0 & SAB(s)c& 0.059 &  513 & 4.80  & TRGB (6)   & 23  & 8.90$\pm$0.19 (15) & \CM \\
24. NGC 5253	    & 13:39:55.90, $-$31:38:24.0 & Im pec & 0.049 &  407 & 3.32  & TRGB (2)   & 16  & 8.19$\pm$0.04 (16) & \CM \\
25. NGC 5457        & 14:03:12.50, $+$54:20:56.0 &SAB(RS)cd& 0.007&  241 & 6.13  & TRGB (2) & 30  &   8.78$\pm$0.04 (17)     &     \\  
26. NGC 5996        & 15:46:58.90, $+$17:53:03.0 & SBc    & 0.030 & 3297 & 53.4  & HF (1)   & 259 &   9.08$\pm$0.14 (18*)     &     \\
27. TOL 1924-416    & 19:27:58.20, $-$41:34:32.0 & Pec    & 0.076 & 2834 & 42.40 & HF (1)     & 206 & 7.99$\pm$0.02 (4)  & \CM \\
28. NGC 7552	    & 20:16:10.70, $-$42:35:05.0 &(R')SB(s)ab&0.013&1608 & 22.50 & HF (1)     & 109 & 8.93$\pm$0.13 (6*)  & \CM \\    
29. NGC 7714	    & 23:36:14.10, $+$02:09:19.0 & SB(s)b & 0.046 & 2798 & 38.50 & HF (1)     & 187 & 8.26$\pm$0.10 (1)  & \CM
\enddata
\tablecomments{
Properties of the present sample.
The galaxies in this work are UV bright, nearby galaxies with high quality UV spectral observations from both
the {\it IUE} and {\it HST} and cover a range of properties.
Column~1 of this table gives the galaxy name. 
The R.A. and Decl. coordinates are listed in Column~2 and the morphological type is given in Column~3.
The Galactic foreground extinctions in Column~4 are taken from \citet{schlafly11} who utilized SDSS data to recalibrate the earlier dust-emission based values of \citet{Schlegel98}.
Heliocentric radial velocities ($v_{\rm rad}$) are listed in Column~5. 
Adopted distances for the sample are listed in Column~6, with corresponding distance determination method and reference
listed in Column~7.
Note that TRGB is used for the tip of the red giant branch method and HF is used for the Hubble flow. 
Column~8 in this table gives the resulting linear scales for each galaxy. 
The spatial scales covered by the COS and IUE entrance apertures, therefore, range from tens of pc to several kpc. 
Columns~9 and 10 list the luminosity and best measurement of the gas-phase oxygen abundance, with
corresponding references.
Finally, Column~11 indicates which galaxies also have COS spectra and are thus part of the IUE-COS Sample. \\
Distance references:
 (1) NED, 
 (2) \citet{sabbi18},
 (3) \citet{tully13},
 (4) \citet{aloisi07},
 (5) \citet{drozdovsky02} 
 (6) \citet{radburn-smith11}. \\
Metallicity references:
 (1) \citet{engelbracht08},
 (2) \citet{annibali15}, 
 (3) \citet{pena-guerrero17},
 (4) \citet{esteban14},
 (5) \citet{vacca92},
 (6) \citet{errozferrer19},
 (7) \citet{zhao10},
 (8) \citet{pastoriza93}
 (9) \citet{pilyugin14}
 (10) \citet{IzotovThuan04},
 (11) This work,
 (12) \citet{hancock06},
 (13) \citet{izotov97},
 (14) \citet{marble10},
 (15) \citet{bresolin16},
 (16) \citet{lopez-sanchez10},
 (17) \citet{berg20},
 (18) \citet{shirazi12}.
*~Metallicity derived using strong-line methods; see Section~\ref{sec:2.5} for more details.}
\end{deluxetable*} 


\section{UV \& Optical
 Spectral Observations}\label{sec:2}


\subsection{Sample definition and basic properties}\label{sec:2.1}

The Mikulski Archive for Space Telescopes (MAST) hosts data from numerous space missions focusing on the UV, optical, and near infrared, including the products of the IUE and HST. We queried MAST for existing UV spectroscopic data from both the IUE and HST missions. As IUE is the much more restricted data set in terms of number and quality of spectra, we initiated our query  with the IUE set, guided by the earlier works of \citet{Kinney93} and \citet{heckman98}. Next, we queried MAST for existing co-spatial HST/COS and HST/STIS spectra of all galaxies with IUE spectra in order to create a sample having both IUE and HST spectra. The science driver for this requirement is the goal to investigate line profiles in the HST data and compare spectral resolution effects between the IUE and HST data. The resulting 29 galaxies are hereafter referred to as the ``IUE Sample''. Of these 29 galaxies, four galaxies have COS M mode spectra, eight have STIS L mode spectra, and 17 have both COS and STIS spectra. COS M mode and STIS L mode have very different resolving power of $R \sim 15,000$ and $R \sim 1,000$, respectively. As our goal is to study the influence of spectral resolution and since $R \sim 1,000$ is insufficient for separating stellar and interstellar lines, we define a sample of galaxies having COS M mode spectra. As a result, we identify 21 galaxies with both IUE and co-spatial COS M mode spectra. We refer to this sample as the ``IUE-COS Sample''. We keep the larger IUE Sample in order to maximize the statistics for all IUE-related analysis, whereas all comparisons between IUE and COS are based on the IUE-COS sample. 

\begin{figure*}
    \includegraphics[width=0.975\textwidth,trim=0mm 0mm 0mm 0mm,clip]{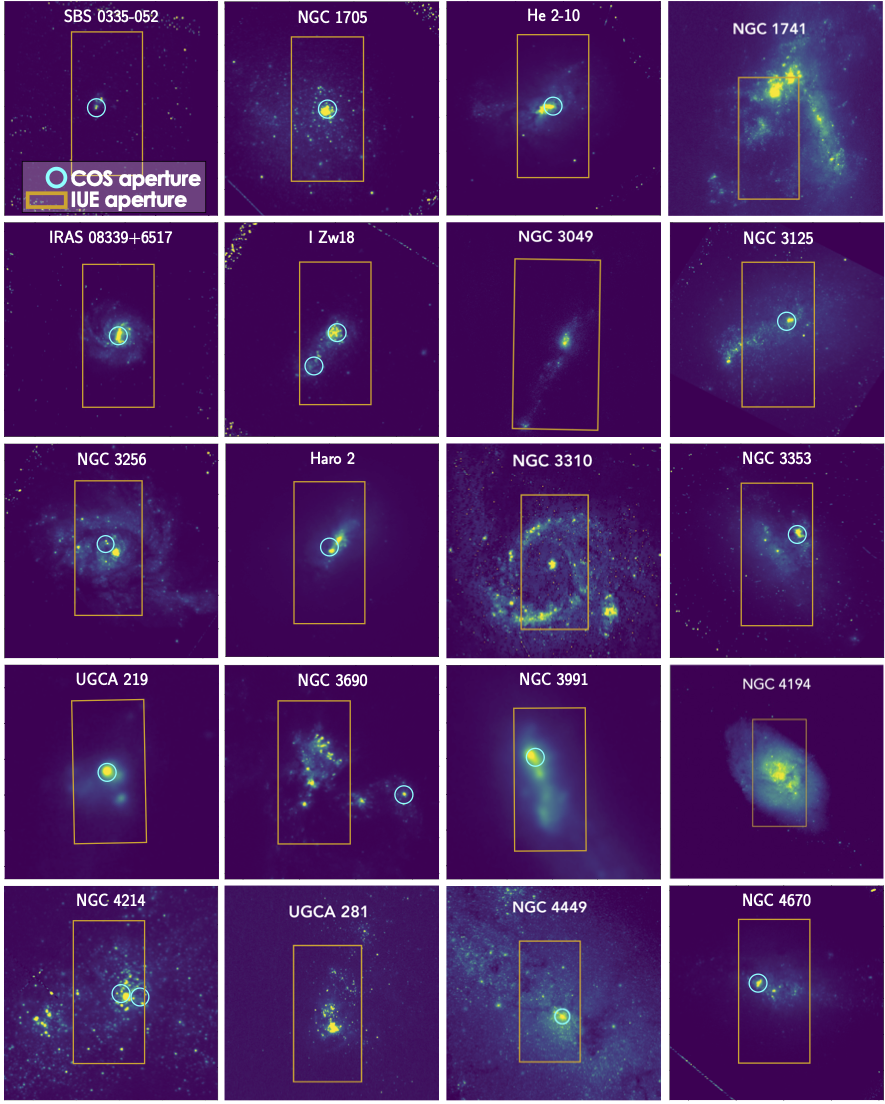}
\caption{Comparison of IUE and COS apertures for reference of the scale. The IUE aperture is represented by the dark yellow 20" x 10" rectangle, and the COS aperture by the 2.5" cyan circle in each image. North is up and east to the left. 
Note that we have chosen to display the IUE aperture vertically, but the true orientation is unknown. 
The instruments and filters for each image are:
SBS 0335-052: ACS/F220W,
NGC 1705: WFC3/F275W,
NGC 1741: ACS/F435W,
He 2-10: ACS/F330W,
IRAS 08339+6517: ACS/F140LP,
I Zw 18: ACS/F125W,
NGC 3049: STIS/F25SRF2,
NGC 3125: ACS/F330W,
NGC 3256: ACS/F140LP,
Haro 2: WFPC2/F336W,
NGC 3310: WFPC2/F300W,
NGC 3351:WFC3/F275W,
NGC 3353: WFPC2/F606W,
UGCA 219: SDSS/g,
NGC 3690: ACS/F125LP,
NGC 3991: FOC/F220W,
NGC 4194: STIS/F25QTZ,
NGC 4214: ACS/F125LP,
UGCA 281: WFC3/F275W,
NGC 4449: ACS/F125LP,
NGC 4670: ACS/F125LP}
\label{fig1}
\end{figure*}

A comparison of the aperture field of views for IUE and COS observations is shown for the full sample in Figure~\ref{fig1}. The IUE apertures have been drawn to indicate the field size; their orientations are arbitrary\footnote{The orientations are not known/provided in MAST or the data headers.}. 
In most cases, the data used in our study are the superposition of multiple spectra obtained at different orientations. 
Therefore the actual area covered can be thought of as a rotating IUE aperture. In almost all cases the COS pointings are near the center of the IUE aperture, and the orientation of the IUE aperture is unimportant. NGC~3690 is an exception: it is unclear whether the COS and IUE spectra are co-spatial. We decided to keep this galaxy in our sample to maximize statistics. 
We searched for archival images 
emphasizing UV wavelengths whenever available. The majority of the images in this figure were obtained with
HST's Advanced Camera for Surveys (ACS), STIS, Wide-Field and Planetary Camera 2 (WFPC2), and Wide-Field Camera 3 (WFC3).
Two galaxies (NGC 3991 and NGC 5996) have spherically aberrated HST Faint Object Camera (FOC) images. 
For UGCA 219 we used a Sloan Digital 
Sky Survey (SDSS) \textit{g}-band image. The image of NGC 5457 was obtained with the Jacobus Kapteyn Telescope (JKT) in the U-band.    

We compile the relevant basic properties for our galaxy sample from the literature and present them 
in Table~\ref{tbl1}.
This table gives important properties, including the galaxy numbering we use throughout this paper,
official galaxy names, the galaxy central coordinates, morphological types, and Galactic foreground extinctions.
The Galactic foreground extinctions were obtained using the recalibrations of \citet{schlafly11} and range between 0.01 and 0.11. 
The heliocentric radial velocities ($v_{\rm rad}$), which fall between 200 and 5700~km~s$^{-1}$, are
relevant for removing and/or deblending Galactic foreground absorption lines from intrinsic spectral lines.
We also performed a literature search in the NASA Extragalactic Database (NED) to obtain individual distance determinations for our galaxy sample. 
We prioritized high-quality distances, $D$, based on the tip of the red giant branch (TRGB) for all galaxies with $D < 13$~Mpc. 
Galaxies at larger distances were assumed to be in the Hubble Flow and, thus, their distances were derived from 
$v_{\rm rad}$ and the local velocity field model of \citet{Mould00} using the terms for the influence of the Virgo Cluster, 
the Great Attractor, and the Shapley Super Cluster. 


\setcounter{figure}{0}
\begin{figure*}
\includegraphics[width=0.975\textwidth,trim=0mm 0mm 0mm 0mm,clip]{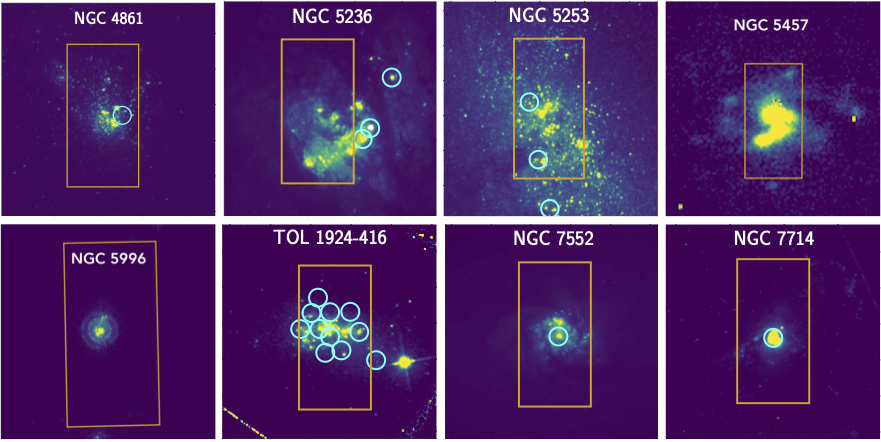}
\caption{Comparison of IUE and COS apertures for reference of the scale continued. The IUE aperture is represented by the dark yellow 20\arcsec$\times$10\arcsec rectangle and the COS aperture by the 2\farcs5 cyan circle in each image. North is up and east to the left. 
Note that we have chosen to display the IUE aperture vertically, but the true orientation is unknown.
The instruments and filters for each image are:
NGC 4861: STIS/F25SRF2,
NGC 5236: ACS/F125LP,
NGC 5253: STIS/F25QTZ,
NGC 5457: JKT/U,
NGC 5996: FOC/F220W,
TOL 1924-416: ACS/F122M,
NGC 7552: WFC3/F218W,
NGC 7714: WFC3/F300X.}
\end{figure*}


\subsection{\it{IUE} Spectra}\label{sec:2.2}
Here we present the IUE spectra for the galaxies listed in Table~\ref{tbl1}.
We retrieved the merged, extracted MXLO spectral files for the IUE Sample;
the properties of the observations are listed in Table~\ref{tbl2}.  
The MXLO files are the one-dimensional spectral tables extracted from the two-dimensional SILO image files 
processed with the NEWSIPS pipeline \citep{Garhart97}. 
The MXLO tables contain the absolutely calibrated fluxes, wavelengths, data quality flags, 
and Poisson error spectra. 
Spectra processed with NEWSIPS show an increase in S/N of 10\% -- 50\% in comparison with the 
previous IUESIPS processing \citep{Nichols96}. 
The retrieved spectra were resampled to equal wavelength steps of 1.2~\AA, i.e., 
five pixels within the nominal resolution of 6~\AA. 
When multiple co-spatial exposures exist, the individual spectra were coadded using a weighting 
factor scaled by the square root of the exposure time, i.e., by the Poisson noise. 
We combined the SWP and LWP/R (when available) exposures for the full far- and near-UV wavelength range. 
The short-wavelength end of the LWR spectra between 1950 and 2200~\AA\ have lower S/N due to the 
SiO$_{\rm 2}$ annihilation coating of the LWR optics \citep[Figure~13 of ][]{Bohlin80}. 
This spectral region has noticeably lower S/N in the combined spectra. 
The spectra were then truncated at starting and ending wavelengths of 1150~\AA\ and 3300~\AA, respectively. 
In cases when only SWP data were available, the long-wavelength truncation was set to 1980~\AA.


\begin{center}
\setlength{\tabcolsep}{2pt}
\begin{deluxetable}{lcc}
     \tabletypesize{\scriptsize}
     \tablewidth{0pt}
     \tablecaption{IUE Observations \label{tbl2}}
\tablehead{ 
\CH{Galaxy}         & \CH{SWP}            & \CH{LWP/LWR}}
\startdata
\ 1. SBS 0335-052   & 44070, 44075, 44078     & --                    \\ \hline
\ 2. NGC 1705       & 25906, 26187            & 3621, 3624, 5949, 6230\\ \hline
\ 3. NGC 1741       & 43481, 56125            & 31656                 \\ \hline
\ 4. He 2-10	    & 43458, 44132            & --                    \\ \hline
\ 5. IRAS 08339+6517& 26088, 35547            & 15031, 15239          \\ \hline
\ 6. I Zw 18	    & 6739, 7133, 27818, 27826&                       \\ 
                    & 27850, 27859, 27862     & 5743, 6150, 7746      \\ \hline
\ 7. NGC 3049       & 30593, 30927            & 10405                 \\ \hline
\ 8. NGC 3125	    & 13693, 13701            &  9895                 \\ \hline
\ 9. NGC 3256	    & 28041                   & --                    \\ \hline
10. Haro 2	        & 10471                   & 10111, 12578          \\ \hline
11. NGC 3310        & 13529, 16429, 17323     & 13568                 \\ \hline
12. NGC 3351        & 18628                   & 14695                 \\ \hline
13. NGC 3353	    & 10483                   & 10112                 \\ \hline
14. UGCA 219	    & 17314                   & --                    \\ \hline
15. NGC 3690	    & 18935,19341,44449       & 15387                 \\ \hline
16. NGC 3991	    & 7199, 9181              & 7939                  \\ \hline
17. NGC 4194        & 22784                   & 3412                  \\ \hline
18. NGC 4214	    & 13534, 33174            & 12941                 \\ \hline
19. UGCA 281        & 10484, 10495            &     --                \\ \hline
20. NGC 4449	    & 49911                   & 27313                 \\ \hline
21. NGC 4670	    & 13533, 32895            & 12640                 \\ \hline
22. NGC 4861	    & 5691, 5692              & 4937                  \\ \hline
23. NGC 5236	    & 17507, 46996            & 13787, 24968          \\ \hline
24. NGC 5253	    & 6066, 6084, 14542       & 5251, 5252, 10094     \\ \hline
25. NGC 5457        & 7422                    &     --                \\ \hline
26. NGC 5996        & 19219                   & 15208                 \\ \hline
27. TOL 1924-416    & 20200, 29517, 30962,    & 9401, 10750, 13306,   \\ 
                    & 33634, 34494, 37326     & 16158, 16566          \\ \hline
28. NGC 7552	    & 10696, 14453            & 9400                  \\ \hline  
29. NGC 7714	    & 928, 3953, 42048,       &                       \\
                    & 42156                   & 3499 
\enddata
\tablecomments{
For each galaxy, we list the identifiers of the retrieved spectra obtained with the Short Wavelength 
Prime (SWP), the Long Wavelength Prime (LWP) and the Long Wavelength Redundant (LWR) cameras. 
See MAST for a description of the SWP, LWP, and LWR cameras. 
All spectra are approximately co-spatial but do not necessarily have the same orientation.}
\end{deluxetable} 
\end{center}


\begin{center}
\setlength{\tabcolsep}{2pt}
\begin{deluxetable}{lccc}
     \tabletypesize{\scriptsize}
     \tablewidth{0pt}
     \tablecaption{HST/COS Observations  \label{tbl3}}
\tablehead{ 
\CH{Galaxy}         &\CH{Grating}&\CH{Setting}&\CH{Program}}
\startdata
\ 1. SBS 0335-052   & G130M & 1222       & 15193 \\
\                   & G160M & 1611, 1623 & 15193 \\
\                   & G185M & 1835       & 13788 \\ \hline
\ 2. NGC 1705       & G130M & 1222, 1291 & 13697 \\ \hline
\ 4. He 2-10	    & G130M & 1222, 1291 & 13697 \\ \hline
\ 5. IRAS 08339+6517& G130M & 1291, 1300 & 12173 \\ \hline
\ 6. I Zw 18-NW     & G130M & 1222, 1291 & 15193, 11523, 11579 \\
\                   & G160M & 1589, 1600, 1611, 1623 & 11523 \\
\                   & G185M & 1900, 1913, 1921 & 11523 \\
\ \ I Zw 18-SE      & G160M & 1611       & 13788 \\
\                   & G185M & 1817       & 13788 \\ \hline
\ 8. NGC 3125	    & G130M & 1300, 1318 & 12172 \\
\                   & G160M & 1600, 1623 & 15828 \\ \hline
\ 9. NGC 3256	    & G130M & 1291, 1300 & 12173 \\ \hline
10. Haro 2	        & G130M & 1222, 1291 & 13697 \\ \hline
12. NGC 3353	    & G130M & 1222, 1291 & 13697 \\ \hline
14. UGCA 219	    & G130M & 1222, 1291 & 13697, 11579, 15193 \\ \hline
15. NGC 3690        & G160M & 1623       & 15193 \\ \hline
16. NGC 3991	    & G130M & 1291       & 15840 \\
\                   & G160M & 1589       & 14120 \\
\                   & G185M & 1817       & 15840 \\ \hline
18. NGC 4214	    & G130M & 1291       & 11579 \\
\                   & G160M & 1623       & 15193 \\ \hline
20. NGC 4449	    & G130M & 1291       & 11579 \\
\                   & G160M & 1623       & 15193$^1$ \\ \hline
21. NGC 4670	    & G130M & 1291       & 11579 \\
\                   & G160M & 1623       & 15193 \\ \hline
22. NGC 4861	    & G130M & 1222, 1291 & 13697 \\\hline
23. NGC 5236-2      & G130M & 1291       & 14861 \\
\                   & G160M & 1623       & 14861 \\
\ \ NGC 5236-3      & G130M & 1291       & 14861 \\
\                   & G160M & 1623       & 14861 \\
\ \ NGC 5236-4      & G130M & 1291       & 14861 \\
\                   & G160M & 1623       & 14861 \\
\ \ NGC 5236-5      & G130M & 1291       & 14861 \\
\                   & G160M & 1623       & 14861 \\ \hline
24. NGC 5253-1      & G130M & 1222, 1291 & 11579, 15193 \\
\                   & G160M & 1623       & 15193 \\
\ \ NGC 5253-2      & G130M & 1222       & 11579 \\
\                   & G130M & 1291       & 15193 \\
\                   & G160M & 1623       & 15193 \\
\ \ NGC 5253-12     & G130M & 1291       & 16045 \\
\                   & G160M & 1600       & 16045 \\ \hline
27. TOL 1924-416-BA2& G130M & 1300       & 14806 \\
\ \ TOL 1924-416-BA3& G130M & 1300       & 14806 \\
\ \ TOL 1924-416-BA4& G130M & 1300       & 14806 \\
\ \ TOL 1924-416-BA5& G130M & 1300       & 14806 \\ \hline
28. NGC 7552	    & G130M & 1300, 1318 & 12173 \\ \hline
29. NGC 7714	    & G130M & 1291       & 12604 \\
\                   & G160M & 1589       & 12604 
\enddata
\tablecomments{
Summary of the {\it HST}/COS observations for the IUE-COS Sample presented in this work.
Columns~2 and 3 list the spectral grating and central wavelength used.
Column~4 lists the program IDs (PIDs) that correspond to the archival observations.\\
$^1$Note that the G160M grating was not used for NGC~4449 due to very low S/N.}
\end{deluxetable} 
\end{center}


\begin{figure*}
\begin{center}
    \includegraphics[width=1.0\textwidth,trim=40mm 5mm 40mm 0mm,clip]{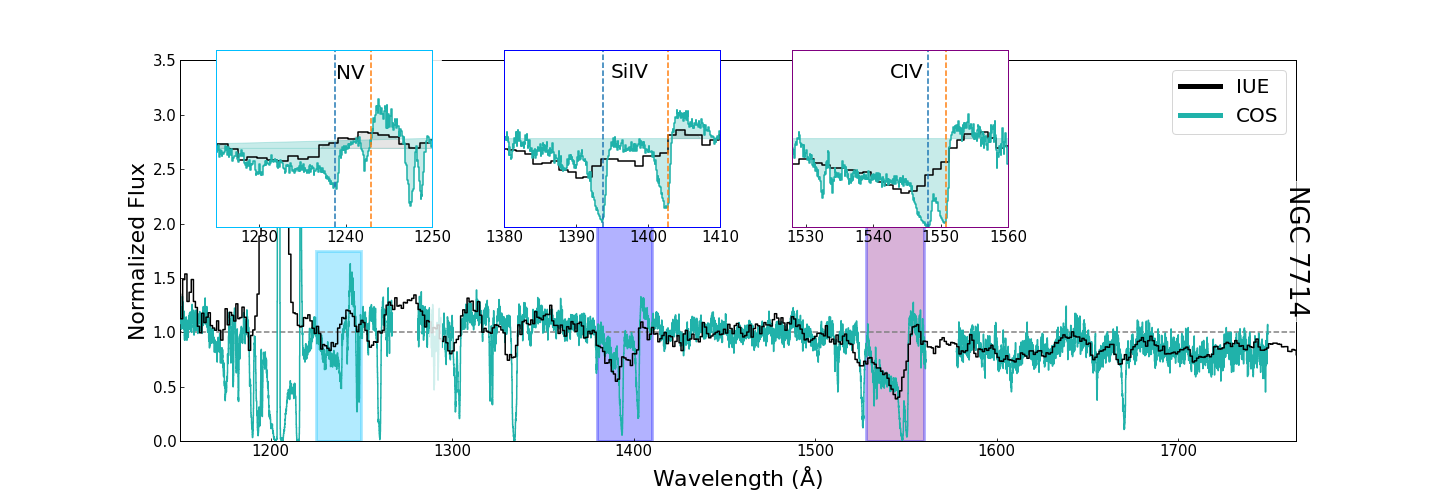}
\end{center}
\caption{
Comparison of IUE and HST/COS rest-frame FUV spectra for NGC 7714.
The IUE spectrum is relatively low resolution (\W/$\Delta$\W$\sim250$) and consists of the 
integrated light within a large aperture (10\arcsec$\times$20\arcsec).
In contrast, the COS spectrum has significantly higher spectral resolution 
(\W/$\Delta$\W$=15,000$) and higher S/N, but only within the much smaller 2\farcs5 COS aperture.
Both spectra are normalized at 1450~\AA.
\label{fig2}}
\end{figure*}


\subsection{\it{HST}/COS Spectra}\label{sec:2.3}
HST/COS spectra were retrieved from MAST for the 21 galaxies in the IUE-COS Sample, 
as indicated by check-marks in Table~\ref{tbl1}.
In particular, we retrieved the highest resolution data available, 
which comes from the medium-resolution gratings of G130M, G160M, and G185M, and
considered all COS apertures that are co-spatial with the IUE aperture.
The HST/COS datasets used in this work are summarized in Table~\ref{tbl3}. 
Note that NGC~5236, NGC~5253, and TOl~1924-416 have COS spectra from multiple pointings, 
not all of which are located within the IUE aperture. 
The pointings outside the IUE aperture were used for studying spectral variations across the 
surface of the galaxies, but not for comparison with the IUE data.  
G130M spectra exist for all galaxies in the IUE-COS Sample, 
while 10 of these galaxies also have G160M spectra, and three galaxies have G160M$+$G185M data. 

All raw data were reduced with the CALCOS pipeline (v3.3.10) using the standard twozone extraction technique.
However, each HST/COS grating has a different spectral resolution and different observations
are taken at different position angles and lifetime positions,
all of which must be accounted for when combining COS data. 
We address these issues by following the coaddition method laid out in \citet{berg22}.
In short, this method involves several steps of 
(1) joining the segments/stripes of individual grating datasets, 
(2) coadding any multiples of individual grating datasets, including all cenwave configurations, 
(3) coadding datasets across gratings, 
(4) binning the spectra by the nominal 6 pixels, and 
(5) correcting for Galactic contamination. 
While the initial flux calibration was performed for each dataset during the initial reduction by CALCOS, 
relative fluxing was also performed during the coadding process when more than one grating existed. 
To do so, the G160M spectrum was treated as the flux anchor of each spectrum and the continuum of 
all other datasets were fit and scaled to G160M at the intercept of their wavelength coverage 
(or average of the grating separation) when neighboring gratings overlaped (were disjoint) prior to coadding. 
Coadding steps used a combined normalized data quality weight (using the DQ\_WGT array; to filter out or 
de-weight photons correlated with anomalies/bad data) and exposure-time-weighted calibration curve 
(to preserve the Poisson count statistics). 
This weighting method was used for all instances where coadding was performed.
For further details of the coadding method, see \citet{berg22} and \citet{james22}.

Both the IUE and HST/COS spectra were corrected for Galactic foreground extinction using the values in Table~\ref{tbl1} 
and the extinction law of \citet{cardelli89}. 
Given the small values of $E(B-V)_{\rm MW}$, the corrections are rather minor. 
Finally, the spectra were transformed into the rest-frame using the heliocentric radial velocities 
listed in Table~\ref{tbl1}.


\subsection{Comparison of IUE and COS Spectra}\label{sec:2.4}
The IUE-COS Sample contains galaxies with both IUE and COS observations, 
allowing visual comparison of the datasets. 
We find the COS apertures focus on the star-forming knots in each galaxy, 
while the IUE probes the wider galactic environment. 
In more than half of our sample the IUE aperture fully covers the entire galaxy, 
including diffuse gas.

As an example, the HST/COS and IUE spectra for one galaxy in our sample, NGC 7714, are 
plotted in Figure~\ref{fig2}, showing substantial, but expected differences. 
The complete figure set showing the spectra of all 21 galaxies with IUE and
HST/COS spectra is available in the online journal.
The smaller aperture of COS relative to the IUE 10\arcsec$\times$20\arcsec aperture 
records lower UV continuum fluxes.
To account for this, we have normalized both spectra at 1450 \AA.
Now in direct comparison, the higher resolution of the HST/COS spectrum ($\sim0.1$ \AA\ 
resolution compared to the $\sim6$ \AA\ of the IUE spectrum)
reveals a significant number of features that are not present in the IUE spectrum, and 
also more complex line profiles.
Specifically, the \ion{N}{5} \W\W1238,1242, \ion{C}{4} \W\W1548,1550, and \ion{Si}{4} \W\W1393,1402 
lines are shown in the inset windows of Figure~\ref{fig2} to have combination profiles of stellar 
P-Cygni absorption$+$emission and extended ISM absorption.
These high-resolution profiles allow the stellar continuum to be fit and removed so that uncontaminated ISM absorption can be studied.

On the other hand, the high-resolution data permit studies of the stellar-wind profiles after the removal of the interstellar contribution. The comparison in Figure~\ref{fig2} demonstrates the need for sufficiently 
high spectral resolution when interpreting stellar and interstellar features. The stellar P~Cygni wind lines have a more or less significant interstellar contribution that must be accounted for when modeling the stellar population. Both Milky Way foreground and intrinsic lines may be important, depending on the ion. Lower ionization levels, such as \ion{Si}{4} have a stronger, or even dominant interstellar contribution, whereas higher levels, such as \ion{N}{5} or \ion{C}{4} are mostly shaped by stellar winds.

\begin{center}
\setlength{\tabcolsep}{4pt}
\begin{deluxetable*}{lCCcCCcCC}
     \tabletypesize{\small}
    \tablewidth{0pt}
    \caption{IUE and COS Spectral Properties\label{tbl4}}
\tablehead{
\CH{}               &  \multicolumn{2}{c}{log($f_{\lambda1500}$)} 
                    && \multicolumn{2}{c}{$\beta-$slopes}  
                    && \multicolumn{2}{c}{$E(B-V)$} \\
                    \cline{2-3} \cline{5-6} \cline{8-9} \\ [-3ex]
\CH{Galaxy}         & \CH{COS}      & \CH{IUE}      && \CH{IUE}       & \CH{COS}         && \CH{IUE}    & \CH{COS}}    
\startdata
\ \ 1. SBS 0335-052 & -14.16\pm0.42 & -13.94\pm0.42 && -2.289\pm0.132 & -2.188\pm0.133 && 0.111 & 0.091 \\
\ \ 2. NGC 1705     & -12.93\pm0.25 & -12.78\pm0.19 && -2.294\pm0.060 & -2.202\pm0.079 && 0.011 & 0.088 \\
\ \ 3. NGC 1741     & --            & -13.42\pm0.38 && -0.832\pm0.127 & -0.727\pm0.311 && 0.245 & -     \\
\ \ 4. He 2-10	    & -13.43\pm0.99 & -13.65\pm0.40 && -1.434\pm0.110 & --             && 0.372 & 0.403 \\
\ \ 5. IRAS 08339+6517
                    & -13.62\pm0.30 & -13.19\pm0.46 && -1.032\pm0.145 & -1.817\pm0.094 && 0.308 & 0.171 \\ 
\ \ 6. I Zw 18      & -13.91\pm0.13 & -13.73\pm0.18 && -2.493\pm0.058 & -2.463\pm0.038 && 0.086 & 0.033 \\
\ \ a. I Zw 18 NW   & -13.82\pm0.28 & -13.73\pm0.18 && --             & -2.816\pm0.088 && --    & --    \\
\ \ b. I Zw 18 SE   & -14.88\pm1.34 & -13.73\pm0.18 && --             & -1.647\pm0.423 && --    & --    \\
\ \ 7. NGC 3049     & --            & -13.75\pm0.38 && -1.284\pm0.121 & --             && 0.280 & --    \\
\ \ 8. NGC 3125	    & -14.34\pm0.68 & -13.35\pm0.47 && -0.730\pm0.149 & +0.444\pm0.215 && 0.342 & 0.652 \\ 
\ \ 9. NGC 3256	    & -14.95\pm2.39 & -13.49\pm0.40 && +0.143\pm0.125 & -0.216\pm0.753 && 0.610 & 0.512 \\
10. Haro 2	        & -13.81\pm0.95 & -13.30\pm0.47 && -1.182\pm0.148 & -1.210\pm0.298 && 0.300 & 0.300 \\
11. NGC 3310        & --            & -12.94\pm0.40 && -0.739\pm0.125 & --             && 0.365 & --    \\
12. NGC 3351        & --            & -13.61\pm0.41 && +0.048\pm0.128 & --             && 0.575 & --    \\
13. NGC 3353	    & -13.88\pm1.23 & -13.41\pm0.40 && -1.369\pm0.125 & -1.358\pm0.386 && 0.248 & 0.269 \\ 
14. UGCA 219	    & -13.95\pm1.43 & -13.46\pm0.45 && -2.442\pm0.142 & -2.492\pm0.450 && 0.067 & 0.026 \\
15. NGC 3690	    & -14.29\pm0.14 & -13.63\pm0.17 && -1.164\pm0.053 & -1.701\pm0.044 && 0.423 & 0.195 \\
16. NGC 3991	    & -13.86\pm0.70 & -13.23\pm0.41 && -2.019\pm0.129 & -2.080\pm0.220 && 0.123 & 0.114 \\
17. NGC 4194        & --            & -13.75\pm0.41 && -0.153\pm0.128 & --             && 0.526 & --    \\  
18. NGC 4214	    & -13.35\pm0.58 & -12.95\pm0.44 && -1.616\pm0.138 & -0.787\pm0.184 && 0.193 & 0.390 \\
19. UGCA 281        & --            & -13.65\pm0.41 && -2.091\pm0.129 & --             && 0.114 & --    \\
20. NGC 4449	    & -13.67\pm0.22 & -13.01\pm0.69 && -1.072\pm0.216 & -0.912\pm0.069 && 0.244 & 0.364 \\ 
21. NGC 4670	    & -13.68\pm0.15 & -13.17\pm0.40 && -1.449\pm0.127 & -1.447\pm0.048 && 0.232 & 0.250 \\
22. NGC 4861	    & -13.54\pm0.53 & -13.00\pm0.43 && -2.253\pm0.136 & -2.219\pm0.166 && 0.042 & 0.085 \\
23. NGC 5236	    & -13.59\pm0.14 & -12.44\pm0.41 && -0.804\pm0.130 & -1.372\pm0.043 && 0.384 & 0.266 \\
\ \ \ \ a. NGC 5236-2   
                    & -14.69\pm0.35 & -12.44\pm0.41 && --             & -1.246\pm0.111 && --    & --    \\
\ \ \ \ b. NGC 5236-3   
                    & -13.30\pm0.10 & -12.44\pm0.41 && --             & -1.495\pm0.033 && --    & --    \\
\ \ \ \ c. NGC 5236-4   
                    & -13.82\pm0.27 & -12.44\pm0.41 && --             & -0.696\pm0.084 && --    & --    \\
\ \ \ \ d. NGC 5236-5   
                    & -14.89\pm0.62 & -12.44\pm0.41 && --             & +0.598\pm0.195 && --    & --    \\
24. NGC 5253        & -14.32\pm1.41 & -12.69\pm0.42 && -1.197\pm0.131 & -1.031\pm0.444 && 0.295 & 0.338 \\
\ \ \ \ a. NGC 5253-1   
                    & -13.86\pm0.10 & -12.69\pm0.42 && --             & -1.767\pm0.031 && --    & --    \\
\ \ \ \ b. NGC 5253-2   
                    & -14.04\pm0.14 & -12.69\pm0.42 && --             & -2.356\pm0.043 && --    & --    \\
\ \ \ \ c. NGC 5253-12  
                    & -14.12\pm0.07 & -12.69\pm0.42 && --             & -1.718\pm0.023 && --    & --    \\
25. NGC 5457        & --            & -13.97\pm0.39 && -1.480\pm0.123 & --             && 0.269 & --    \\
26. NGC 5996        & --            & -13.72\pm0.42 && -0.654\pm0.133 & --             && 0.323 & --    \\
27. TOL 1924-416    & -13.94\pm0.46 & -13.18\pm0.41 && -2.057\pm0.129 & -1.987\pm0.146 && 0.095 & 0.134 \\
\ \ \ \ a. TOL 1924-416-BA2 
                    & -14.21\pm0.52 & -13.18\pm0.41 && --             & -2.211\pm0.164 && --    & --    \\
\ \ \ \ b. TOL 1924-416-BA3 
                    & -13.91\pm0.33 & -13.18\pm0.41 && --             & -2.631\pm0.104 && --    & --    \\
\ \ \ \ c. TOL 1924-416-BA4 
                    & -13.99\pm0.80 & -13.18\pm0.41 && --             & -1.850\pm0.253 && --    & --    \\
\ \ \ \ d. TOL 1924-416-BA5 
                    & -14.42\pm0.59 & -13.18\pm0.41 && --             & -1.989\pm0.186 && --    & --    \\
28. NGC 7552    	& -15.03\pm0.71 & -13.70\pm0.39 && +0.557\pm0.123 & +0.445\pm0.224 && 0.668 & 0.654 \\
29. NGC 7714    	& -13.40\pm0.06 & -13.28\pm0.41 && -1.181\pm0.128 & -1.257\pm0.166 && 0.297 & 0.290
\enddata
\tablecomments{
We compare properties derived from the COS and IUE spectra for each of the 29 galaxies.
Columns 2--3 show the COS and IUE continuum flux values measured at 1500 \AA. 
For galaxies with more than one COS aperture, we include the continuum for the resulting combined spectrum
as the default comparison to the IUE, as well as the continuum for each individual aperture, which are listed
below each galaxy. 
Columns 4--5 list the measured $\beta$-slope values and uncertainties derived from a boot-strap
Monte Carlo least squares linear fit to the COS and IUE stellar continua.
Stellar continuum windows free of contamination in the range of 1250 \AA\ $< \lambda < $ 1850 \AA\ 
were used to avoid the stellar continuum turnover at bluer wavelengths and possible contamination from 
the broad 2200 \AA\ dust bump, which can affect the continuum as blue as $\sim1850$ \AA.
Columns 6--7 list the E(B-V) measurements that were obtained using the relationship between $\beta$ and 
E(B-V) from \cite{reddy18} (see Equation \ref{eq1}) and the $\beta$-slopes from
Columns 4 \& 5.
\label{tbl8}}
\end{deluxetable*} 
\end{center}


\subsection{Ancillary Optical Spectra}\label{sec:2.5}

Optical spectra of star-forming galaxies can be used to derive the physical conditions 
of the nebular gas, the total chemical abundances, and current conditions such as star formation rate (SFR). 
In this work, we are particularly concerned with how UV spectral properties change as a function of
gas-phase metallicity, where the metallicity (or oxygen abundance) is used to trace the metal enrichment 
of the ionized gas as a proxy for galaxy evolution in star-forming regions.
Accurate oxygen abundances are derived via the so-called the direct method or $T_{e}$-method,
which requires the detection of an inherently-faint auroral line
(e.g., [\ion{O}{3}] \W4363, [\ion{N}{2}] \W5755, or [\ion{S}{3}] \W6312).
We searched the literature for oxygen abundances derived via the direct-method (or $T_{\rm e}$-sensitive method) for each galaxy. The compilation of the metallicities for the IUE Sample is included in Table~\ref{tbl1} and will be used for our interpretation of the observed UV spectra. In summary,  23 of the 29 galaxies in the IUE Sample have published direct-metallicity determinations. For two galaxies, NGC\,3390 and NGC\, 3991, we  used archival optical spectra  
from the SDSS to derive the metallicity. This provides the opportunity to improve the measurements of the metallicity of these galaxies previously derived from strong-line methods \citep{heckman98}. The SDSS spectra of NGC\,3690 and NGC\,3991 show the presence of [\ion{N}{2}]~\W5755 and [\ion{S}{3}]~\W6312, respectively, allowing the computation of the electron temperature. Therefore, we have determined the physical conditions and metallicity following the procedure described in \citet{arellanocordova22}, which includes the correction of Galactic extinction, the subtraction of the underlying population, the emission line fitting, and the reddening correction using the \citet{cardelli89} reddening law. For a summary of this procedure, we have fitted the emission lines with Gaussian profiles using the \texttt{Python} package \texttt{LMFIT}\footnote{Non-Linear Least-Squares Minimization and Curve-Fitting: \url{https://github.com/lmfit/lmfit-py}}. We constrained the offset from line centers and the line width. To calculate the flux error, we used the expression reported in \citet{berg13} and \citet{Rogers21}. Finally, to derive the electron density and temperature, and metallicity, we use the \texttt{PyNeb} package (version 1.1.14) \citep{luridiana15} with the atomic data also reported in \citet{arellanocordova22}.

For the six remaining galaxies NGC\,3049, NGC\, 3256, NGC\, 3310, NGC\, 3351, NGC\, 5996 and NGC\,7552 we need to rely on alternative methods to derive the metallicity using the emission spectra reported in the literature. For three of those galaxies the metallicity was determined using the calibration of \citet{dopita16}. Three galaxies have COS spectra available (NGC\,3049, NGC\,3256, and NGC\,7552). This calibration is based on photoionization models and uses a linear fit between [\ion{N}{2}]~\W6584 and [\ion{S}{2}]~\W\W6717, 6731 to estimate the metallicity. 

For NGC\,3310, NGC\,3351, and  NGC\,5996, we report metallicities derived using the C-method of \citet{pilyugin12}, the combination of the [\ion{N}{2}]~\W6584/H$\alpha$ (N2) and [\ion{O}{3}]~\W5007/H$\beta\times$ H$\alpha$/[\ion{N}{2}]~\W6584 (O3N2) methods of \citet{pettini04}, and the  method  using [\ion{O}{2}] \W3726,29, [\ion{O}{3}] \W5007 and H$\beta$ emission lines ($R_{23}$) of \citet{kobulnicky99}, respectively. The resulting metallicities of all galaxies are listed in Table~\ref{tbl1}; they cover a range of $7.19<$ 12+log(O/H) $< 9.08$. 

Determining a characteristic metallicity of massive galaxies is challenging.
However, \citet{moustakas06b} showed that the metallicities inferred from integrated spectra of disk
galaxies correlates well with the characteristic gas-phase abundance, as determined by the \ion{H}{2} 
region abundance measured at 0.4$R_{25}$\footnote{$R_{25}$ is the radius of the major axis at a 
surface brightness of 25 mag arcsec$^{-2}$ in the $B-$band.}.
Moreover, 86\% of our sample with COS and IUE spectra are dwarf galaxies with a relatively homogeneous spatial distribution of metals within ~1~kpc scales \citep{anto22}.
Thus, we assume the integrated abundances adopted here are representative of our galaxy sample
and allow for a safe comparison with other galaxies properties.


\section{UV Spectral Measurements}\label{sec:3}

Now that we have established the different physical scales and spectra resolution probed 
by the IUE and HST/COS spectra, we can begin to investigate their effects on properties
measured from the UV spectra.
Below we describe our uniform measurements of the UV $\beta-$slope and absorption
feature equivalent widths (EWs).

\subsection{Beta-Slope Measurements}\label{sec:3.1}
An important property in characterizing UV spectra is the slope of the FUV stellar continuum, in a given wavelength interval,
otherwise referred to as the $\beta-$slope.
The $\beta-$slope is only weakly sensitive to the {\em stellar} properties of a young population, whose spectral energy distribution is in the Raleigh-Jeans regime at these wavelengths. In contrast,
dust attenuation strongly affects the UV continuum. Therefore, the $\beta-$slope is commonly used to derive the dust reddening.
Additionally, the $\beta-$slope has also been
theoretically predicted to correlate with the escape of ionizing continuum photons 
\citep{zackrisson13}, which was recently observationally confirmed by \citet{chisholm22}.

Typically, the $\beta-$slope is measured over a sufficiently wide UV wavelength range, 
centered at a wavelength around 1500~\AA.
Assuming a power-law model fit to the continuum such that
$F_\lambda \propto \lambda^\beta$, we measure the $\beta-$slope using a
least-squared first-degree polynomial fit to log-wave versus log-flux data.
Specifically, we use the featureless continuum-windows 
recommended by \citet{calzetti94} of
1268--1274, 1309--1316, 1342--1371, 1407--1420, 1563--1583,
1677--1740, 1760--1833, 1866--1890, 1930--1950, and 2400-2580 \AA\ to mask out
undesirable portions of the spectra.
Given that most of our COS spectra only have G130M coverage,
most of the COS spectra are fit with the first four windows only. 
Note that it is important to exclude continuum blueward of 1250 \AA\ 
in the $\beta-$slope fit because this wavelength regime can probe the peak flux 
(and thus turnover) of massive stars in the FUV. 
On the other hand, wavelengths longward of $\sim1800$ \AA\ should also be excluded for metal-rich
galaxies where there can be significant contributions from the broad 2200 \AA\ dust feature. Longward of 2500~\AA\ additional emission from older, less massive stars may contribute, and the $\beta-$slope may become sensitive to the details of the star-formation history. 
Therefore the wavelength range between 1250~\AA\ and 1850~\AA\ is the ``sweet spot'' for determining 
the $\beta-$slope.  
We use a bootstrap Monte Carlo method with 3000 iterations 
of adding a normal distribution of the error fluctuations to the observed data.
The final $\beta-$slope fits to the COS and IUE spectra are included in the plots shown 
in Figure~\ref{fig2} (see online version for the complete figure set).

The derived $\beta$-slope values for both the IUE and COS spectra are listed in Table~\ref{tbl4}.
We test the reliability of the IUE measurements by comparing our measurements with 
results published in the literature. 
\citet{Kinney93}, \citet{calzetti94}, and \citet{heckman98} measured $\beta$ in samples 
with  significant overlap with our sample. 
We calculated the mean differences between the $\beta$-slope values measured by them and by us for the common 
galaxies and found $\Delta\beta$ (this work)-(literature) = $-0.213 \pm 0.096$, $0.108 \pm 0.059$,
and $0.098 \pm 0.249$ for the differences with Kinney, Calzetti, and Heckman, respectively. 
These small differences suggest that our measurements are consistent with the previous studies and are, therefore, robust. 
The resulting $\beta-$slopes range from $-2.6$ to $+0.5$, which is consistent with our sample 
containing young stellar populations with dust attenuation. 

\begin{figure*}
    \centering
    \begin{tabular}{cc}
    \includegraphics[height=3.15in,trim=23mm 7mm 198mm 10mm,clip]{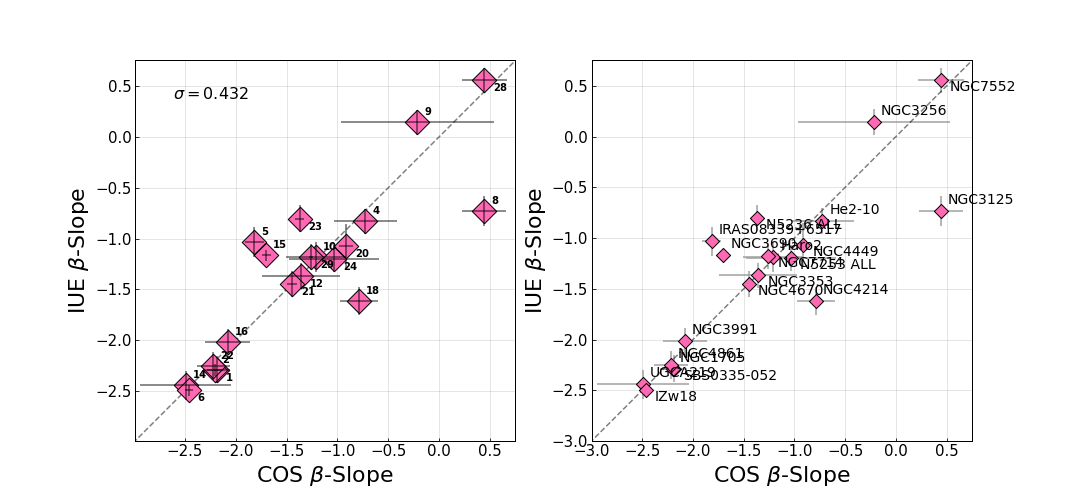} &
    \includegraphics[height=3.15in,trim=0mm 7mm 1mm 10mm,clip]{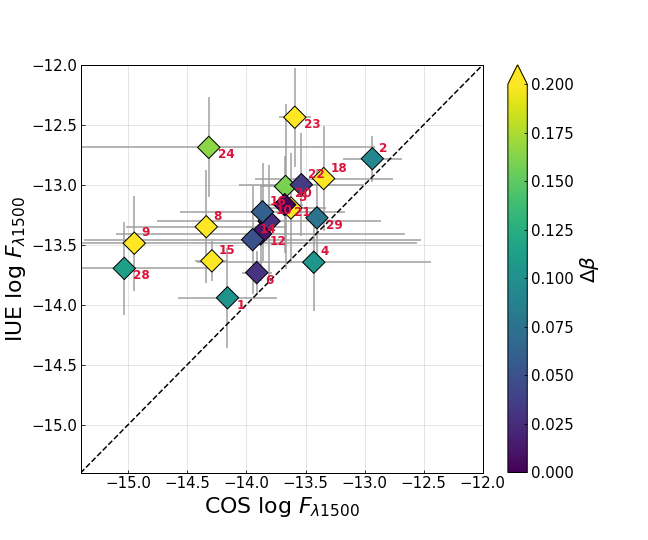}
    \end{tabular}
      \caption{ 
  {\it Left:} Comparison of the IUE and HST/COS $\beta-$slope measurements. 
  There is relatively little scatter between the two sample despite their large
  differences in aperture sizes.
  {\it Right:} The flux at 1500 \AA\ in units of $10^{-15}$ ergs s$^{-1}$ cm$^{-2}$ \AA$^{-1}$,
  for both the IUE and HST COS spectra.
  As expected, the integrated UV flux through the IUE apertures is larger 
  than for the physically smaller COS apertures. The color coding indicates the differences between $\beta$ measured for the IUE and the HST/COS data. Note that all data with $\Delta\beta > 0.2$ have yellow colors. The labels next to the data points are the galaxy identifiers used in Table~\ref{tbl1}.
  \label{fig3}}
\end{figure*}

We compare the IUE and COS $\beta-$slopes in the left panel of Figure~\ref{fig3}.
In general, the two samples are consistent within the errors of the measurements,
with a standard deviation of $\sigma = 0.43$.
We find that $\beta-$slopes measured from the bluest windows only are nearly 
identical across most of the sample. 
This suggests that we can measure robust $\beta-$slopes, regardless of the aperture size,
when considering only the youngest stellar populations 
(which are best probed by the bluer FUV wavelengths).
This is because the COS apertures are centered on the young clusters, and the full light of 
these clusters are easily contained within the IUE aperture.

There are a few strong outliers from the $\beta_{\rm IUE}$ vs $\beta_{\rm COS}$ trend
in Figure~\ref{fig3}, namely (5) IRAS~08339+6517, (8) NGC~3125, (15) NGC~3690, 
(18) NGC~4214, and (23) NGC~5236.
The COS spectra for both (5) IRAS~08339+651 and (15) NGC~3690 show bluer slopes. Because IRAS~08339+651 is the most distant galaxy in our sample,  the IUE aperture captures the extended light profile of the galaxy, while the COS aperture contains the bright center. Thus, the physical regions covered by IUE and COS are among the most divergent. As for NGC~3690, the COS pointing is only in one of the star-forming knots outside of the other bright regions in the center of this galaxy. This offset of the COS aperture towards the edge of NGC~3690 may explain the different $\beta$--slope we obtain for this galaxy.
The coadded COS spectrum of (23) NGC 5236 includes four individual
pointings that probe significantly different $\beta$-slopes ($\Delta\beta \approx 2.1$).
On the other hand, the COS spectra for (8) NGC 3125 and (18) NGC 4214 seem to have 
redder slopes than their corresponding IUE spectra.
The COS and IUE spectra for NGC 4214 are somewhat different visually but agree within the 
uncertainties of the two spectra.
However, the COS and IUE spectra for NGC 3125 differ by a larger amount.
Given that we do not know the true orientation of the IUE aperture, it seems likely 
that the IUE aperture for NGC 3125 shown in Figure~\ref{fig1} was actually rotated 
$\sim45~\deg$ clockwise such that it captured multiple young star-forming clusters 
and resulting in a bluer integrated continuum slope.

In the right panel of Figure \ref{fig3} we show a comparison of the average flux at 1500~\AA, as determined from our $\beta$-slopes using $y = \beta\times x + \alpha$. 
The data for both COS and IUE are listed in Table~\ref{tbl4}. 
Despite the general agreement in $\beta-$slopes, the relative fluxes between the COS and IUE
spectra are significantly biased to higher values in the IUE spectra.
This trend is expected given that the area encompassed by the IUE aperture is 
40$\times$ larger than that of COS, and so collects much more light from these extended galaxies.
This skew is consistent with our assertion that the IUE spectra are capturing most of the 
light from the galaxies in our sample. 
The points that deviate the most from the 1:1 line also show the largest differences between IUE and COS $\beta$-slope measurements ($\Delta\beta$). This suggests a trend of increasing $\Delta\beta$ for larger differences of $F_{\lambda 1500}$.

\begin{figure*}
\begin{center}
    \includegraphics[width=0.975\textwidth,trim=30mm 10mm 30mm 10mm,clip]{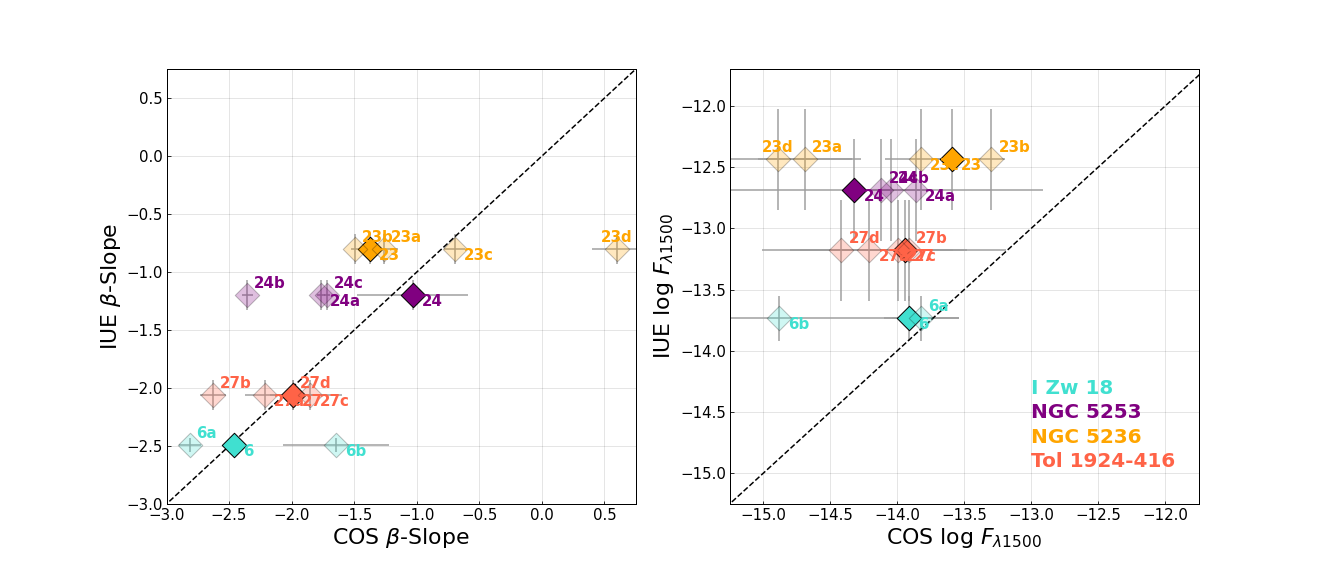} 
    \caption{
    Comparison of the stellar continuum properties measured from IUE spectra versus HST/COS 
    spectra for galaxies with multiple COS pointings. 
    {\it Left:} Comparison of $\beta-$slopes. 
    The points labeled by a number only are the coadded spectra of all the COS pointings from an 
    individual galaxy. 
    The points labeled by a number$+$letter are the individual COS pointings. 
    All points are color-coded by the galaxies as identified in the insert to the right panel.
    Individual COS pointings show significant scatter in their $\beta-$slopes, 
    but the coadded COS spectra all agree more closely with the IUE $\beta-$slope values.
    {\it Right:} Comparison of the flux at 1500~\AA\ is in units of $10^{-15}$ erg s$^{-1}$ 
    cm$^{-2}$ \AA$^{-1}$.
    Similar to Figure~\ref{fig3}, the larger aperture of the IUE spectra results in 
    larger continuum fluxes at 1500 \AA.
    \label{fig4}}
\end{center}
\end{figure*}

In Figure \ref{fig4} we plot the $\beta-$slopes versus $F_{\lambda 1500}$ values for the galaxies 
with multiple COS aperture measures (I~Zw~18, NGC~5236, NGC~5253, and TOL~1924-416)
to investigate how this trend varies between stellar clusters of a
given galaxy.
We plot the individual measurements as semi-transparent points and the measurements 
from the coadded spectra as solid points. 
Overall we see that the individual $\beta-$slopes show a large range (up to $\Delta\beta \approx 2$), and
so fall far off the 1:1 relation with the IUE measurements. 
On the other hand, the measurements from the coadded spectra are in much closer agreement with 
the IUE measurements for all four galaxies.
This aligns with our previous analysis of the $\beta$-slopes and highlights that the IUE slopes are different due to the separate stellar clusters going into the integrated light.
In the right hand panel of Figure~\ref{fig4} we plot the continuum flux at 1500~\AA\ 
for the four galaxies with multiple HST/COS apertures and this again skews 
towards the high IUE values, as expected.


\subsection{Stellar Reddening}\label{sec:3.2}
As discussed above, the slope of the UV continuum is primarily sensitive to dust reddening.
Therefore, we use $\beta$-slope to derive the dust reddening for both the COS and IUE spectra and evaluate any differences between them.
We determine the stellar reddening with two methods:
(1) using empirical relationships between $\beta$-slopes and $E(B-V)$ and
(2) using stellar population synthesis (SPS) models to fit the continuum.
We use two different empirical relationships. 
First, we use the fit derived by \citet{reddy18}:
\begin{equation}
    E(B-V) = 0.558 + 0.215\times\beta,
    \label{eq1}
\end{equation}
and then use the \texttt{BPASS} continuous star formation models with the $\beta$-slope 
values derived in Section~\ref{sec:3.1}.
We also use the relationship derived by \citet{chisholm22} using a linear combination of single-burst \texttt{Starburst99} models
\begin{equation}
    E(B-V) = 0.470 + 0.171\times\beta.
    \label{eq2}
\end{equation}
Note that both relations were originally derived for galaxies at low and high redshift using data from COS, but it is important to note that in Equation \ref{eq2} the galaxies used we further away so the apertures contained all the light from the galaxy. Therefore, the assumed attenuation laws used in these works are applicable when both dust absorption and scattering are important. This may not always be the case for very nearby galaxies that fill the $2.5''$ COS aperture. However, for the sake of consistency, we opt against switching between different attenuation laws for different galaxies.  

For the second method, we split our sample into two groups: 
metal-poor galaxies with 12+log(O/H) $<$ 8.40 and 
metal-rich galaxies with 12+log(O/H) $\geq$ 8.40.
We then create a grid of SPS models, adopting a pair of \texttt{Starburst99} models \citep{leitherer14}, 
one metal-rich ($Z=0.014$ or $Z_{\odot}$) and one metal-poor ($Z=0.002$ or $0.14 Z_{\odot}$), and applying the
reddening curve of \citet{calzetti00} with a range of $E(B-V)$ values between 0 and 1.
We resample the 
\texttt{Starburst99} models to match the dispersion of the IUE and COS spectra. 
Once this is done, we perform a $\chi^2$ minimization between the model and observed
spectra, using windows that only contain continuum, and adopt the $E(B-V)$ value
that produced the smallest value.

\begin{figure}
\begin{center}
    \includegraphics[width=0.45\textwidth,trim=1mm 1mm 1mm 1mm,clip]{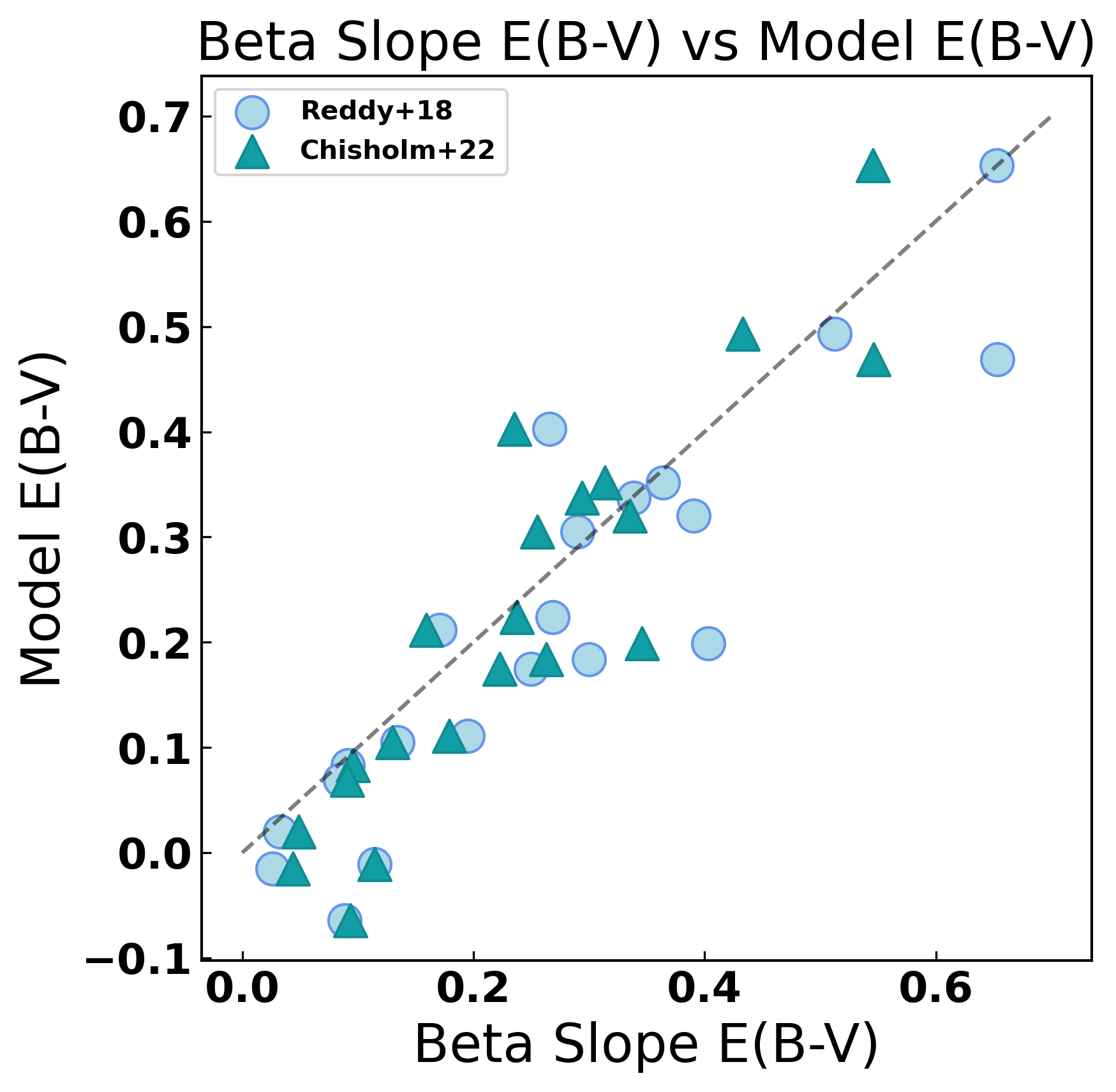} 
    \includegraphics[width=0.45\textwidth,trim=1mm 1mm 1mm 1mm,clip]{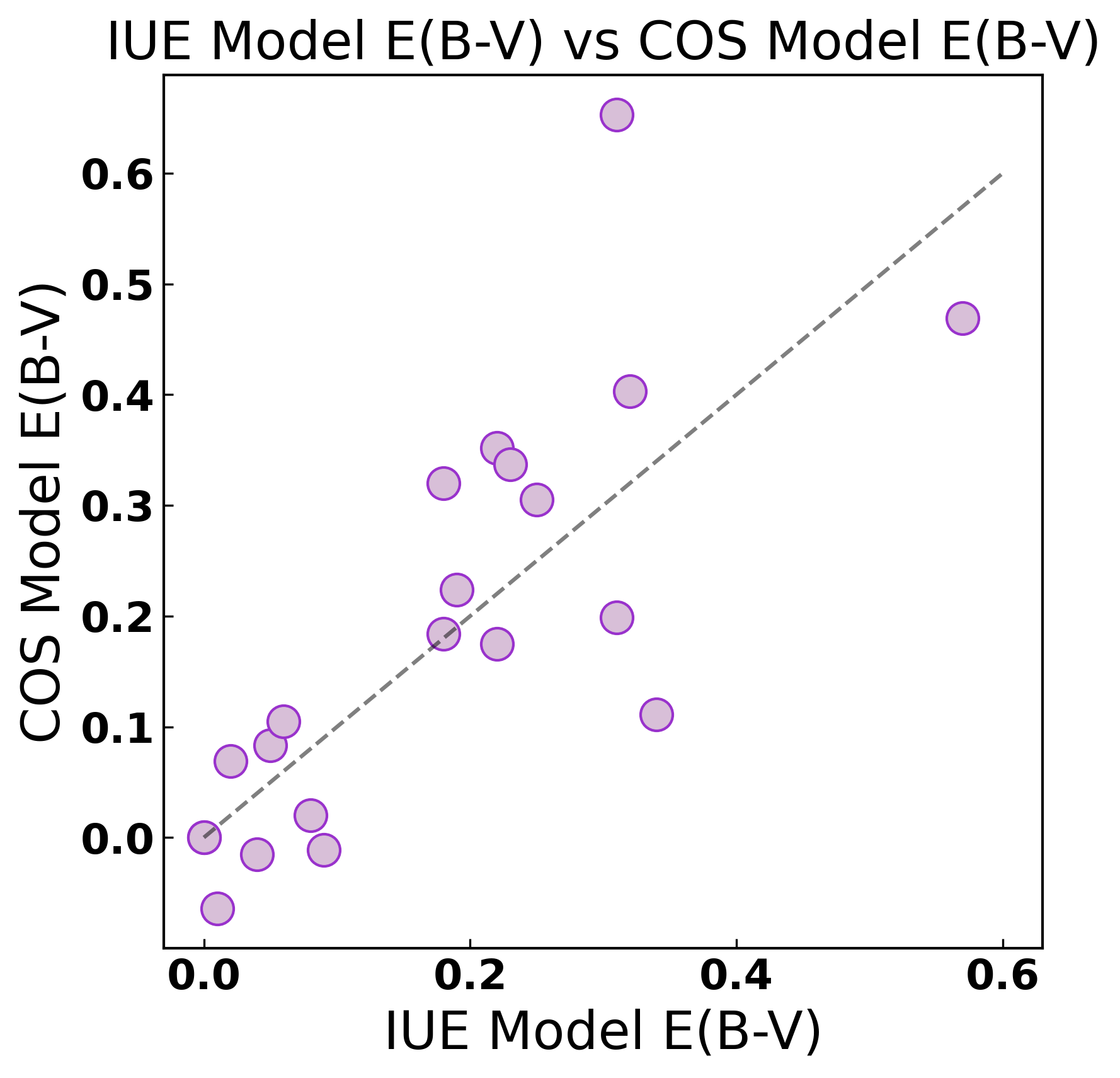} 
\caption{
{\it Top:} $E(B-V)$ values derived for the COS spectra from the 
SPS continuum-fitting method (Model) versus the empirical $\beta-$slope method. "The $\beta$-slope values derived using \citet{reddy18} are plotted as light blue circles and the values derived using \citet{chisholm22} are plotted as teal triangles.
The two methods
show a tight agreement, demonstrating that these methods are consistent. 
{\it Bottom:} Comparison of the SPS continuum-fitting derived $E(B-V)$ values 
for the COS spectra versus the IUE spectra.
The trend is consistent with a 1:1 relationship but
shows some scatter between the two apertures.
\label{fig5}}
\end{center}
\end{figure}

In Table~\ref{tbl4} we summarize the results found from the continuum analysis of the IUE and COS spectra. The values in this table are derived by using Eq.~(\ref{eq1}).
A comparison of the resulting $E(B-V)$ values for the COS spectra from the empirical 
$\beta-$slope method (column~7 of Table~\ref{tbl4}) and the SPS continuum-fitting method is shown in the top panel 
of Figure~\ref{fig5}. This figure shows the $E(B-V)$ values obtained with both Eqs.~(\ref{eq1}) and (\ref{eq2}). Only the former values are listed in Table~\ref{tbl4}. We list only one set of measurements because there is very little difference between the sets of values. The trends in this figure indicate that all three of the $E(B-V)$ determinations used in this work are consistent. We find a strong 1:1 correlation between the $\beta$-slope and model $E(B-V)$, suggesting the two methods are equivalent and relatively insensitive to the model assumptions of each method.

The bottom panel of Figure~\ref{fig5} shows the comparison of the $E(B-V)$ values derived from SPS continuum-fitting for the COS and IUE spectra.
The overall trend is still in 1:1 agreement, but with a significantly larger scatter
of $\sigma = 0.116$. This again reflects the effects of the different aperture sizes collecting light from different galaxy areas and, thus, the spectra having different shapes. Figure~\ref{fig5} (bottom) mirrors the trend in Figure~\ref{fig3} (left).
If the dispersion is in fact due to physical differences in the spatial distribution of the dust and/or stellar populations
observed, then the dispersion informs the potential uncertainty in $E(B-V)$
values derived with different apertures.


\subsection{Equivalent Width Measurements}\label{sec:3.3}

In this section, we test how the wavelength resolution and aperture size differences between the IUE and COS spectra affect our ability to characterize the strengths of ISM and stellar spectral features. To do so, we measure EW values of 11 different spectral lines (when available):
\ion{S}{2} \W1253, \ion{Si}{2} \W1260, \ion{O}{1} \W1302, \ion{C}{2} \W1335, 
\ion{Si}{4} \W\W1393,1402, \ion{C}{4} \W\W1548,1550, \ion{Fe}{2} \W\W1608,1611, 
and \ion{Al}{2} \W1671. 
Since the IUE spectra have lower spectral resolution than those of COS,
it is often not possible to disentangle absorption lines that are close together, including contamination from Milky Way lines in the lowest-redshift galaxies. We correct the IUE absorption line measurements for Milky Way contamination using a hybrid approach. First, we measure the EW of the Milky Way foreground lines in the COS spectra (where they are sufficiently separated) and then subtract them from the corresponding IUE measurement. This hybrid approach has the distinct advantage of being able to improve our measurements from low-resolution spectra. We correct for this by measuring the EW of the Milky Way foreground lines in the COS spectra and then subtracting that value in the corresponding IUE measurement. This is an advantage of the hybrid approach, with high- and low-resolution spectra that we are using in this work. 

\begin{figure}
\begin{center}
    \includegraphics[width=0.45\textwidth,trim=0mm 0mm 0mm 0mm,clip]{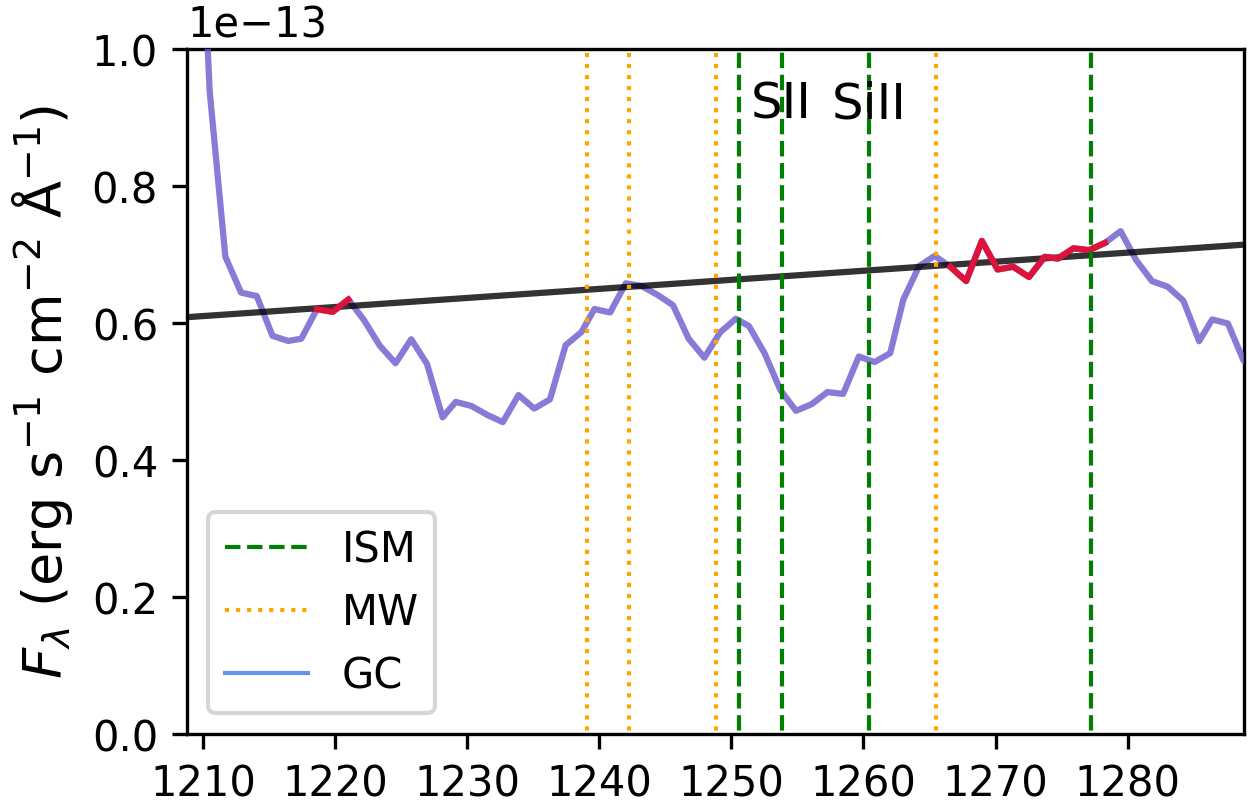} \\
    \includegraphics[width=0.45\textwidth,trim=0mm 0mm 0mm 0mm,clip]{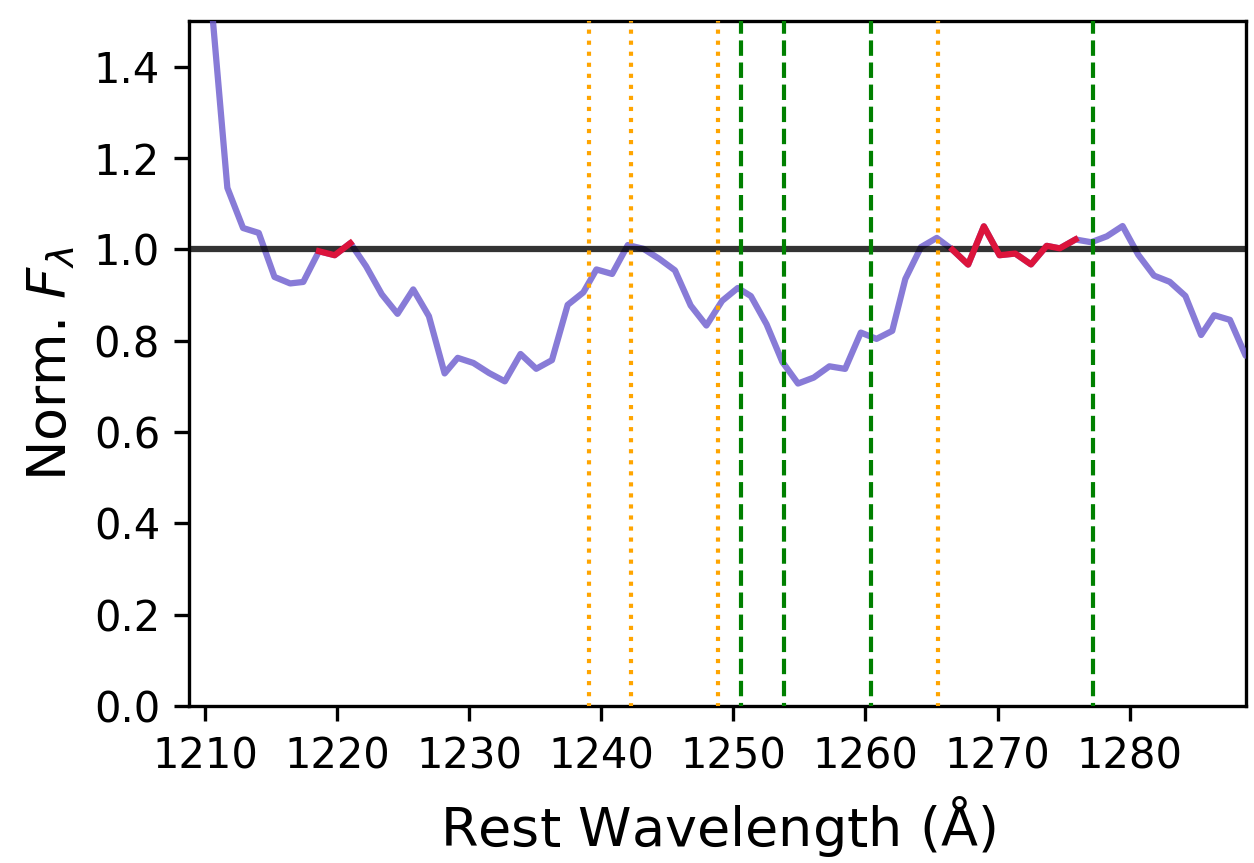}    
\caption{
A demonstration of the continuum normalization process for the IUE spectrum of NGC 7714. 
The top panel shows the initial unnormalized spectrum around \ion{S}{2} \W1253 and \ion{Si}{2} \W1260.
The red portions of the spectra are the windows used to fit a linear continuum around each absorption 
feature, with the resulting fit plotted in black. 
The bottom panel shows the subsequent normalized spectrum. \label{fig6}}
\end{center}
\end{figure}

In measuring the EWs, it is first necessary to normalize the continuum.
Given the complex nature of these spectra, we choose to do local normalizations
around each line of interest rather than attempt a global normalization. 
Specifically, we carefully select windows of the continuum on both sides of 
a given absorption line and characterize it with a least-squares linear fit.
An example of this fit and the subsequently normalized continuum of the IUE spectrum of NGC~7714 are shown in Figure~\ref{fig6}.

In order to perform an appropriate comparison with IUE and take advantage of the superior resolution of COS, we generate two sets of EW measurements: 
\begin{enumerate}[noitemsep,partopsep=2pt,leftmargin=*]
    \item {\it Broad Sample:} measurements of both COS and IUE absorption features using broad integration windows appropriate for the low-resolution IUE spectra.
    \item {\it Narrow Sample:} measurements of the COS absorption features only using narrow integration windows customized to the individual lines in each spectrum.
\end{enumerate}


For the Broad Sample, we use broad integration windows designed to capture the full
extent of the line wings at IUE resolution.
We list the line complexes of interest below, with their line centers and integration window widths: 
\begin{itemize}[noitemsep,partopsep=2pt,leftmargin=*]
    \item \ion{S}{2} \W1253$+$\ion{Si}{2} \W1260: \W$_{cen}=1260$, $\Delta$\W$=\pm10$
    \item \ion{O}{1} \W1302$+$\ion{Si}{2} \W1304: \W$_{cen}=1300$, $\Delta$\W$=\pm10$
    \item \ion{C}{2} \W1335: \W$_{cen}=1335$, $\Delta$\W$=\pm10$
    \item \ion{Si}{4} \W\W1393,1402: \W$_{cen}=1400$, $\Delta$\W$=^{+10}_{-25}$
    \item \ion{C}{4} \W\W1548,1550: \W$_{cen}=1550$, $\Delta$\W$=^{+20}_{-40}$ 
    \item \ion{Fe}{2} \W\W1608,1611: \W$_{cen}=1607$, $\Delta$\W=$\pm20$
    \item \ion{Al}{2} \W1671: \W$_{cen}=1670$, $\Delta$\W=$\pm15$ 
\end{itemize}
For the Narrow Sample, we defined the limits of integration as the point where 
the absorption line returned back to the normalized continuum ($F_\lambda = 1$). 


The Broad Sample EWs in the COS and IUE spectra are measured using a straight integration technique in the \texttt{Interactive Data Language} (IDL) software. 
Broad Sample errors $\Delta$EW are estimated using the method from \citet{stetson08}, 
which is based on \citet{cayrel88}:
\begin{equation}
\Delta EW = 1.6 \times
\frac{\sqrt {\delta \lambda \times EW}}{S/N},
\label{eq3}
\end{equation}
where $\delta \lambda$ is the spectral resolution and 
S/N is the average S/N over the whole spectrum, obtained from performing the 
$\beta$-slope fit in Section~\ref{sec:3.1}. 

As an initial test, we compared the IUE to the COS data smoothed to the same lower spectral resolution of IUE in order to determine the effects of aperture size for those 21 galaxies with both IUE and COS spectra. Then we compared the original high-resolution COS spectra to the smoothed COS spectra of these galaxies in order to determine the effects of spectral resolution. While we find significant scatter between the measurements for individual galaxies, there is no systematic trend. As an example, we give the measurements for the detected lines in the spectra of NGC~7714. The quoted values are EWs in \AA\ for IUE, COS original, and COS smoothed to the resolution of IUE, respectively: \ion{S}{2} \W1253$+$\ion{Si}{2} \W1260:  ($2.23 \pm 0.40, 2.53 \pm 0.01, 2.32\pm0.01$), \ion{O}{1} \W1302$+$\ion{Si}{2} \W1304: ($2.70 \pm 0.44, 3.75 \pm 0.01, 3.18 \pm 0.01$), \ion{C}{2} \W1335: ($2.83 \pm 0.54, 3.18 \pm 0.01, 3.22 \pm 0.01$), \ion{Si}{4} \W\W1393,1402: ($6.96 \pm 0.70, 6.00 \pm 0.01, 5.89 \pm 0.01$), and \ion{C}{4} \W\W1548,1550: ($6.91 \pm 0.70, 11.3 \pm 0.10, 9.45 \pm 0.01$). As we find no benefit in utilizing the smoothed COS spectra, we proceed with the analysis of the original high-resolution COS spectra for comparison with the IUE EWs. The Broad Sample EWs for both IUE and COS thus obtained are reported in Table~\ref{tbl6}.

EWs are measured for the Narrow Sample using a Bootstrap Monte Carlo simulation 
with 3000 iterations.
For each iteration, a new spectrum is generated, drawn from the normal 
distribution of values with a center and width corresponding 
to the flux and 1-$\sigma$ uncertainty, respectively, at each wavelength. 
The EW of each iteration is measured using the \texttt{Numpy.trapz} 
function in \texttt{PYTHON},
which integrates along the wavelength axis using the composite trapezoidal rule.
The final EW value and uncertainty is taken as the average and standard deviation 
of the resulting distribution calculated, respectively.
The Narrow Sample EWs for COS are reported in Table~\ref{tbl7}.


\startlongtable
\begin{deluxetable*}{rcccccccc}
    \begin{center}
    \setlength{\tabcolsep}{3pt}
    \tablewidth{0pt}
    \tablecaption{Broad Sample Equivalent Widths}
    \end{center}
\tablehead{
\CH{} & \CH{} & \CH{\ion{S}{2} \W1253,} & \CH{\ion{O}{1} \W1302,} & \CH{} & \CH{\ion{Si}{4} \W1393} 
& \CH{\ion{C}{4} \W1548,} & \CH{\ion{Fe}{2} \W1608,} & \CH{} \vspace{-2ex} \\ 
\CH{Galaxy} & \CH{Instrument} & \CH{\ion{Si}{2} \W1260} & \CH{\ion{Si}{2} \W1304} & \CH{\ion{C}{2} \W1335} 
& \CH{\ion{Si}{4} \W1402} & \CH{\ion{C}{4} \W1550} & \CH{\ion{Fe}{2} \W1611}   & \CH{\ion{Al}{2} \W1671}}
\startdata
SBS 0335-052    & COS & $0.65\pm0.01$ & $0.64\pm0.01$ & $0.77\pm0.01$ & \nodata       & \nodata       & $0.71\pm0.06$ & \nodata       \\ 
                & IUE & $0.43\pm0.26$ & $< 1.30$      & $< 0.86$      & $-0.19\pm0.17$ & $-4.63\pm0.84$ & $0.13\pm0.14$ & $-0.80\pm0.35$ \\ [1ex] 
NGC 1705        & COS & $2.00\pm0.01$ & $1.06\pm0.01$ & $2.47\pm0.01$ & $3.26\pm0.01$ & \nodata       & \nodata       & \nodata       \\ 
                & IUE & $1.09\pm0.24$ & $1.55\pm0.28$ & $0.77\pm0.20$ & $3.76\pm0.44$ & $5.79\pm0.54$ & $4.14\pm0.46$ & $2.50\pm0.36$ \\ [1ex] 
NGC 1741        & COS & \nodata       & \nodata       & \nodata       & \nodata       & \nodata       & \nodata       & \nodata       \\ 
                & IUE & $< 0.67$      & $2.44\pm0.58$ & $1.64\pm0.47$ & $5.04\pm0.83$ & $1.31\pm0.42$ & $-1.05\pm0.38$ & $2.47\pm0.58$ \\ [1ex] 
He 2-10         & COS & $4.92\pm0.02$ & \nodata       & $5.20\pm0.03$ & $7.48\pm0.09$ & \nodata       & \nodata       & \nodata       \\ 
                & IUE & $2.53\pm0.80$ & $3.36\pm0.92$ & $< 1.13$      & $4.51\pm1.07$ & $8.23\pm1.44$ & $4.98\pm1.12$ & $5.91\pm1.22$ \\ [1ex] 
IRAS 08339      & COS & $1.75\pm0.01$ & $3.77\pm0.01$ & $3.31\pm0.01$ & $4.86\pm0.01$ & \nodata       & \nodata       & \nodata       \\ 
+6517           & IUE & $< 2.70$      & $< 3.89$      & $< 3.45$      & $< 5.39$      & $2.43\pm1.22$ & $< 3.35$      & $< 2.80$      \\ [1ex] 
I Zw18          & COS & $1.71\pm0.01$ & \nodata       & $1.72\pm0.01$ & $0.85\pm0.01$ & $< -0.57$   & $0.39\pm0.07$ & $0.38\pm0.06$ \\ 
                & IUE & $1.45\pm0.39$ & $1.82\pm0.44$ & $< 0.66$      & $3.19\pm0.58$ & $-5.35\pm0.75$ & $1.93\pm0.45$ & $-0.61\pm0.25$ \\ [1ex]
NGC 3049        & COS & \nodata       & \nodata       & \nodata       & \nodata       & \nodata       & \nodata       & \nodata       \\ 
                & IUE & $1.88\pm0.60$ & $2.42\pm0.68$ & $1.76\pm0.58$ & $7.72\pm1.22$ & $12.06\pm1.52$ & $9.46\pm1.35$ & $3.16\pm0.78$ \\ [1ex] 
NGC 3125        & COS & $4.25\pm0.01$ & \nodata       & $3.09\pm0.01$ & $3.56\pm0.02$ & $7.96\pm0.04$ & \nodata       & $0.81\pm0.10$ \\ 
                & IUE & $< 0.49$      & $3.58\pm0.59$ & $< 0.60$      & $4.69\pm0.67$ & $2.16\pm0.46$ & $1.33\pm0.36$ & $-0.76\pm0.27$ \\ [1ex] 
NGC 3256        & COS & $2.12\pm0.31$ & $6.18\pm0.17$ & $6.61\pm0.24$ & $12.67\pm0.29$ & \nodata      & \nodata       & \nodata       \\ 
                & IUE & $3.26\pm1.00$ & $3.88\pm1.09$ & $4.46\pm1.16$ & $7.95\pm1.55$ & $11.25\pm1.85$ & $6.79\pm1.44$ & $1.21\pm0.61$ \\ [1ex] 
Haro 2          & COS & $4.36\pm0.01$ & \nodata       & $3.97\pm0.03$ & $5.97\pm0.12$ & \nodata       & \nodata       & \nodata       \\ 
                & IUE & $0.01\pm0.05$ & $< 1.18$      & $< 1.65$      & $5.34\pm1.06$ & $5.99\pm1.12$ & $6.08\pm1.13$ & $< 1.82$      \\ [1ex] 
NGC 3310        & COS & \nodata       & \nodata       & \nodata       & \nodata       & \nodata       & \nodata       & \nodata       \\ 
                & IUE & $4.66\pm0.70$ & $4.95\pm0.72$ & $2.78\pm0.54$ & $6.40\pm0.82$ & $6.55\pm0.83$ & $5.20\pm0.74$ & $2.74\pm0.54$ \\ [1ex] 
NGC 3351        & COS & \nodata       & \nodata       & \nodata       & \nodata       & \nodata       & \nodata       & \nodata       \\  
                & IUE & $0.45\pm0.39$ & $< 2.27$      & $1.14\pm0.61$ & $7.01\pm1.52$ & $15.06\pm2.23$ & $8.07\pm1.63$ & $< 2.20$      \\ [1ex] 
NGC 3353        & COS & $3.68\pm0.03$ & \nodata       & $3.45\pm0.05$ & $3.46\pm0.17$ & \nodata       & \nodata       & \nodata       \\ 
                & IUE & $2.05\pm0.51$ & $2.82\pm0.60$ & $1.56\pm0.45$ & $5.00\pm0.80$ & $1.59\pm0.45$ & $3.98\pm0.71$ & $< 0.97$      \\ [1ex] 
UGCA 219        & COS & $1.68\pm0.04$ & $0.84\pm0.04$ & $1.68\pm0.05$ & \nodata       & \nodata       & \nodata       & \nodata       \\ 
                & IUE & \nodata       & $1.87\pm0.47$ & $1.12\pm0.36$ & $5.47\pm0.80$ & $-2.88\pm0.58$ & $4.54\pm0.73$ & $2.67\pm0.56$ \\ [1ex]
NGC 3690        & COS & $2.63\pm0.01$ & $4.05\pm0.01$ & $3.37\pm0.01$ & $4.59\pm0.02$ & $6.08\pm0.02$ & \nodata       & $1.17\pm0.04$ \\ 
                & IUE & $2.58\pm0.76$ & $4.09\pm0.96$ & $3.11\pm0.84$ & $3.97\pm0.95$ & $8.85\pm1.41$ & $3.92\pm0.94$ & $3.00\pm0.82$ \\ [1ex] 
NGC 3991        & COS & \nodata       & $1.96\pm0.01$ & $1.34\pm0.02$ & $3.14\pm0.02$ & $6.93\pm0.01$ & $0.81\pm0.01$ & \nodata       \\ 
                & IUE & $< 0.47$      & $2.13\pm0.34$ & $0.80\pm0.21$ & $4.35\pm0.49$ & $4.74\pm0.51$ & $4.73\pm0.51$ & $1.40\pm0.28$ \\ [1ex]
NGC 4194        & COS & \nodata       & \nodata       & \nodata       & \nodata       & \nodata       & \nodata       & \nodata       \\ 
                & IUE & $4.30\pm1.11$ & $6.47\pm1.37$ & $2.95\pm0.92$ & $6.76\pm1.40$ & $5.74\pm1.29$ & $3.75\pm1.04$ & $1.03\pm0.54$ \\ [1ex] 
NGC 4214        & COS & $2.87\pm0.01$ & \nodata       & $2.30\pm0.01$ & $1.91\pm0.03$ & $4.80\pm0.01$ & \nodata       & $1.81\pm0.01$ \\ 
                & IUE & $1.18\pm0.25$ & $1.85\pm0.32$ & $1.35\pm0.27$ & $3.92\pm0.46$ & $3.55\pm0.44$ & $3.70\pm0.45$ & $0.75\pm0.20$ \\ [1ex] 
UGCA 281        & COS & \nodata       & \nodata       & \nodata       & \nodata       & \nodata       & \nodata       & \nodata       \\
                & IUE & $< 1.13$      & $0.75\pm0.39$ & $< 1.14$      & $< 1.12$      & $-2.64\pm0.74$ & $< 1.82$      & $-2.12\pm0.66$ \\ [1ex] 
NGC 4449        & COS & $3.54\pm0.01$ & $4.44\pm0.01$ & $3.35\pm0.01$ & $4.34\pm0.01$ & \nodata       & \nodata       & \nodata       \\ 
                & IUE & $2.48\pm0.65$ & $4.75\pm0.90$ & $< 1.48$      & $2.37\pm0.63$ & $7.11\pm1.10$ & $4.10\pm0.83$ & $< 0.96$      \\ [1ex] 
NGC 4670        & COS & $3.32\pm0.01$ & $2.95\pm0.01$ & $2.74\pm0.01$ & $4.83\pm0.01$ & $6.08\pm0.02$ & \nodata       & $1.08\pm0.05$ \\ 
                & IUE & $2.00\pm0.37$ & $2.82\pm0.44$ & $1.23\pm0.29$ & $4.89\pm0.58$ & $2.06\pm0.38$ & $4.09\pm0.53$ & $2.21\pm0.39$ \\ [1ex] 
NGC 4861        & COS & $2.54\pm0.01$ & \nodata       & $1.90\pm0.01$ & $1.79\pm0.04$ & \nodata       & \nodata       & \nodata       \\ 
                & IUE & $0.06\pm0.07$ & $0.17\pm0.11$ & $0.21\pm0.12$ & $4.42\pm0.56$ & $2.43\pm0.42$ & $4.05\pm0.54$ & $1.97\pm0.38$ \\ [1ex]
NGC 5236        & COS & $2.82\pm0.01$ & $5.13\pm0.01$ & $1.87\pm0.01$ & $8.27\pm0.01$ & $8.56\pm0.02$ & \nodata       & $2.08\pm0.04$ \\ 
                & IUE & $2.07\pm0.52$ & $4.82\pm0.80$ & $1.47\pm0.44$ & $5.54\pm0.86$ & $10.48\pm1.18$ & $4.70\pm0.79$ & $0.10\pm0.12$ \\ [1ex]
NGC 5253        & COS & $3.17\pm0.01$ & $1.63\pm0.01$ & $2.88\pm0.01$ & $2.84\pm0.01$ & $6.04\pm0.01$ & $1.49\pm0.03$ & $1.33\pm0.03$ \\ 
                & IUE & $1.66\pm0.22$ & $2.29\pm0.26$ & $0.79\pm0.15$ & $3.99\pm0.35$ & $3.90\pm0.34$ & $4.05\pm0.35$ & $1.91\pm0.24$ \\ [1ex]
NGC 5457        & COS & \nodata       & \nodata       & \nodata       & \nodata       & \nodata       & \nodata       & \nodata       \\
                & IUE & $< 3.44$      & $5.88\pm1.74$ & $< 2.59$      & $7.07\pm1.90$ & $11.09\pm2.38$ & $5.68\pm1.71$ & $0.96\pm0.70$ \\ [1ex] 
NGC 5996        & COS & \nodata       & \nodata       & \nodata       & \nodata       & \nodata       & \nodata       & \nodata       \\
                & IUE & $< 3.14$      & $4.10\pm1.29$ & $< 3.25$      & $9.36\pm1.95$ & $11.71\pm2.18$ & $11.34\pm2.15$ & $< 2.49$      \\ [1ex]
TOL 1924-416    & COS & $1.63\pm0.01$ & $1.48\pm0.01$ & $0.99\pm0.01$ & $2.69\pm0.02$ & \nodata       & \nodata       & \nodata       \\ 
                & IUE & $-0.35\pm0.12$ & $0.76\pm0.18$ & $0.98\pm0.21$ & $4.67\pm0.45$ & $3.89\pm0.41$ & $3.89\pm0.41$ & $0.85\pm0.19$ \\ [1ex] 
NGC 7552        & COS & $4.18\pm0.02$ & \nodata       & $5.68\pm0.03$ & $8.66\pm0.05$ & \nodata       & \nodata       & \nodata       \\ 
                & IUE & $3.05\pm0.80$ & $5.89\pm1.11$ & $3.67\pm0.88$ & $9.48\pm1.41$ & $14.98\pm1.77$ & $9.83\pm1.44$ & $1.91\pm0.63$ \\ [1ex]
NGC 7714        & COS & $2.53\pm0.01$ & $3.75\pm0.01$ & $3.71\pm0.01$ & $6.00\pm0.01$ & $11.3\pm0.1$  & \nodata       & $1.89\pm0.01$ \\ 
                & IUE & $2.26\pm0.40$ & $2.70\pm0.44$ & $2.83\pm0.45$ & $6.96\pm0.70$ & $6.91\pm0.70$ & $3.86\pm0.52$ & $< 0.62$      \\ [1ex] 
\hline
Window (\AA)    &     & 1260$\pm$10   & 1300$\pm$10   & 1335$\pm$10   & $1400^{+10}_{-25}$ & $1550^{+20}_{-40}$ & 1607$\pm$20 & 1670$\pm$15 
\enddata
\tablecomments{
Comparison of the FUV EWs measured from the small-aperture ($2\farcs5$) COS data and the large-aperture ($10\arcsec\times20\arcsec$) IUE data. Given the low spectral-resolution of the IUE spectra ($\sim6$ \AA), we use the broad integration windows listed in the bottom row for both the COS and IUE spectra. The limits set in place in the table are as follows:  if the error was less than 0.01 we made the error measurement 0.01, if the flux was greater than  3 $\sigma$, we kept the original measurement and flux, and if the flux was greater than only 2$\sigma$, we set this as a lower limit.\label{tbl6}}
\end{deluxetable*}


\begin{figure}
  \centering
\begin{tabular}{cc}
\includegraphics[width=0.45\textwidth,trim=3mm 4mm 10mm 5mm,clip]{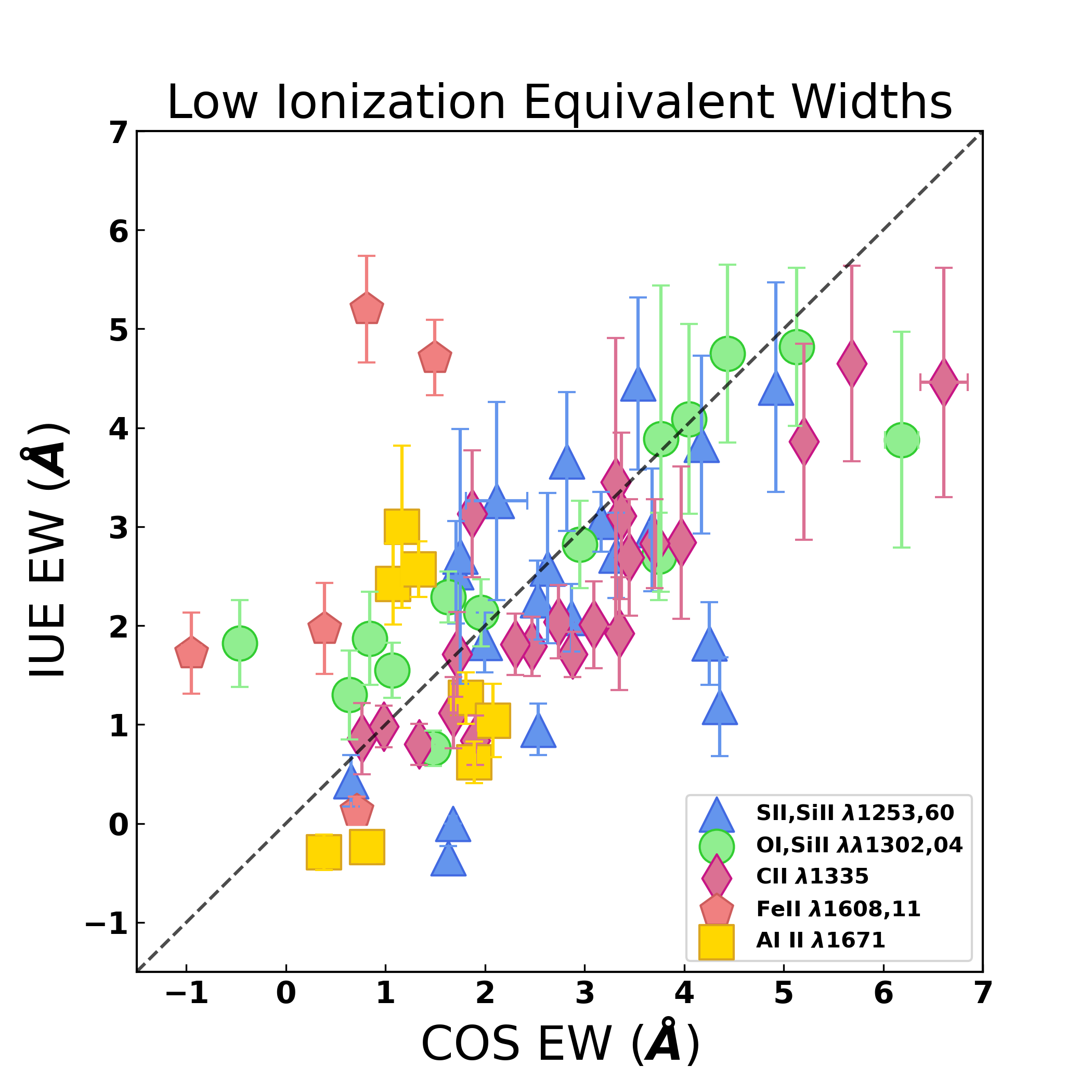} \\
\includegraphics[width=0.45\textwidth,trim=3mm 4mm 10mm 10mm,clip]{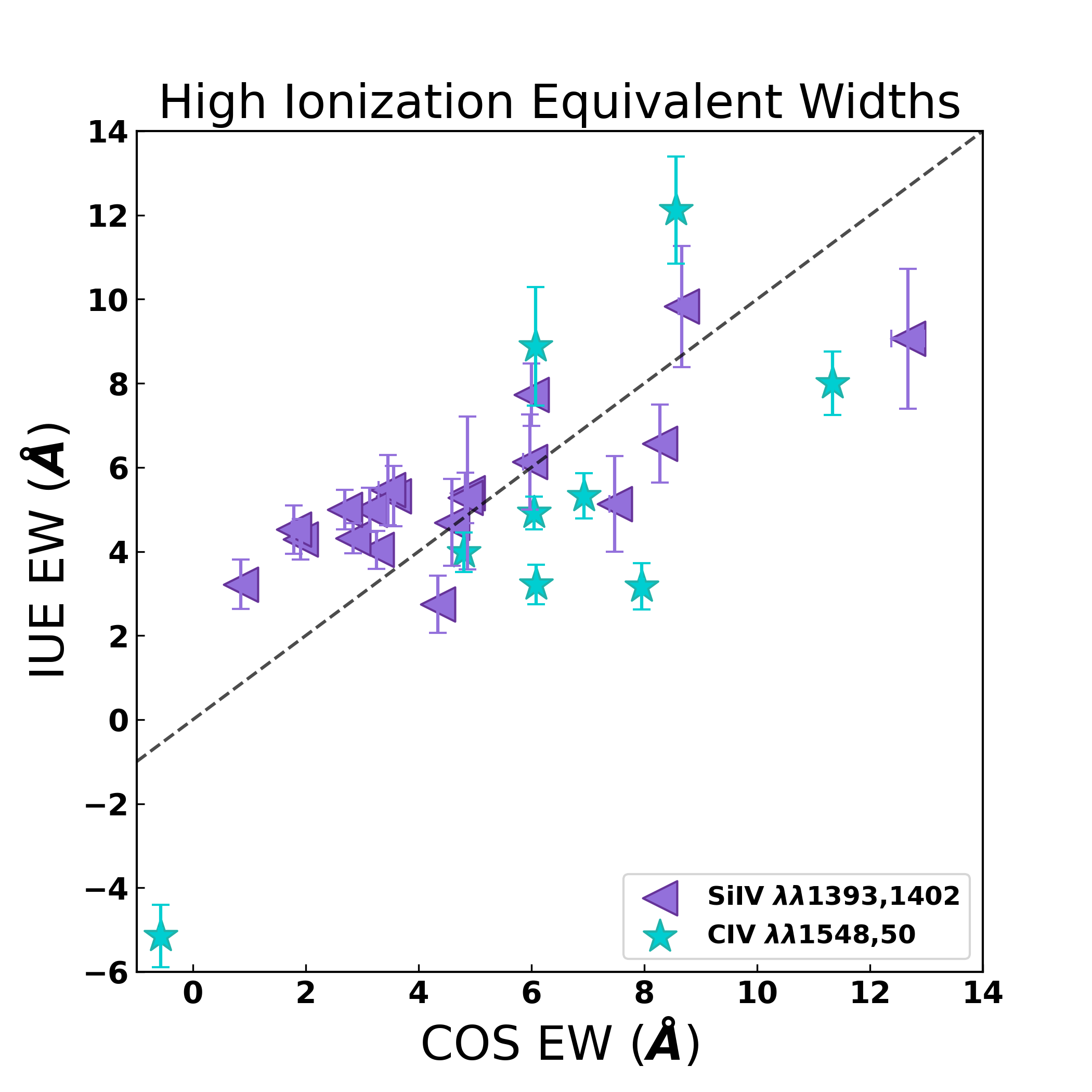} \\
\end{tabular}
    \caption{Comparison of the broad EW measurements from the IUE and COS spectra. 
    The top panel shows the low-ionization state ions, which include \ion{S}{2}$+$\ion{Si}{2} \W\W1253.60, 
    \ion{O}{1}$+$\ion{Si}{2} \W\W1302,04, \ion{C}{2} \W1335, \ion{Fe}{2} \W\W1608,11, and \ion{Al}{2} \W1671. 
    The lower panel shows the high-ionization state ions: \ion{Si}{4} \W\W1393,1402 and \ion{C}{4} 
    \W\W1548,1550. 
    \label{fig7}}
\end{figure}


\begin{center}
\setlength{\tabcolsep}{4pt}
\begin{deluxetable*}{lccccccc}
    \tablewidth{0pt}
    \tablecaption{Equivalent Widths of the COS Narrow Sample\label{tbl7}}
\tablehead{
Line    & SBS 0335-052 & NGC 1705 & He 2-10 & IRAS 08339+6517 & I Zw18 & NGC 3125 & NGC 3256 }
\startdata
\ion{S}{2} \W1253.81    & $0.11\pm0.01$ & $0.11\pm0.01$ & $0.62\pm0.01$ & \nodata       & \nodata       & $0.56\pm0.02$ & $0.49\pm0.01$ \\ 
\ion{Si}{2} \W1260.42   & $0.66\pm0.01$ & $1.81\pm0.01$ & $3.59\pm0.02$ & $1.47\pm0.08$ & $0.56\pm0.04$ & $1.21\pm0.02$ & $2.76\pm0.01$ \\ 
\ion{O}{1} \W1302.17    & $0.35\pm0.01$ & \nodata       & \nodata       & $2.14\pm0.02$ & \nodata       & \nodata       & \nodata       \\ 
\ion{Si}{2} \W1304.37   & $0.21\pm0.01$ & \nodata       & \nodata       & $2.19\pm0.09$ & $0.28\pm0.01$ & \nodata       & \nodata       \\ 
\ion{C}{2} \W1334.53    & $0.66\pm0.01$ & $2.25\pm0.01$ & $4.45\pm0.02$ & $3.04\pm0.01$ & $0.55\pm0.01$ & $1.36\pm0.01$ & $4.55\pm0.01$ \\ 
\ion{Fe}{2} \W1608.45   & $0.30\pm0.02$ & \nodata       & \nodata       & \nodata       & $0.26\pm0.01$ & $0.48\pm0.03$ & \nodata       \\ 
\ion{Fe}{2} \W1611.20   & $0.16\pm0.04$ & \nodata       & \nodata       & \nodata       & \nodata       & $< 0.06$      & \nodata       \\ 
\ion{Al}{2} \W1670.79   & \nodata       & \nodata       & \nodata       & \nodata       & $0.30\pm0.02$ & $0.67\pm0.02$ & $2.342\pm0.01$      \\ 
\\ \hline\hline \\ [-1ex]
Line                     & Haro 2 & NGC 3353 & UGCA 219 & NGC 3690 & NGC 3991 & NGC 4214 & NGC 4449 \\ 
[0.5ex]
\hline
\ion{S}{2} \W1253.81    & $1.56\pm0.03$ & $< 0.55$      & $0.04\pm0.01$ & $0.24\pm0.01$ & \nodata       & $0.69\pm0.01$ & $1.32\pm0.01$ \\ 
\ion{Si}{2} \W1260.42   & $2.11\pm0.37$ & $1.34\pm0.06$ & $0.92\pm0.01$ & $1.83\pm0.01$ & \nodata       & $1.93\pm0.01$ & $2.02\pm0.01$ \\ 
\ion{O}{1} \W1302.17    & \nodata       & \nodata       & $1.07\pm0.24$ & $4.52\pm0.01$ & $0.94\pm0.12$ & \nodata       & $3.10\pm0.01$ \\ 
\ion{Si}{2} \W1304.37   & \nodata       & $0.63\pm0.02$ & $< 0.96$      & $4.51\pm0.01$ & $0.95\pm0.07$ & \nodata       & $3.10\pm0.01$ \\ 
\ion{C}{2} \W1334.53    & $2.43\pm0.01$ & $0.95\pm0.04$ & $1.07\pm0.05$ & $2.44\pm0.01$ & $1.29\pm0.01$ & $1.91\pm0.01$ & $2.60\pm0.01$ \\ 
\ion{Fe}{2} \W1608.45   & \nodata       & \nodata       & \nodata       & \nodata       & \nodata       & \nodata       & \nodata       \\ 
\ion{Fe}{2} \W1611.20   & \nodata       & \nodata       & \nodata       & \nodata       & $0.19\pm0.01$ & \nodata       & \nodata       \\ 
\ion{Al}{2} \W1670.79   & \nodata       & \nodata       & \nodata       & $1.10\pm0.07$ & \nodata       & $1.17\pm0.26$ & \nodata       \\ 
\\ \hline\hline \\ [-1ex]
Line                     & NGC 4670 & NGC 4861 & NGC 5236 & NGC 5253 & TOL 1924-416 & NGC 7552 & NGC 7714 \\ 
[0.5ex]
\hline
\ion{S}{2} \W1253.81    & $1.36\pm0.01$ & $0.87\pm0.26$ & $0.49\pm0.01$ & $0.93\pm0.21$ & $1.76\pm0.08$ & $1.70\pm0.32$ & $0.29\pm0.01$ \\ 
\ion{Si}{2} \W1260.42   & $1.35\pm0.01$ & $< 1.30$      & $2.76\pm0.01$ & $2.36\pm0.01$ & $1.11\pm0.16$ & $4.31\pm0.22$ & $1.86\pm0.01$ \\ 
\ion{O}{1} \W1302.17    & $1.76\pm0.01$ & \nodata       & \nodata       & \nodata       & $0.91\pm0.01$ & \nodata       & $2.37\pm0.01$ \\ 
\ion{Si}{2} \W1304.37   & $1.76\pm0.01$ & \nodata       & \nodata       & \nodata       & $0.91\pm0.01$ & \nodata       & $2.36\pm0.01$ \\ 
\ion{C}{2} \W1334.53    & $1.53\pm0.01$ & $0.95\pm0.01$ & $4.55\pm0.01$ & $2.57\pm0.01$ & $1.06\pm0.01$ & $4.53\pm0.08$ & $3.09\pm0.01$ \\ 
\ion{Fe}{2} \W1608.45   & \nodata       & \nodata       & \nodata       & $0.92\pm0.28$ & \nodata       & \nodata       & \nodata       \\ 
\ion{Fe}{2} \W1611.20   & \nodata       & \nodata       & \nodata       & $0.10\pm0.01$ & \nodata       & \nodata       & \nodata       \\ 
\ion{Al}{2} \W1670.79   & $0.89\pm0.01$ & \nodata       & $2.34\pm0.01$ & $1.74\pm0.06$ & \nodata       & \nodata       & $2.04\pm0.11$ \\ 
\enddata
\tablecomments{
It is important to note that some lines could not be measured separately, particularly for \ion{O}{1}$+$\ion{Si}{2} \W\W1302,04. In the case of blended lines both measurements will have around the same value as we are measuring approximately the same EW. The limits set in place in the table are as follows:  if the error was less than 0.01 we made the error measurement 0.01, if the flux was greater than  3 $\sigma$, we kept the original measurement and flux, and if the flux was greater than only 2 $\sigma$, we set this as a lower limit.}
\end{deluxetable*} 
\end{center}


\subsection{EWs of High-vs Low-Ionization Lines}\label{sec:3.4}

Comparing the velocity structure and equivalent widths of low- and high-stellar ionization lines provide important diagnostics of the physical gas conditions, such as the
ionization structure and relative gas flows.
In this sense, ions with similar ionization potentials are expected to be
entrained in the same gas and so their absorption profiles should scale together
\citep[see, e.g.,][]{chisholm16}.
Additionally, low- and high-ionization lines have been observed to trace one another
kinematically \citep[e.g.,][]{chisholm16}, but do not necessarily have to.
However, the interpretation of ISM and stellar absorption features can be 
impacted by low spectral resolution that washes out and blends fine details.

We, therefore, now turn our attention to the differences between
the EW measurements of ISM absorption features in the IUE and COS spectra.
In order to provide a consistent comparison between galaxy-scale and cluster-scale
measurements, we use the Broad Sample measurements in this analysis.
In these broad integration windows, most of the measured absorption features are
blended line complexes.
Specifically, we investigate blends of \ion{S}{2} \W1253$+$\ion{Si}{2} \W1260, 
\ion{O}{1} \W1302$+$\ion{Si}{2} \W1304, \ion{Si}{4} \W\W1393,1402, \ion{C}{4} \W\W1548,1550,  
and \ion{Fe}{2} \W\W1608,1611.
Additionally, the high-ionization \ion{Si}{4} and \ion{C}{4} lines 
are combinations of ISM and stellar-wind features. 

In the top panel of Figure~\ref{fig7} we analyze the galaxy-scale versus cluster-scale
EW measurements of low-ionization lines.
In general, we find relatively good agreement between both the IUE and COS EWs, 
with an average scatter of $\sigma = $ 0.23 \AA. 
However, the individual complexes show a range of trends, with \ion{Fe}{2} \W\W1608,1611 
showing the largest dispersion and values generally skewed to larger IUE values. 
Interestingly, the high-EW end of the \ion{C}{2} \W1335 and \ion{O}{1} \W1302$+$\ion{Si}{2} \W1304 
trends are generally skewed to larger COS EWs relative to the IUE values. 
On the other hand, both \ion{Fe}{2} \W\W1608,1611 and \ion{O}{1} \W1302$+$\ion{Si}{2} \W1304 EWs are skewed 
toward higher IUE values at the low EW end of the trend \footnote{There are only four measurements in our sample for the \ion{Fe}{2} \W\W1608,1611 lines so this skew towards the IUE at low EWs most likely comes from the lack of COS spectral coverage for some of our galaxies at this wavelength.}. 
This could result from poorer detections of faint low-ionization lines in IUE spectra, where the lower 
spectral resolution tends to broaden and wash out weak absorption features.


In the bottom panel of Figure~\ref{fig7} we plot the high-ionization ions, \ion{Si}{4} \W\W1393,1402 and 
\ion{C}{4} \W\W1548,1550, measured from COS and IUE.
We find an increased scatter for the high-ionization trends relative to the low-ionization trends,
with an average scatter of $\sigma = $ 0.38 \AA. 
Similar to the trends observed for some of the low-ionization ions, we find that the \ion{Si}{4} 
trend deviates from the 1:1 line with a flatter slope. 
However, while we expect the high-ionization ions should trace the same gas, the \ion{C}{4} profile is 
markedly different from the \ion{Si}{4} trend. 
Our \ion{C}{4} EW measurements only sample values greater than $\sim4$\AA, 
so we are not able to access the low-EW trend, but see large scatter about the 1:1 line
at the high-EW end. 
This may be due to the strong P-Cygni stellar-wind features observed in many of the COS spectra.
As a result, these complex profiles are smeared out by IUE, reducing the integrated absorption profile,
and skewing the trend towards larger COS values. 
In our sample of galaxies with \ion{C}{4} we see one galaxy that stands out with an uncharacteristic EW compared to the rest of the sample. 
With an IUE measurement of EW(\ion{C}{4})$\sim-5$~\AA\ and a COS measurement of $\sim-0.5$ \AA\
(where negative EWs correspond to emission), I~Zw~18 is where we observe the highest offset between 
the two apertures.

Overall, we find that the high-ionization ions, with their more complex line profiles, have greater 
dispersion between the IUE and COS measurements than the low-ionization states, which emphasizes the
importance of high spectral resolution for robust EW measurements.
We also find a generally flatter than 1:1 trend present for both high- and low-ionization ions that
divides the skew towards higher IUE or COS EW values around 4 \AA. 
However, the statistical significance of this trend is not high enough to permit firm conclusions.

\begin{center}
\setlength{\tabcolsep}{4pt}
\begin{deluxetable}{crrr}
    \tabletypesize{\normalsize}
    \tablecaption{EW vs. Metallicity Relationships\label{tbl:EW vs Metal}}
    
\tablehead{
\CH{Absorption Line} & \CH{Fit/Instrument} & \CH{$p_0$}    & \CH{$p_1$}  }
\startdata
\ion{S}{2} \W1253     & Narrow COS  & $-3.573 $   & $+0.522 $  \\ 
             & Broad COS   & $-7.378$     & $1.253$    \\
             & IUE         & \nodata     & \nodata     \\ 
[1ex] 
\ion{Si}{2} \W1260   & Narrow COS & $-11.974$   & $+1.681$  
\\ 
             & Broad COS   & $-9.748 $  & $+1.527  $   \\
             & IUE         & $-11.600 $  & $+1.599 $   \\ [1ex] 
\ion{O}{1} \W1302 $+$ & \multirow{2}{*}{Narrow COS} & \multirow{2}{*}{$-21.145$} & \multirow{2}{*}{$+3.003$} \\ 
\ion{Si}{2} \W1304    &           &            &       \\  
             & Broad COS   & $-25.280 $    & $+3.423$    \\
             & IUE         & $-16.247$     & $+2.302$     \\ 
[1ex] 
\ion{C}{2} \W1335     & Narrow COS & $-11.158 $   & $+1.598 $
\\   
             & Broad COS   & $-14.920$   & $+2.170$    \\
             & IUE         & $-9.274$     & $+1.322$     \\ 
[1ex] 
\ion{Si}{4} \W\W1393,1402 & Narrow COS & \nodata  & \nodata \\  
             & Broad COS   & $-41.923$   & +$5.612$    \\
             & IUE         & $-25.663$     & $+3.690$     \\ 
[1ex] 
\ion{C}{4} \W\W1548,50    & Narrow COS  & \nodata  & \nodata \\
             & Broad COS   & $-38.192$   & $+5.394$    \\
             & IUE         & $-77.457$     & $+9.904$     \\ 
[1ex] 
\ion{Fe}{2} \W1608    & Narrow COS & $-0.177 $   & $+0.0722 $   \\
             & Broad COS   & $-2.586$  & +$0.441$    \\
             & IUE         & $-29.292$     & $+4.051$     \\ 
[1ex] 
\ion{Al}{2} \W1670     & Narrow COS & $-7.115$  & $+1.0202$ \\
             & Broad COS   & $-6.453$  & $+0.945$    \\
             & IUE         & $-10.273$     & $+1.412$     \\ 
[1ex] 
\enddata
\tablecomments{
Polynomial fits to the trends between EW and metallicity 
shown in Figures~\ref{fig8}, \ref{fig9} and \ref{fig10}.
The fits used are first-order polynomials 
of the function EW $= p_0 + p_1 Z$.}
\end{deluxetable} 
\end{center}


\begin{figure*}
\centering
\begin{minipage}{0.07cm}
\rotatebox{90}{\textcolor{black}{\textbf{EW (\AA)}}}
\end{minipage}%
\begin{tabular}{cc}
\includegraphics[width=0.40\textwidth,trim=1mm 0mm 0mm 0mm,clip]{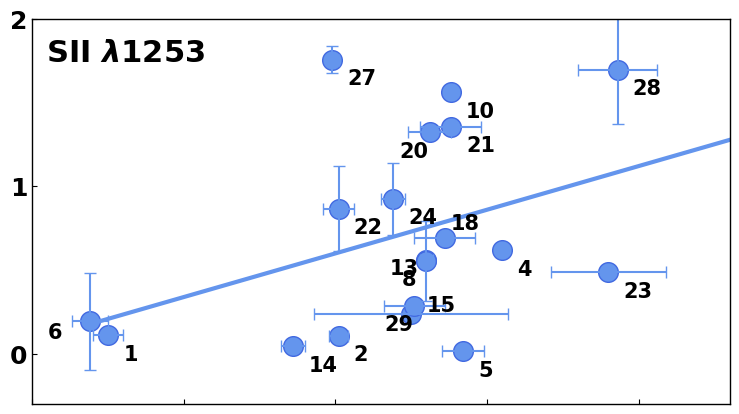} & 
\includegraphics[width=0.40\textwidth,trim=1mm 0mm 0mm 0mm,clip]{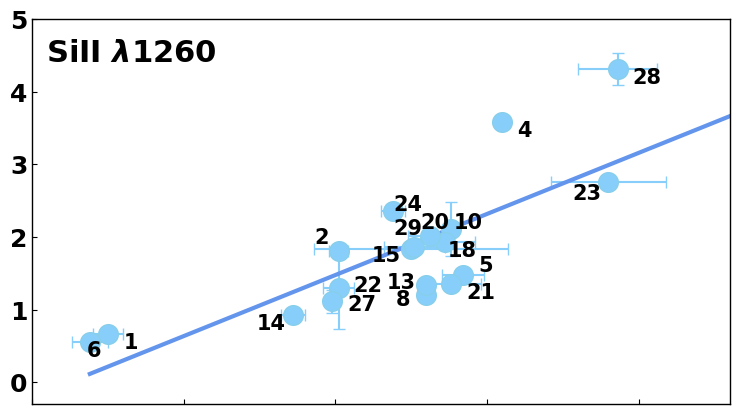} \\
\includegraphics[width=0.40\textwidth,trim=1mm 0mm 0mm 0mm,clip]{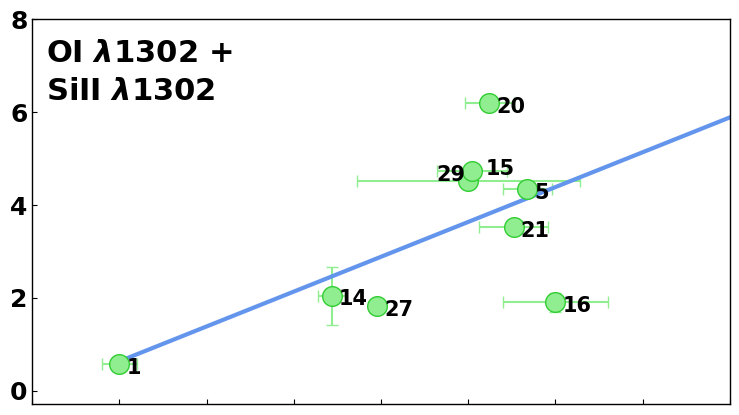} &
\includegraphics[width=0.40\textwidth,trim=1mm 0mm 0mm 0mm,clip]{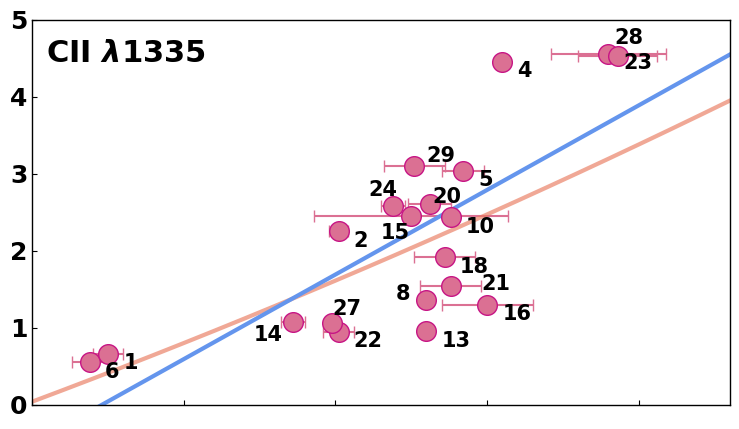} \\
\includegraphics[width=0.415\textwidth,trim=1mm 0mm 0mm 0mm,clip]{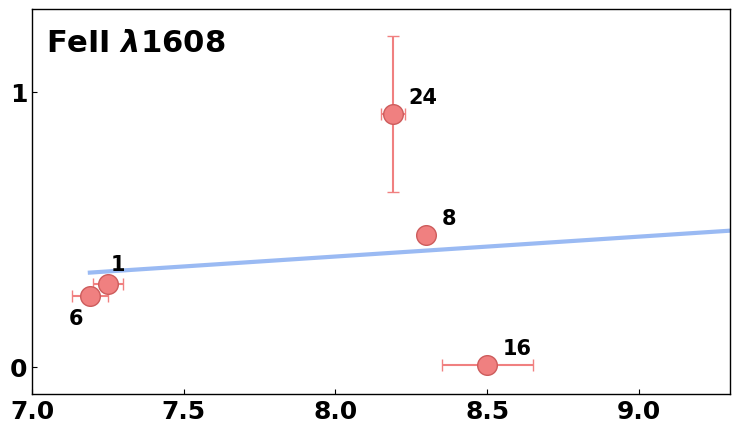} & 
\includegraphics[width=0.415\textwidth,trim=1mm 0mm 0mm 0mm,clip]{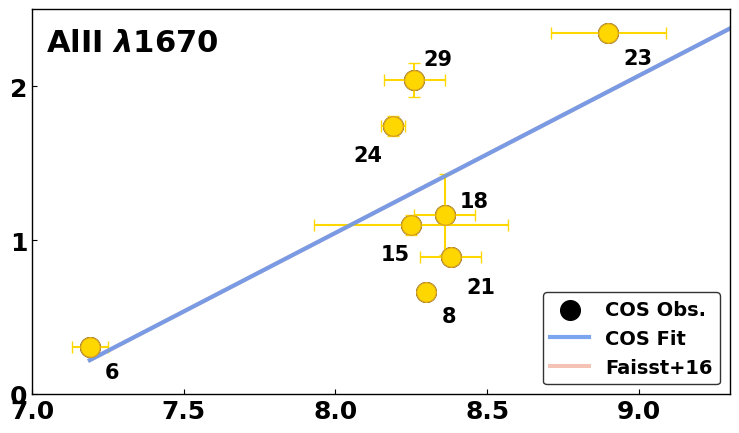} \\
\end{tabular}
    \vspace*{0.1cm}\hspace*{1.4cm}\textcolor{black}{\textbf{12+log(O/H)}}
    \caption{
    Equivalent width measurements versus gas-phase metallicity for the Narrow Sample (see Section~\ref{sec:3.3} for description). 
    Equivalent widths were determined using integration windows appropriate to each line in a given COS spectrum. 
    The solid blue lines are our best linear fits to the observed distributions.
    In comparison, our fit for \ion{C}{2} in the middle left panel 
    is in good agreement with the trend of \citet{faisst16} (solid red line),
    demonstrating the increasing trend between absorption line EW and 
    gas-phase metallicity.
    \label{fig8}}
\end{figure*}


\section{Metallicity Dependence}\label{sec:4}

In this section, we investigate the behavior of the measured EWs with metallicity. 
We study the Narrow Sample and the Broad Sample separately. 
The Narrow Sample takes advantage of the superior COS resolution and S/N, which allows one to 
correct for line blending and contamination. 
However, this sample is unsuitable for a direct comparison with EWs measured in the 
low-resolution IUE spectra. 
Therefore, we investigate the metallicity dependence of the EWs measured in the IUE spectra 
in conjunction with the Broad Sample COS measurements.


\begin{figure*}
\centering
\begin{minipage}{0.07cm}
\rotatebox{90}{\textcolor{black}{\textbf{EW (\AA)}}}
\end{minipage}%
\begin{tabular}{cc}
\includegraphics[width=0.40\textwidth,trim=1mm 0mm 0mm 0mm,clip]{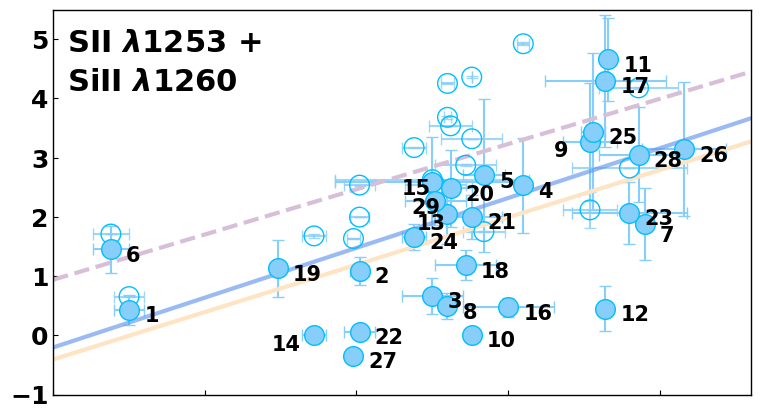} &
\includegraphics[width=0.30\textwidth,trim=2mm 25mm 50mm 2mm,clip]{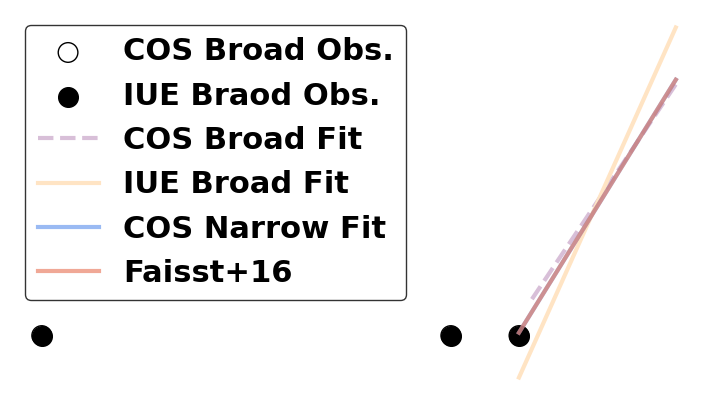} \\
\includegraphics[width=0.40\textwidth,trim=1mm 0mm 0mm 0mm,clip]{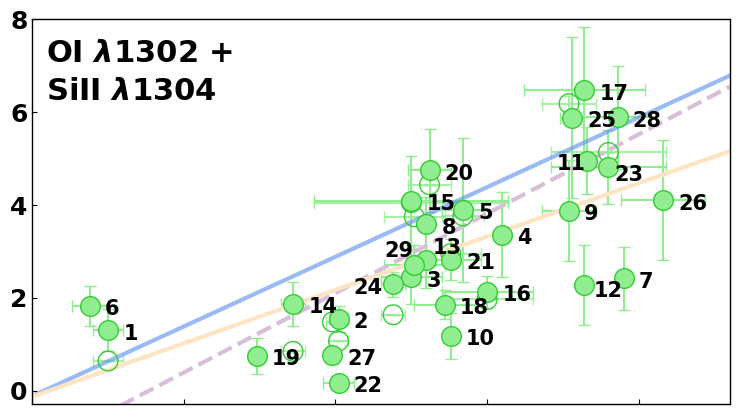} &
\includegraphics[width=0.40\textwidth,trim=1mm 0mm 0mm 0mm,clip]{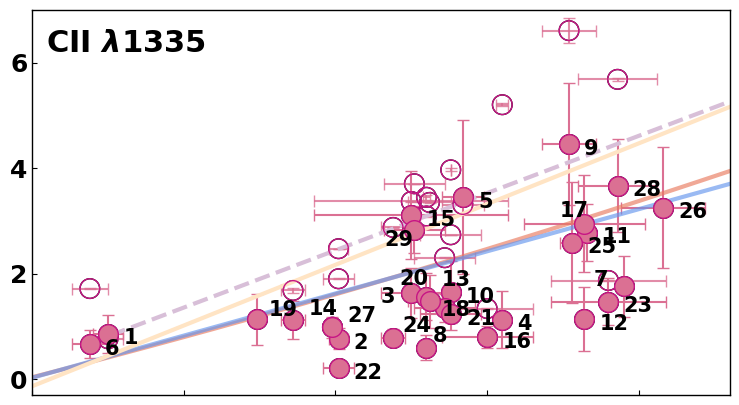} \\
\includegraphics[width=0.40\textwidth,trim=1mm 0mm 0mm 0mm,clip]{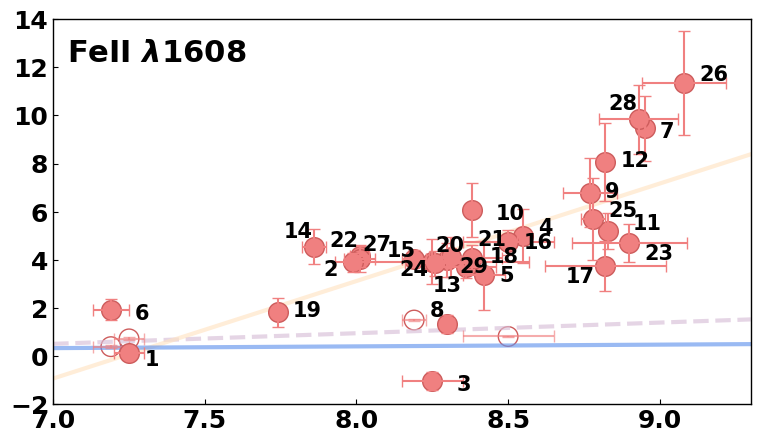} & 
\includegraphics[width=0.40\textwidth,trim=1mm 0mm 0mm 0mm,clip]{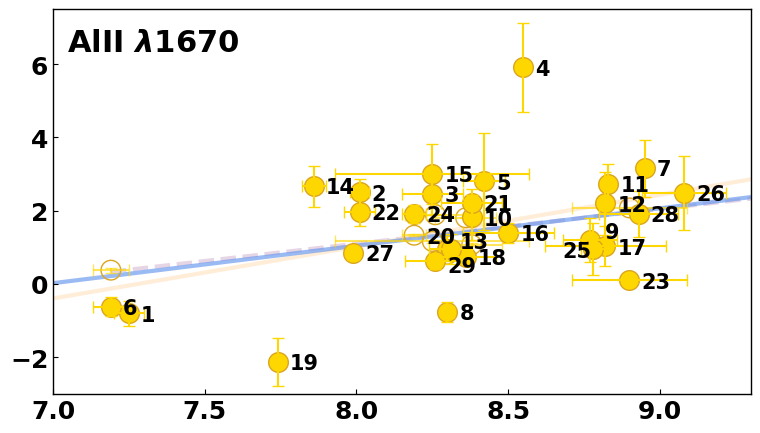} \\
\end{tabular}
    \vspace*{0.1cm}\hspace*{1.4cm}\textcolor{black}{\textbf{12+log(O/H)}}
    \caption{
    Broad Sample equivalent width measurements of low-ionization species from both the COS and IUE 
    spectra versus gas-phase metallicity.
    Equivalent widths were obtained using broad integration windows
    (see Section~\ref{sec:4.2}). 
    The dashed lines are linear fits to the observed distributions. 
    For comparison, the solid blue lines are the best fits to the corresponding Narrow Sample 
    EWs shown in Figure~\ref{fig8}. 
    The increasing trend between EW and metallicity is seen for both the Narrow and Wide
    samples, but with vertical offsets due to the effects of the assumed integration windows. 
    We note that the \ion{C}{2} \W\W1335 trend is in good agreement with that 
    from \citet{faisst16} (red line).
    \label{fig9}}
\end{figure*}


\begin{figure*}
  \centering
\begin{minipage}{0.07cm}
\rotatebox{90}{\textcolor{black}{\textbf{EW (\AA)}}}
\end{minipage}%
\begin{tabular}{cc}
\includegraphics[width=0.40\textwidth,trim=1mm 0mm 0mm 0mm,clip]{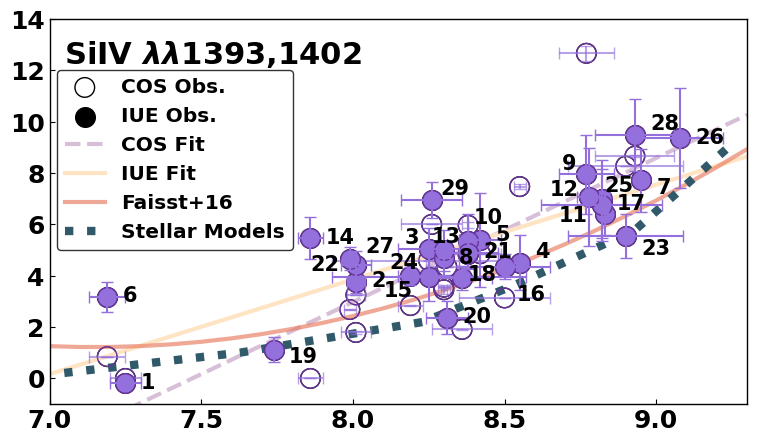} &
\includegraphics[width=0.40\textwidth,trim=1mm 0mm 0mm 0mm,clip]{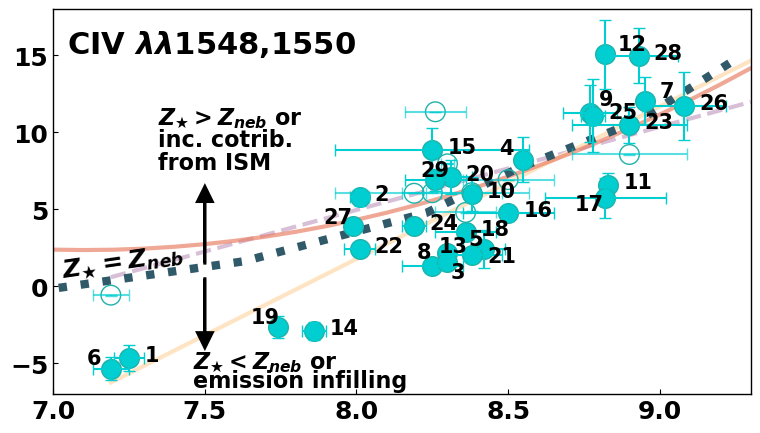}\\
\end{tabular}
    \vspace*{0.1cm}\hspace*{1.4cm}\textcolor{black}{\textbf{12+log(O/H)}}
    \caption{
    Broad Sample equivalent width measurements of high-ionization species from both the COS and IUE 
    spectra versus gas-phase metallicity.
    Equivalent widths were obtained using broad integration windows
    (see Section~\ref{sec:4.2}). 
    The dashed lines are linear fits to the observed distributions. 
    For comparison, the solid blue lines are the best fits to the corresponding Narrow Sample 
    EWs shown in Figure~\ref{fig8}. 
    We also compare to the trends from \citet{faisst16} (red lines), with generally good agreement.
    Theoretical expectations from \texttt{Starburst99} stellar-wind models are shown  
    (dotted lines), revealing a stronger ISM contribution relative to the stellar 
    contribution for \ion{Si}{4} than \ion{C}{4}.
    Interestingly, some of the small \ion{C}{4} EWs may indicate gas-phase metallicities that
    are enhanced relative to the stars.
    \label{fig10}} 
\end{figure*}


\subsection{The Narrow Sample}\label{sec:4.1}
In Figure~\ref{fig8} we analyze the trend between the gas-phase metallicity and EW for each ion. 
The values for 12+log(O/H) are from Table~\ref{tbl1} (see Section~\ref{sec:2.5}).
In order to isolate trends for individual ion features, we use the Narrow Sample, which allows us 
to minimize the contamination by neighboring lines, as well as by Milky Way and geocoronal features. 
For the case of the \ion{O}{1}$+$\ion{Si}{2} \W\W1302,04 line complex, 
the lines are blended and cannot be measured separately. 
For such blended lines, we use a single integrated EW measurement of the blended profile. 
For ions with isolated multiplet lines, we plot the line with the strongest oscillator strength 
(strongest absorber) in Figure~\ref{fig8}. 
Additionally, the \ion{Si}{4} \W\W1393,1402 and \ion{C}{4} \W\W1548,50 lines show strong stellar P-Cygni 
profiles in many galaxies and so cannot be deblended. 
Therefore we opt against considering these two  line complexes in the Narrow Sample, and defer a study of their 
properties to the discussion of the Broad Sample.

For each trend in Figure~\ref{fig8}, we fit the individual relations with a first-order
polynomial using the \texttt{NumPy.polyfit} function in \texttt{Python}.
The best fits are shown as solid blue lines and the resulting polynomial coefficient values are 
given in Table~\ref{tbl:EW vs Metal}. 
We perform the fits for all spectral lines considered. 
While we find correlations for some lines (e.g., \ion{C}{2} \W1335), none are found for others, such as \ion{Fe}{2} \W1608, which we attribute to the small number of data points. 
In general, we find EW increases with metallicity, as expected from the increase in the respective elemental abundances.

In order to test our EW versus nebular metallicity trends, we compare them to the results 
from \citet{faisst16}. 
Specifically, we prioritize the comparison of \ion{C}{2} \W1335 because it is an isolated 
line and less likely to be saturated. 
We plot the \citet{faisst16} derived relation as a solid red line in the middle right panel 
of Figure~\ref{fig8}.
In general, our fit to the COS data is consistent with their trend
but extends to higher EWs at high metallicities.
This good agreement gives confidence to our other fits measured without literature values to compare to.

\subsection{The Broad Sample}\label{sec:4.2}

\subsubsection{Low-Ionization Species}\label{sec:4.2.1}
In Figure~\ref{fig9} we repeat our analysis of the EWs of low-ionization species 
versus gas-phase metallicity for the Broad Sample measurements for both the COS and IUE spectra.
The IUE dataset has more data points for two reasons: 
(i) There are 29 versus 21 galaxies and 
(ii) the wavelength coverage is larger. 
Several galaxies lack COS G160M spectra where \ion{Fe}{2} \W\W1608,1611 and \ion{Al}{2} \W1671 
are located; this results in significantly stronger trends for the IUE dataset compared to that 
of the COS dataset.
While the data in Figure~\ref{fig9} include blends of multiple absorption features,
the Broad Sample also allows us to examine how larger integration windows and lower spectral
resolution affect the measured EWs for the same instrument aperture. 
As we did for the Narrow Sample in Figure~\ref{fig8}, we fit a first-order polynomial to each 
dataset, and over plot the fits derived for the Narrow Sample. 
All five spectral features considered show a positive correlation with 12+log(O/H). 
The fit coefficients are listed in Table~\ref{tbl:EW vs Metal}.
Overall, the Narrow Sample and Broad Sample COS measurements display the same trends 
with 12+log(O/H), but significant discrepancies are seen between the COS and IUE Broad
Sample Fits.

The top two left column plots of Figure~\ref{fig9} show our most significant trends: 
\ion{S}{2} \W1253$+$\ion{Si}{2} \W1260 (left top) and \ion{C}{2} \W1335 (left middle). 
The Broad Sample is skewed towards higher EW values in both cases. 
In the case of \ion{S}{2} \W1253$+$\ion{Si}{2} \W1260 this trend can be understood in 
terms of the integration window, which includes both lines in the Broad Sample 
measurements but only \ion{Si}{2} \W1260 in the Narrow Sample measurements. 
For the \ion{C}{2} \W1335 trend in the middle right panel of Figure~\ref{fig9}, 
the Broad Sample fit (dashed line) is offset to larger EWs than the Narrow Sample fit 
(solid blue line). 
This offset is likely due to the inclusion of the \ion{C}{2}$^*$ fine-structure line 
in the Broad Sample integration window, but not in the Narrow Sample window. 
There is also significant Milky Way contamination of \ion{C}{2} \W1335 in the Broad Sample 
EW measurements that contribute to this high offset.
We note that Milky Way contamination was particularly difficult to remove from
\ion{C}{2} \W1335 due to the blended nature of this line in the IUE Spectra. 
Therefore, we only remove Milky Way contamination for galaxies where we can disentangle the 
Milky Way component in the low-resolution IUE Spectra.

\subsubsection{High-Ionization Species}\label{sec:4.2.2}
The high-ionization counterpart to Figure~\ref{fig9} is shown in Figure~\ref{fig10},
where strong correlations are seen for \ion{Si}{4} \W\W1393,1402 and \ion{C}{4} \W\W1548,50. 
The latter doublet is stellar-wind dominated, whereas the former has contributions from both 
stellar-wind and interstellar lines. 
We fit a first-order polynomial to both the COS and IUE datasets for both ions. 
Both features were also studied by \citet{faisst16}, whose 2nd-order polynomial 
relations (red solid lines) agree rather well with our best fits. 

To examine the \ion{Si}{4} and \ion{C}{4} trends further, 
we investigate the theoretical \emph{stellar} \ion{Si}{4} \W\W1393,1402 and 
\ion{C}{4} \W\W1548,1550 profiles as a function of stellar metallicity using 
synthetic UV spectra from the \texttt{Starburst99} code \citep{leitherer14}. 
We adopt the library of theoretical spectra derived from WM-Basic model atmospheres 
\citep{leitherer10} and use the same integration windows as in the corresponding 
observed spectra to measure EWs. 
The Geneva 1992-94 evolutionary tracks with high mass loss cover a metallicity range 
of $7.6<$ 12+log(O/H) $<9.2$ \citep{meynet94}. 
This range is consistent with the relevant metallicity range of our sample, where the 
lowest metallicity galaxies do not have significant absorption features. 
We assume a standard young population forming constantly over 20~Myr with a standard 
Kroupa initial mass function (IMF) and power-law exponents of 1.3 and 2.3, producing 
mass boundaries of 0.1~M$_\odot$, 0.5~M$_\odot$, and 100~M$_\odot$, respectively 
\citep{Kroupa08}. 
In order to investigate the impact of IMF variations, we modify the high-mass exponent 
to 1.3 and 3.3 for an IMF more or less skewed towards massive stars, respectively. 
Varying the IMF has little effect on the predicted relation. 
We also test the influence of using different evolutionary tracks available in 
\texttt{Starburst99} and found no significant change. 

The predicted stellar model is shown in Figure~\ref{fig10} as a dark dashed line. 
In general, the theoretical stellar trend underpredicts the observed \ion{Si}{4} 
features, likely due to the larger relative contribution from ISM absorption.
On the other hand, the stellar model is in excellent agreement for large EWs of 
\ion{C}{4}, but the observed data points seem to fall off at lower metallicities.
Below 12+log(O/H) $\approx$ 8.0, nebular emission contributes to \ion{C}{4} 
\W\W1548,50 (and other lines). 
This is reflected in the negative EWs in the figure. 
While a proper comparison with stellar models would require correction for this nebular 
contribution, this trend may still be diagnostically useful.

The observed trend of \ion{C}{4} EWs versus \emph{nebular} oxygen abundance in
Figure~\ref{fig10} follows the predicted \emph{stellar} relation remarkably well 
over the metallicity range for most of the sample (12+log(O/H) $>$ 8).
This trend can be understood in terms of the metallicity-dependent stellar 
wind properties of massive stars, with some deviation, when large contributions 
of ISM absorption are present. 
Specifically, while (29) has the largest \ion{C}{4} EW offset above the theoretical 
trend, it is also offset to larger than average \ion{C}{2} EWs for its metallicity, 
indicating a large ISM absorption component.
In contrast, the observed \ion{Si}{4}~\W\W1393,1402 EWs are systematically higher 
than the theoretical values. 
This offset is not surprising, however, as the \ion{Si}{4} stellar features are 
generally weaker then those of \ion{C}{4}, and so the strong EWs indicate a more 
significant relative ISM contribution.

The strong empirical correlation between EW and 12+log(O/H) and the agreement with 
model predictions suggest that the EW of \ion{C}{4} \W\W1548,50 can be used to estimate 
the gas-phase metallicity.
The empirical relation can be expressed as 
\begin{equation}
12 + {\rm log(O/H)} = (0.075 \pm 0.008) \times EW + (7.956 \pm 0.063),
\end{equation}
where the EW here refers to the \ion{C}{4} EW.
For instance, a galaxy with a measurement of EW(\ion{C}{4}) $\approx$ 5~\AA\ would 
have an estimated gas-phase oxygen abundance of 12+log(O/H) $\approx$ 8.3, i.e., 
similar to that of the Large Magellanic Cloud (LMC). 
We emphasize that this relation has been derived for a local galaxy sample and, 
therefore, application to other galaxy samples, e.g., at high-$z$ would require 
further verification. 

Our broad measurement windows and the low spectral resolution of the data do not 
permit removal of the interstellar components originating within galaxies. 
As opposed to the case of \ion{Si}{4}, the relative interstellar contribution to the 
stellar \ion{C}{4} is small but still not negligible, a point raised by 
\citet{crowther2006}. 
In principle, this effect is accounted for in our empirical calibration, but the 
strength of the interstellar components relative to the stellar \ion{C}{4} may be 
different in a different galaxy sample. 
This issue could be mitigated by utilizing stellar \ion{N}{5} \W\W1238,42; 
owing to its larger ionization potential of 77~eV, the ISM contribution to \ion{N}{5} 
is much smaller. 
However, \ion{N}{5} \W\W1238,42 is only present in very young stellar populations 
\citep[$\lesssim5$ Myr; see, e.g.,][]{chisholm19} and, in the present sample, is blended 
with strong Lyman-$\alpha$ absorption, so is not analyzed here. 

Alternatively, deviations from the stellar model in Figure~\ref{fig10} could
also indicate non-equal stellar and nebular metallicities or non-solar $\alpha$/Fe 
ratios \citep[e.g.,][]{steidel16}.
Since stellar winds are most sensitive to the Fe opacity in their atmospheres,
an enhanced $\alpha$/Fe abundance would divert points below the trend due to seemingly
higher nebular oxygen abundances compared to the inferred stellar abundance.
On the other hand, deficient $\alpha$/Fe abundances would drive points above the trend.
Despite these complications, the dispersion in the \ion{C}{4} EW trend is still 
relatively small for metallicities of 12+log(O/H)$>$8.0, 
indicating that the \ion{C}{4} EW can be used as a gas-phase metallicity diagnostic.


\begin{figure}
\begin{center}
\includegraphics[width=0.465\textwidth,trim=0mm 0mm 0mm 0mm,clip]{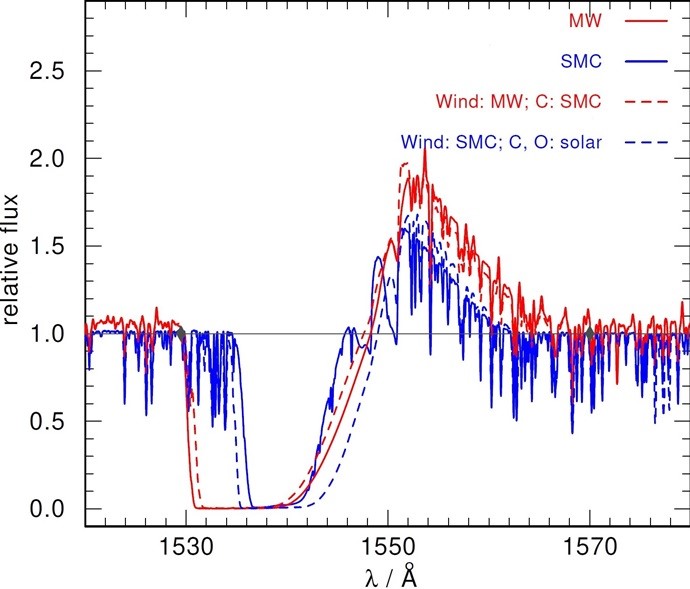} 
    \caption{Theoretical C \,\textsc{iv} 1548, 1550 profiles for a representative O supergiant with $T_\mathrm{eff} = 40,000$~K and log~$g = 4.0$. The profiles were obtained with PoWR model atmospheres using different abundances and abundance ratios. Solid red: standard solar abundances; solid blue: 0.2~$Z_\odot$, approximating SMC abundances; dashed red: $Z_\odot$ for all elements except for carbon, which is 20\% solar; dashed blue: 0.2~$Z_\odot$ for all elements, except carbon and oxygen, which are solar.
    \label{fig11}}
\end{center}
\end{figure}


\subsection{Interpretation of the Observed Relations}\label{sec:4.3}

The trends seen in Figures~\ref{fig9} and \ref{fig10} may seem surprising, 
as these spectral lines are deeply saturated, at which point they become insensitive 
to chemical abundance.  
In the saturation limit, the observed EWs of the interstellar lines lie on the flat part 
of the curve-of-growth,
\begin{equation}
EW \propto b \sqrt{ \ln \left(\frac{N_{\rm{ion}}}{b}\right)}, 
\end{equation}
where $b$ is the Dopper line-broadening parameter and $N_{\rm ion}$ is the column density of 
the corresponding ion.
In this limit, the EW is relatively insensitive to the column density ($N_{\rm{ion}}$) and
becomes mainly dependent on velocity via the Doppler parameter ($b$). 
Therefore, the metallicity dependence of saturated \emph{interstellar} lines, such as, e.g., 
\ion{C}{2} \W1335, can be understood in terms of macroscopic turbulence affecting $b$ 
\citep[e.g.,][]{heckman98}. 
The observed trend of ISM line strengths versus oxygen abundance originates from mechanical 
energy input from powerful stellar-winds and supernovae. 
As a consequence, more metal-rich galaxies have more luminous starbursts with stronger winds 
and higher supernova rates, which cause more macroscopic turbulence. 

The progression of EW with metallicity in Figures~\ref{fig9} and \ref{fig10} also reflects an 
increase in galaxy sizes: metal-rich galaxies tend to be more luminous, more massive, and 
larger in size than metal-poor dwarfs
\citep[i.e., the luminosity-metallicity and mass-metallicity relationships, e.g,][]{skillman89,tremonti04,berg12}.
This may indicate that the underlying cause of the correlation between oxygen abundance and 
EW is broadening by increased galactic rotation with galaxy mass.
Most of our sample galaxies were also studied by \citet{heckman98} who demonstrated that EWs 
of the interstellar lines also correlate with the rotation velocities derived from the 
\ion{H}{1} 21~cm line widths. 
However, the correlation is much weaker than that found in Figures~\ref{fig9} and \ref{fig10}, 
suggesting that galactic rotation is not the prime mechanism responsible for the line broadening. 
More importantly, the measured EWs would require rotation velocities significantly larger than 
those obtained from typical \ion{H}{1} line widths. 
We, therefore, conclude that macroscopic turbulence and galactic-scale outflows are primarily 
responsible for the correlation of EW of the interstellar lines with oxygen abundance.

The trends with abundance for the \emph{stellar}+ISM lines of \ion{Si}{4}~\W\W1393,1402 and 
\ion{C}{4} \W\W1548,1550 are even stronger than those of the interstellar lines. 
Like the interstellar lines, the stellar-wind+ISM lines are deeply saturated. 
Therefore, the abundance dependence cannot be primarily due to {\em direct} changes in the 
wind column densities. 
Further insight can be gained by studying wind models for individual stars. 

In Figure~11 we plot synthetic spectra from \texttt{PoWR} atmosphere models 
\citep[][]{Graefener02,Hamann03,sander15}
for a fiducial O-supergiant ($T_{\rm eff} = 40,000$~K, log~$g~=~4.0$) assuming the wind 
mass-loss rates of \citet{Vink01}.
The \ion{C}{4} profile is stronger for the solar-abundance MW model (solid red line) 
than for the 20\%\ solar SMC profile (solid blue line). 
Going one step further, we might expect the abundance {\it pattern} to deviate from the standard 
solar abundance pattern, where differences in the relative C abundance could affect the observed
\ion{C}{4} profile.
However, variations of the \emph{relative carbon abundance} over a range of solar to 0.2 solar 
leave the \ion{C}{4} \W\W1548,1550 line strength almost unchanged, as demonstrated by the 
C-deficient MW profile (dashed red line: 0.2 solar C abundance, solar abundances for all 
other elements).

Alternatively, we can examine the role of the stellar wind in shaping the \ion{C}{4} line profile.
The blue dashed line in Figure~\ref{fig11} shows a solar abundance profile but with a weaker
SMC-like wind strength that looks similar to the standard SMC profile (solid blue line).
Therefore the profile shape is mostly driven by the wind properties, which are largely determined 
by the Fe opacity for O-supergiants of these metallicities, and not {\em directly} by the 
relative abundances. 
More specifically, such hot-star winds are driven by radiation pressure from numerous strong and 
weak spectral lines predominantly located in the extreme-UV below the Lyman edge at 912 \AA. 
Winds from stars with Milky Way, the LMC, and SMC abundances are mainly driven by spectral lines 
from Fe-group elements \citep[][]{Abbott82,Kudritzki87,Vink22}, which largely determine the 
resulting wind properties.
Since the wind properties have a stronger effect on the line profiles than the relative 
abundances, the \ion{Si}{4} and \ion{C}{4} relations seen in Figures~\ref{fig9} and \ref{fig10} 
could partly, or even mostly, reflect a relation between stellar Fe and nebular O. 

So far, we assume a given mass-loss recipe in our calculations rather than predicting the wind 
parameters self-consistently from the abundances. 
However, even if the absolute scaling of the wind mass-loss rate and terminal velocity should 
change, the general metallicity scaling has been confirmed in various different wind modeling 
approaches \citep[e.g.,][]{Vink22}. 
Our test calculations also show that relative abundances of individual elements have some impact on the derived EWs, so we will need more extensive calculations, including a full population synthesis, to test our interpretation. 
In particular, we emphasize an important underlying assumption made in the theoretical Starburst99 models in Figure 11. All element ratios in the stellar-wind models are solar for all values of 12 + log (O/H). The theoretical relation is expected to change if the abundance of elements driving stellar winds, i.e. Fe, were modified relative to oxygen. The agreement between the models and the data therefore suggests that our galaxy sample has an O/Fe ratio that is consistent with the solar value. On the other hand, \citep{steidel16} found evidence of strongly enhanced O/Fe ratios in a sample of strongly star-forming galaxies at z $\approx$ 2.4 and interpret this result as due to oxygen abundance enhancement by core-collapse supernovae. Their sample of galaxies may not follow the predicted relation in Figure 11. Consistent population synthesis models incorporating stellar models with non-solar abundance ratios may provide an opportunity to study any anomalous O/Fe relation in star-forming galaxies.


\section{Conclusions}\label{sec:5}
We present an analysis of the effects of spectral-resolution and aperture scales on 
derived FUV galaxy properties.
The rest-frame FUV is fundamental to our understanding of star-forming galaxies, as it simultaneously provides 
a unique window on massive stellar populations, chemical evolution, feedback processes, and reionization. 
The recent launch of JWST has already revealed how restframe UV spectroscopy traces galaxy evolution into the early universe,
but we lack a sufficient understanding of how aperture and resolution affects the interpretation of these galactic properties.
We, therefore, constructed an atlas of FUV archival spectra of local star-forming galaxies with multiple
aperture sizes and spectral resolutions for comparison.
In order to examine observations that mimic the anticipated galaxy-scale, low-resolution observations 
of high-redshift galaxies from JWST ($R\sim100-3,500$) we used large-aperture (10$\arcsec\times20\arcsec$) 
spectra from IUE ($R\sim250$) and compared to the stellar 
cluster-scale ($2\farcs5$), high-resolution ($R\sim15,000$) spectra from COS on board the Hubble Space Telescope (HST). 

We examined how FUV-derived properties were affected by the galaxy-scale aperture and low-resolution spectra
of the IUE versus the stellar cluster-scale aperture and high-resolution of the COS spectra. 
We find that the overall effect of the aperture size difference is non-consequential
for galaxies whose light is dominated by a single, bright stellar cluster,
while the effect of spectral resolution is strongest when measuring the EWs of interstellar 
absorption features:
\begin{itemize}
    \item Using featureless regions of the FUV continuum, we measured $\beta$-slopes and found that they
    were generally consistent between the different aperture measurements, except when multiple bright 
    stellar clusters populated the IUE field of view.
    \item We then measured the reddening due to dust, $E(B-V)$, of the stellar continua using two methods:
    (1) converting directly from the $\beta$-slope measurements and 
    (2) using a minimization routine to fit \texttt{starburst99} stellar population synthesis models to the
    observed spectra.
    We found that the two methods were consistent within their uncertainties.
    We also find that the $E(B-V)$ values agreed within their uncertainties across different aperture sizes 
    and spectral resolutions.
    Similar to the $\beta$-slopes, we find little difference between cluster-scale and galaxy-scale
    measurements for galaxies dominated by a single ongoing burst.
    \item Aperture size starts to play a more significant role in the measurement of equivalent widths. 
    We examined trends for the EWs of ions at different ionization states versus metallicity. 
    For both low- and high-ionization states, we measure trends with slopes $< 1$ such that EW measurements skew to 
    higher IUE values at the low-EW end and towards higher COS values at the high-EW end. 
    While we find a weak high-EW trend around 4 \AA, it currently lacks the statistical significance needed to draw a robust conclusion.
    \item We determined oxygen abundances for our sample and established correlations with UV properties. 
    We found significant correlations between the strength of stellar and interstellar lines and 
    the oxygen abundance despite these lines being heavily saturated. 
    These correlations can be understood in terms of metallicity-dependent line-driven stellar-winds 
    and interstellar macroscopic gas flows shaping the stellar and interstellar spectral lines, respectively. 
    The observed line-strength versus metallicity relation of stellar-wind lines agrees with the prediction of 
    population synthesis models for young starbursts. 
    Measurements of the strong \ion{C}{4} \W\W1548,1550 line in particular provides an opportunity to 
    determine stellar abundances as a complement to gas-phase abundances from nebular emission lines.   
\end{itemize}

The application of these results to JWST observations of high-redshift galaxies implies that integrated,
galaxy-scale properties can similarly characterize the overall galaxy and the dominating stellar cluster 
regions well. 
However, care will need to be taken in interpreting interstellar and stellar absorption features from
the low-resolution JWST spectra to avoid confusion from contaminating features, such as determinations of
the massive star population masses and outflow properties. 
This will be crucial for understanding the evolution of galaxies from the early universe extrapolated 
out to the local Universe.

\begin{acknowledgments}
We would like to express our gratitude to the referee for many thoughtful comments
that improved the clarity and impact of this work. 

The HST and IUE data presented in this paper were obtained from the Mikulski Archive 
for Space Telescopes (MAST) at the Space Telescope Science Institute. 
The specific observations analyzed can be accessed via 
\dataset[doi:10.17909/r6n6-3m45]{https://doi.org/doi:10.17909/r6n6-3m45}.

This research has made use of the NASA/IPAC Extragalactic Database (NED),
which is operated by the Jet Propulsion Laboratory, California Institute of Technology,
under contract with the National Aeronautics and Space Administration.

Funding for the creation and distribution of the SDSS Archive has been provided by the Alfred P. Sloan Foundation, the Participating Institutions, the National Aeronautics and Space Administration, the National Science Foundation, the U.S. Department of Energy, the Japanese Monbukagakusho, and the Max Planck Society. The SDSS Web site is http://www.sdss.org/. The Participating Institutions are The University of Chicago, Fermilab, the Institute for Advanced Study, the Japan Participation Group, The Johns Hopkins University, the Max-Planck-Institute for Astronomy (MPIA), the Max-Planck-Institute for Astrophysics (MPA), New Mexico State University, Princeton University, the United States Naval Observatory, and the University of Washington.

Support for this work has
been provided by NASA through grant No. AR-13878 from the
Space Telescope Science Institute, which is operated by
AURA, Inc., under NASA contract NAS5-26555.

AACS is acknowledges funding by the Deutsche Forschungsgemeinschaft (DFG - German Research Foundation) in the form of an Emmy Noether Research Group -- Project-ID 445674056 (SA4064/1-1, PI Sander). AACS is further supported by funding from the Federal Ministry of Education and Research (BMBF) and the Baden-Württemberg Ministry of Science as part of the Excellence Strategy of the German Federal and State Governments.
\end{acknowledgments}

\facilities{HST (COS), IUE}
\software{
\texttt{astropy} \citep{astropy:2013, astropy:2018, astropy:2022},
\texttt{LMFIT} version 1.2.2 \citep{newville15},
\texttt{PyNeb} version 1.1.14 \citep{luridiana15},
\texttt{BPASS} version 2.2 \citep{stanway18},
\texttt{Starburst99} \citep{leitherer14},
\texttt{numpy} version 1.26 \citep{harris20},
\texttt{PoWR} \citep{Graefener02,Hamann03,sander15} 
}


\typeout{} 
\bibliography{mybib}


\appendix
\begin{figure*}[h!]
\begin{center}
    \includegraphics[width=0.75\textwidth,trim=20mm 0mm 20mm 17mm,clip]{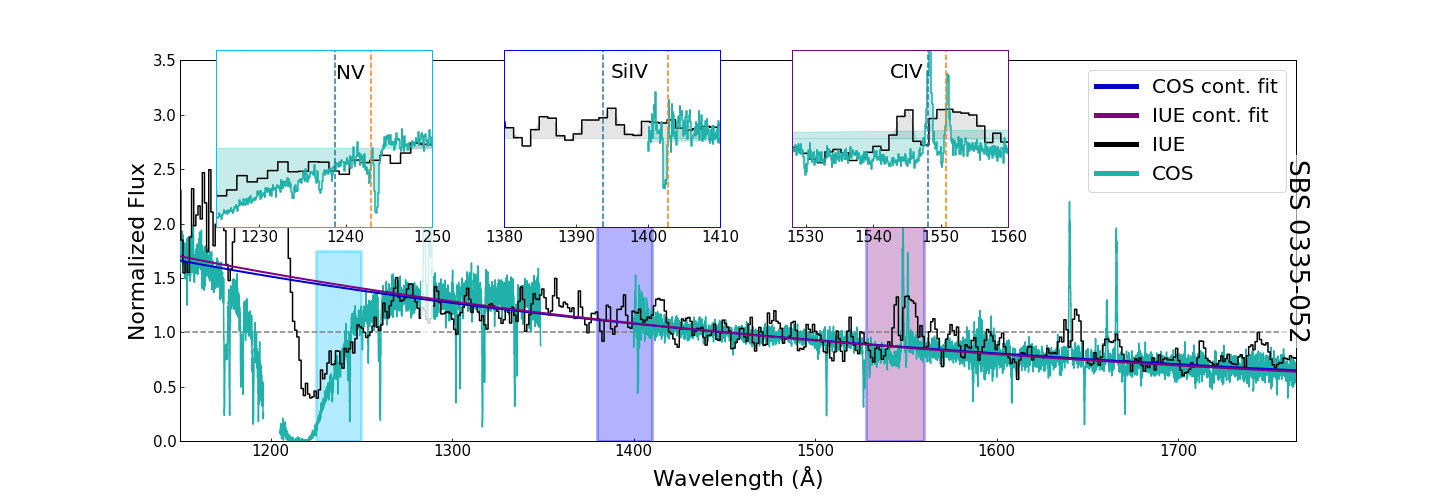} \\
    \includegraphics[width=0.75\textwidth,trim=20mm 0mm 20mm 17mm,clip]{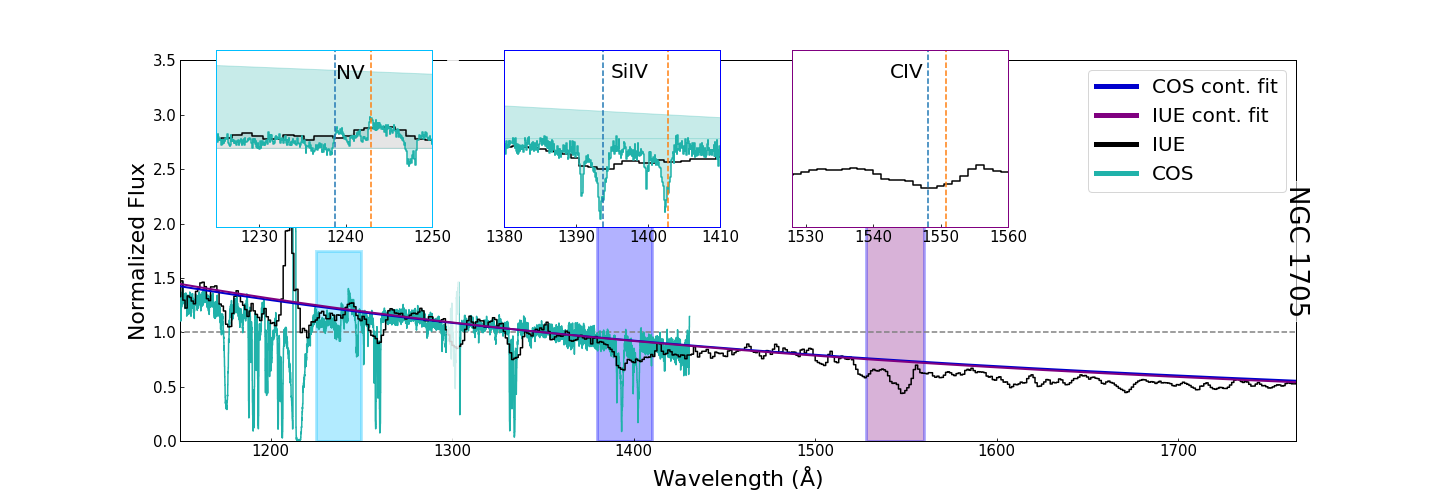}\\
    \includegraphics[width=0.75\textwidth,trim=20mm 0mm 20mm 17mm,clip]{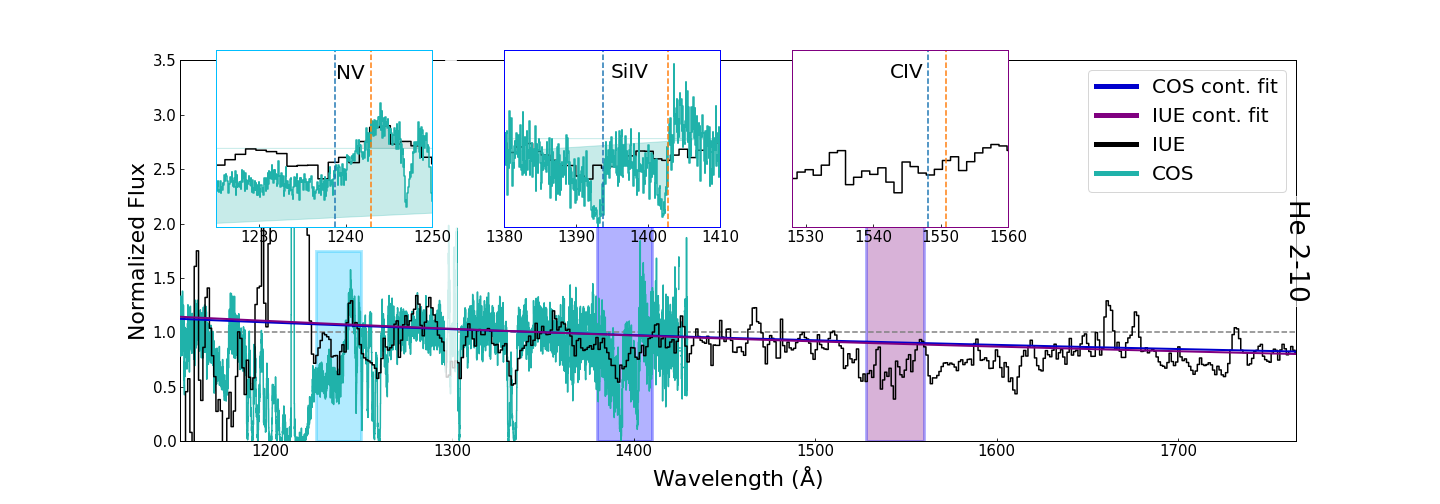} \\ 
    \includegraphics[width=0.75\textwidth,trim=20mm 0mm 20mm 17mm,clip]{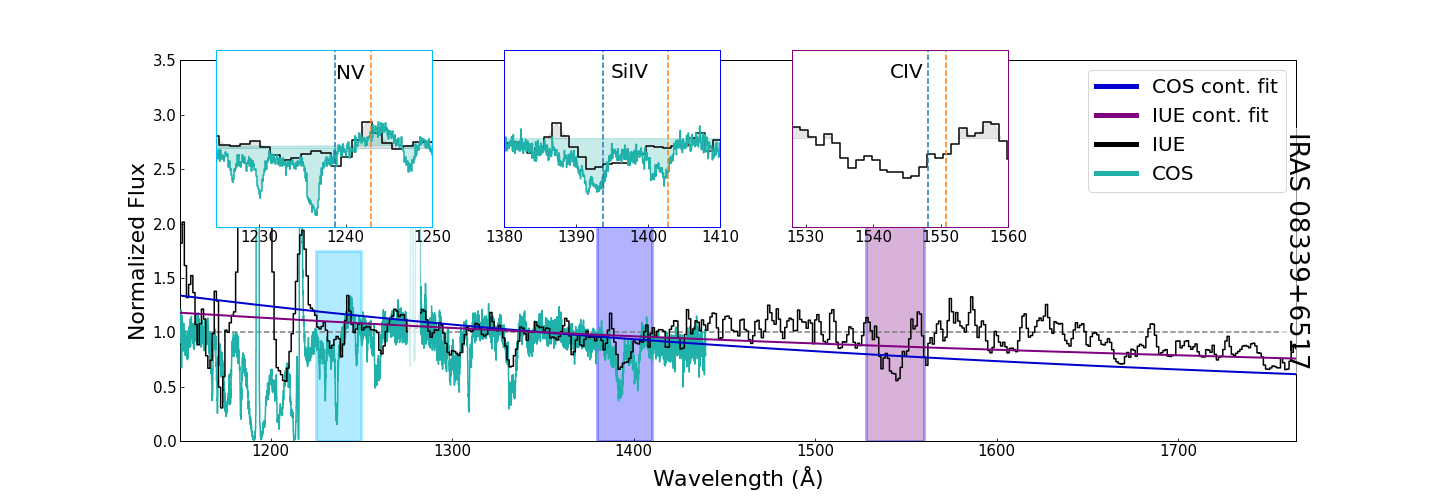} 
 \end{center}
\end{figure*}

\begin{figure*}
\begin{center}
    \includegraphics[width=0.7\textwidth,trim=20mm 0mm 20mm 17mm,clip]{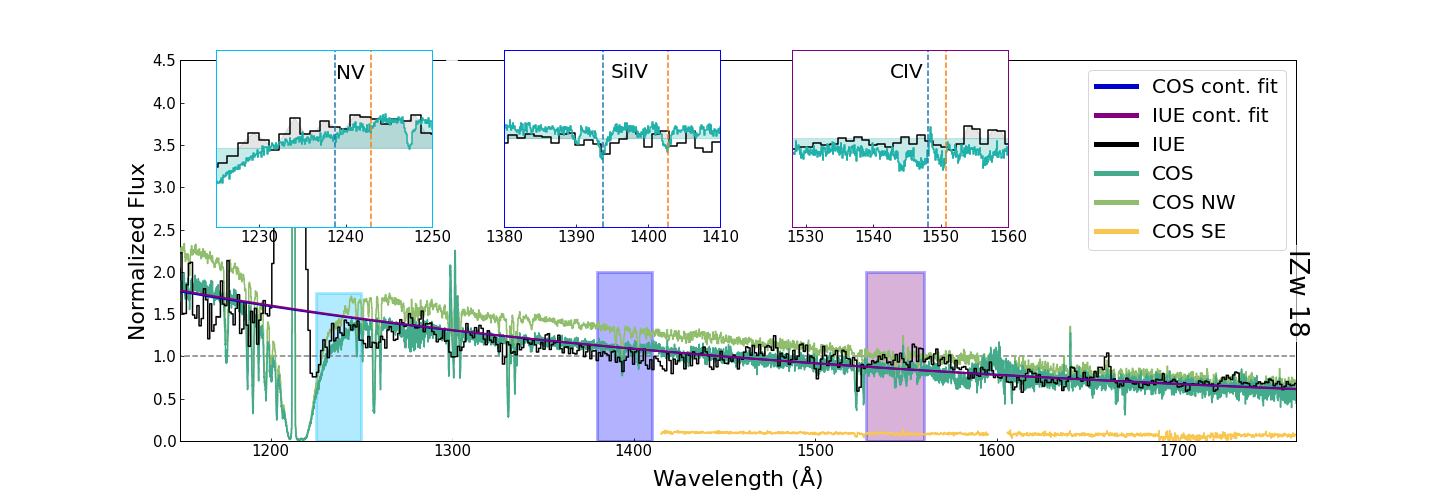} \\
    \includegraphics[width=0.75\textwidth,trim=20mm 0mm 20mm 17mm,clip]{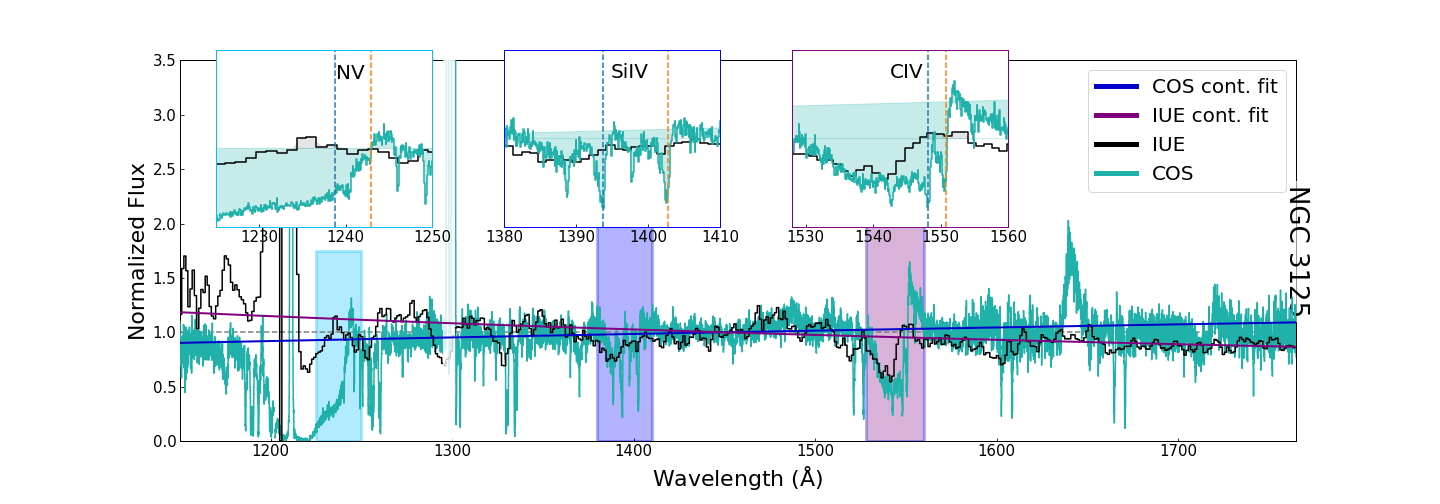} \\ 
    \includegraphics[width=0.75\textwidth,trim=20mm 0mm 20mm 17mm,clip]{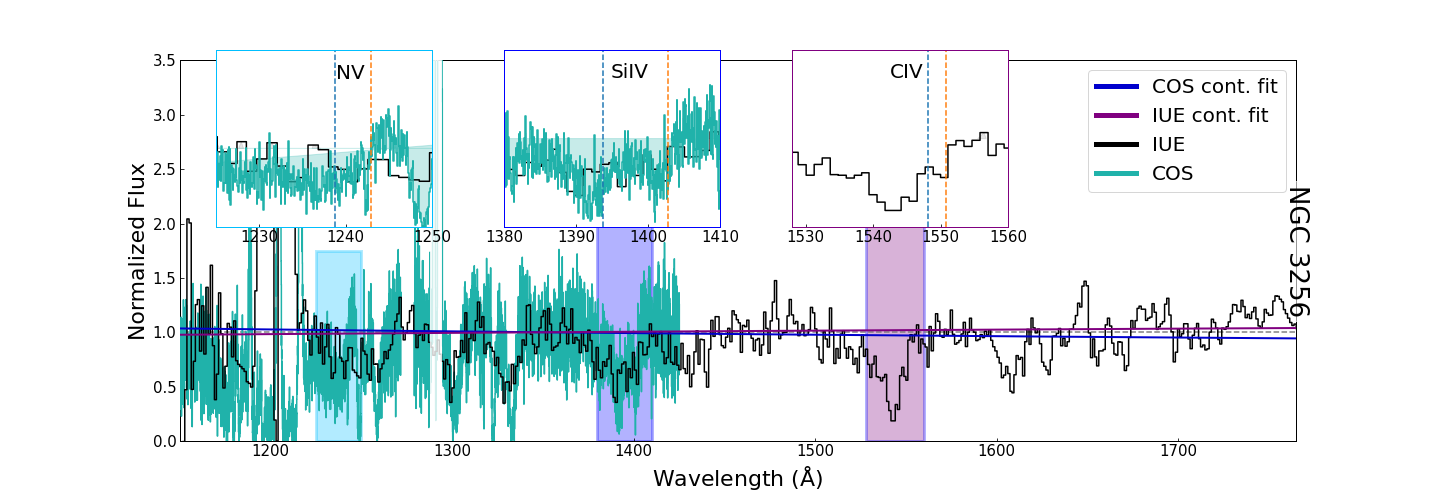} \\ 
    \includegraphics[width=0.75\textwidth,trim=20mm 0mm 20mm 17mm,clip]{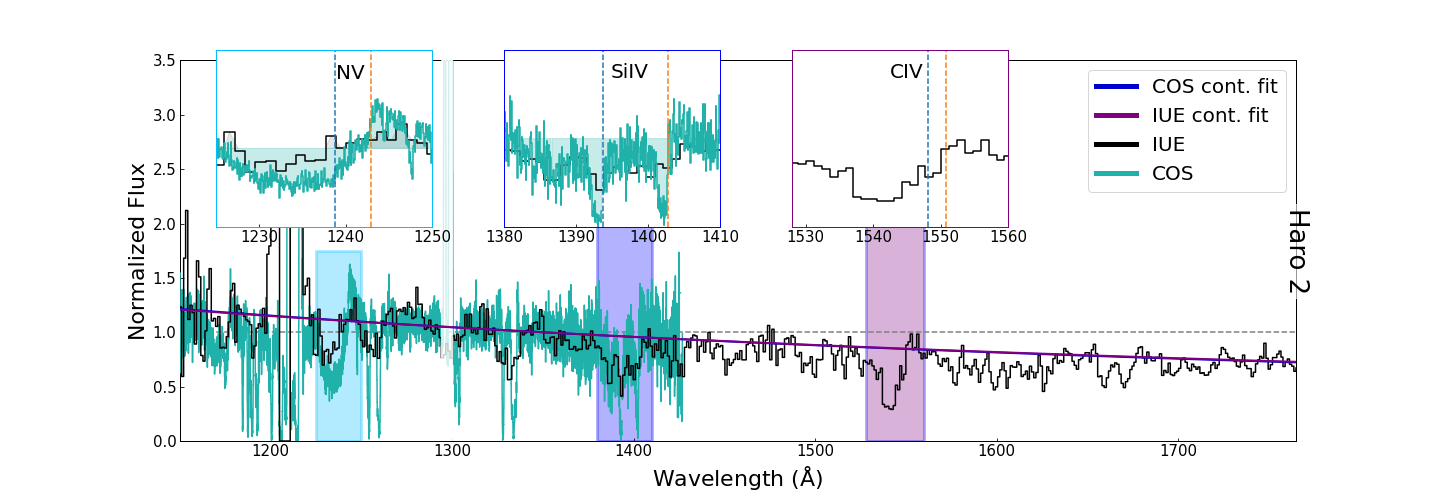} \\ 
 \end{center}
\end{figure*}

\begin{figure*}
\begin{center}
    \includegraphics[width=0.75\textwidth,trim=20mm 0mm 20mm 17mm,clip]{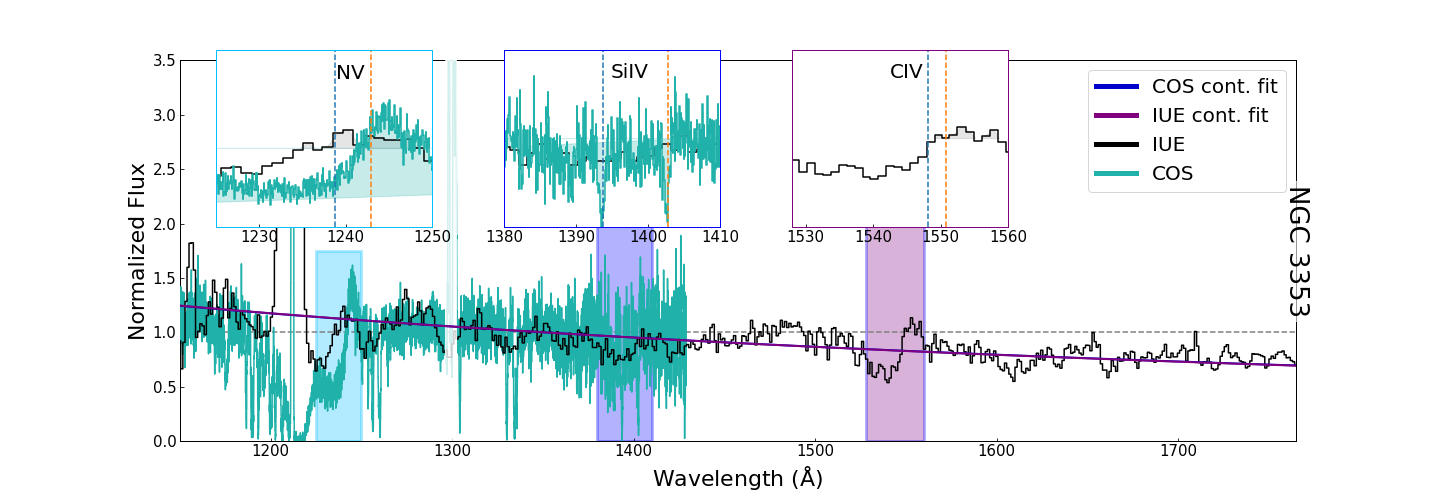} \\ 
    \includegraphics[width=0.75\textwidth,trim=20mm 0mm 20mm 17mm,clip]{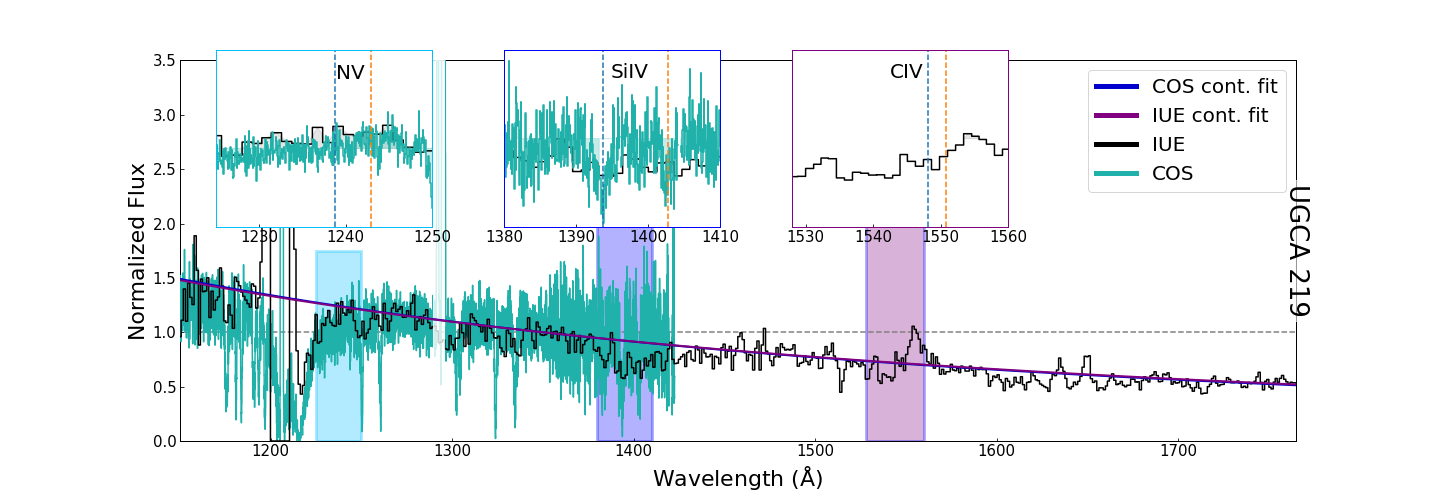} \\ 
    \includegraphics[width=0.75\textwidth,trim=20mm 0mm 20mm 17mm,clip]{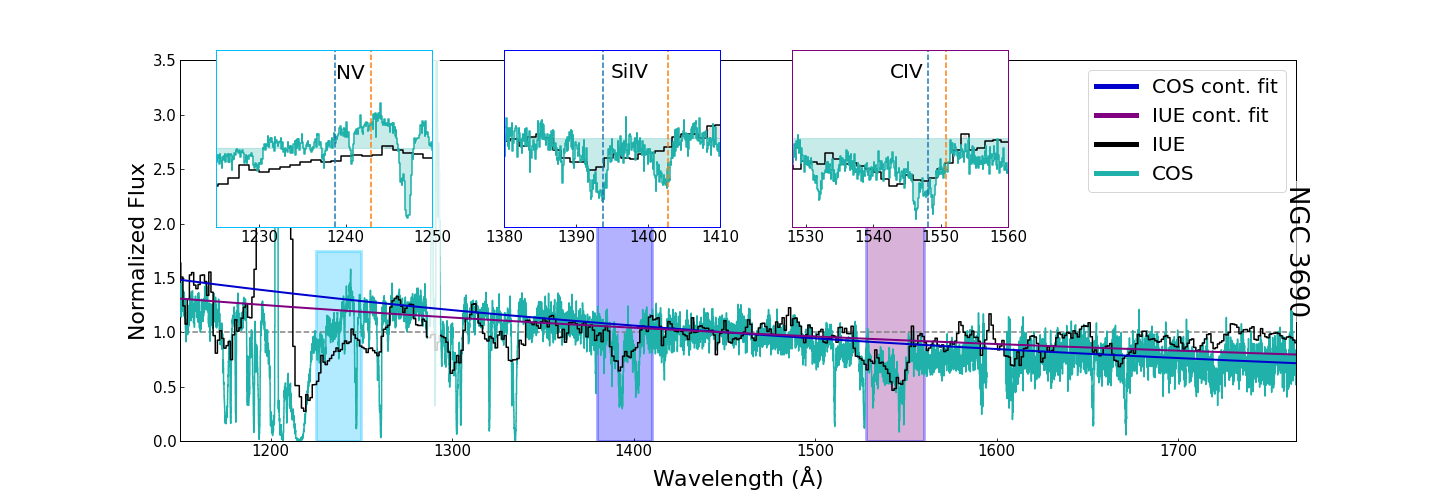} \\
    \includegraphics[width=0.75\textwidth,trim=20mm 0mm 20mm 17mm,clip]{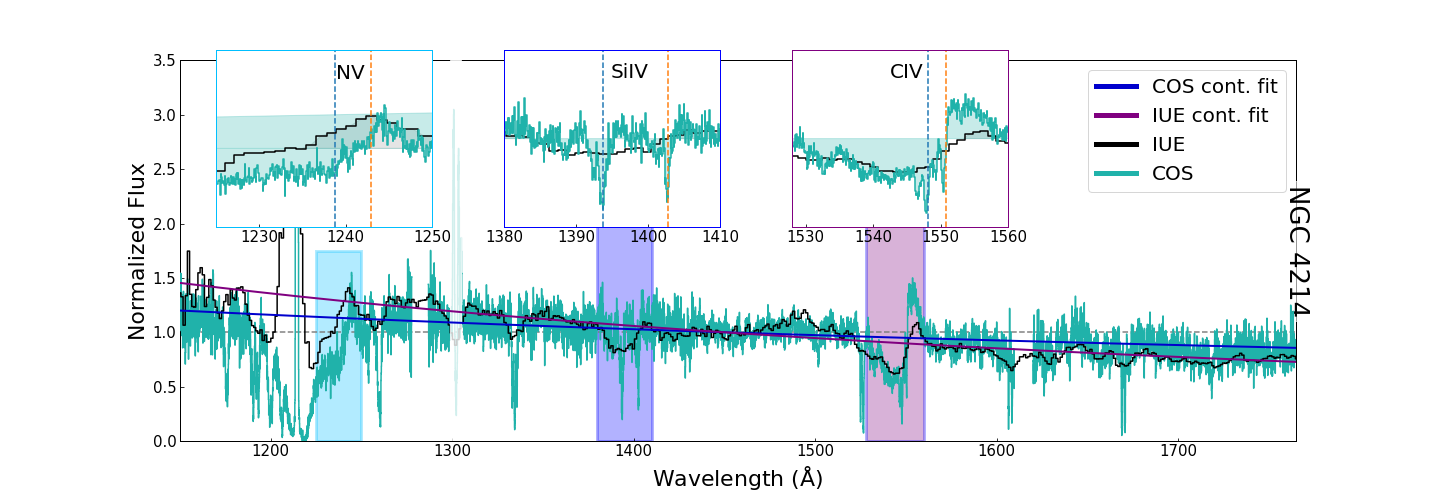} \\  
 \end{center}
\end{figure*}

\begin{figure*}
\begin{center}
    \includegraphics[width=0.75\textwidth,trim=20mm 0mm 20mm 17mm,clip]{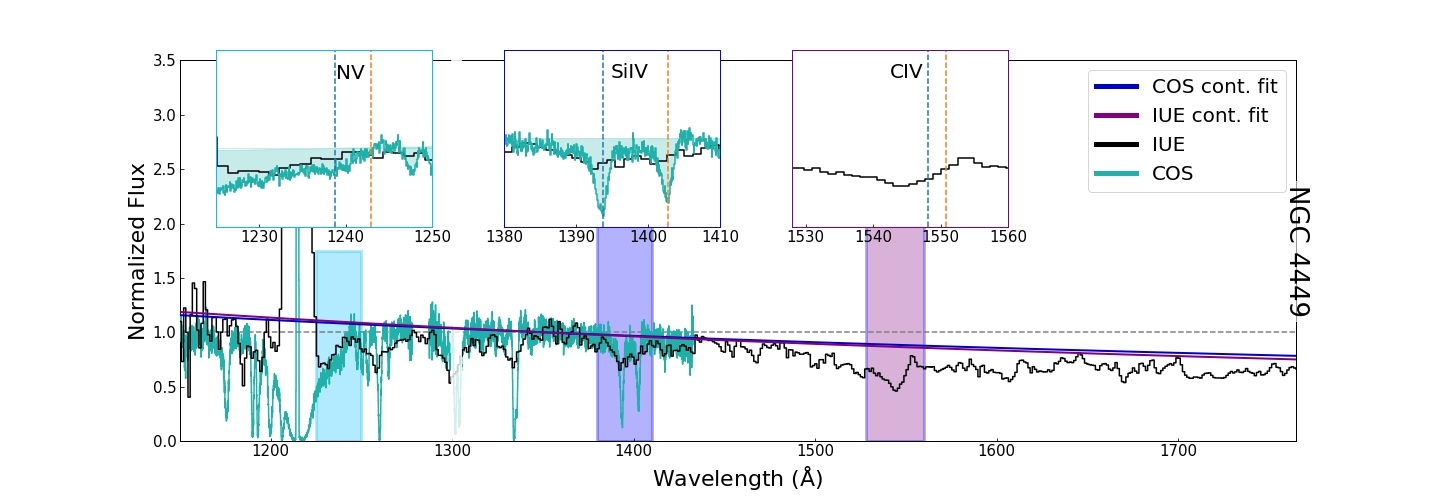} \\ 
    \includegraphics[width=0.75\textwidth,trim=20mm 0mm 20mm 17mm,clip]{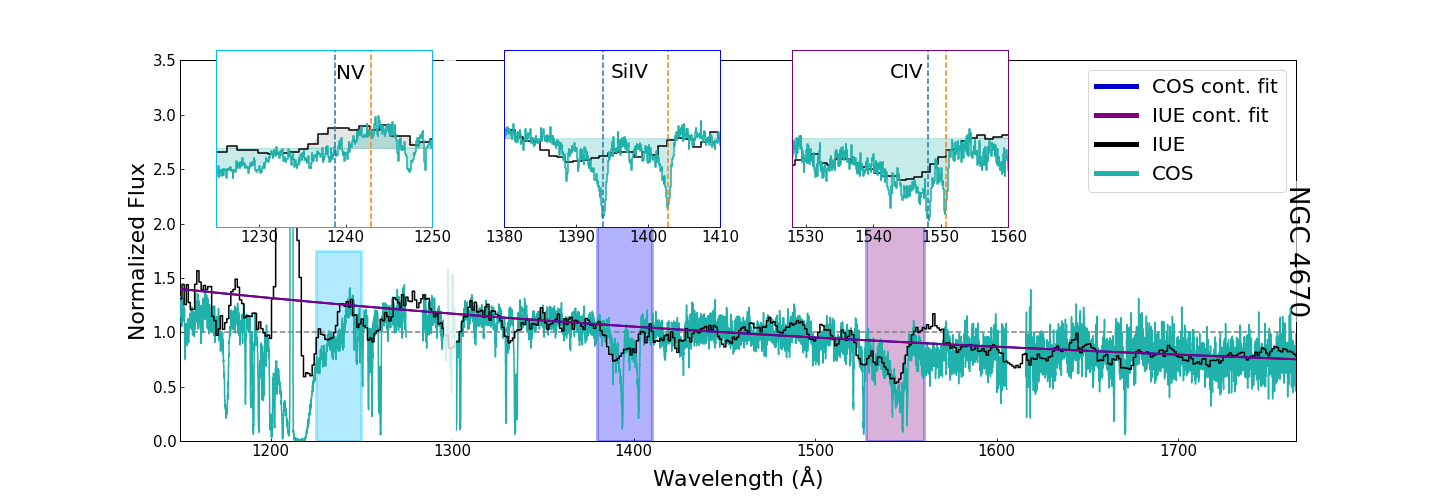} \\ 
    \includegraphics[width=0.75\textwidth,trim=20mm 0mm 20mm 17mm,clip]{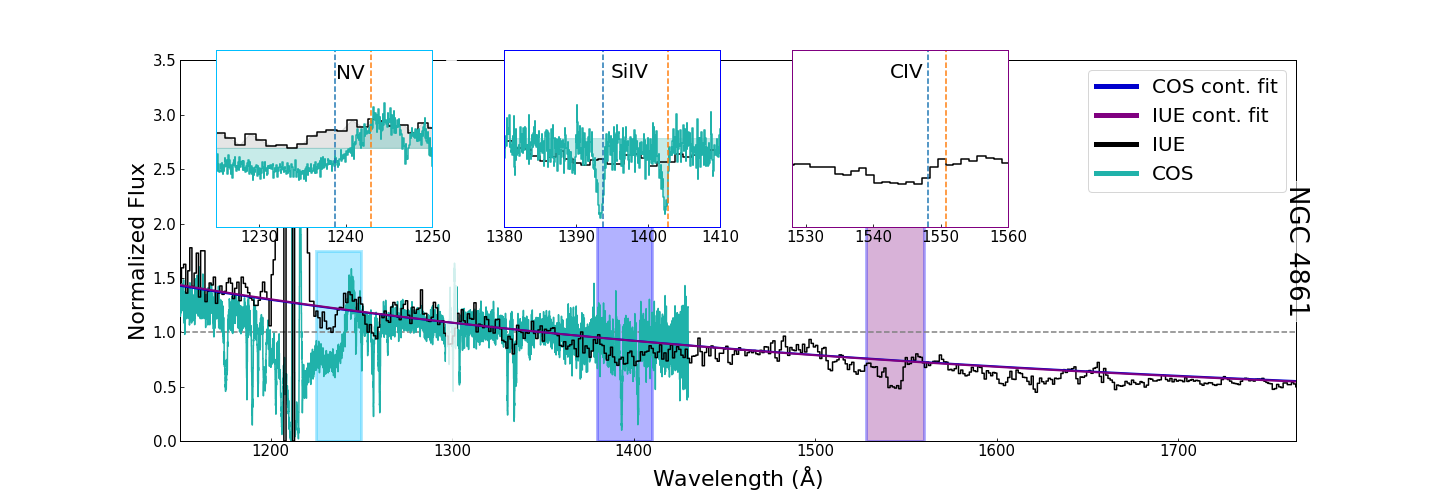} \\
    \includegraphics[width=0.75\textwidth,trim=20mm 0mm 20mm 17mm,clip]{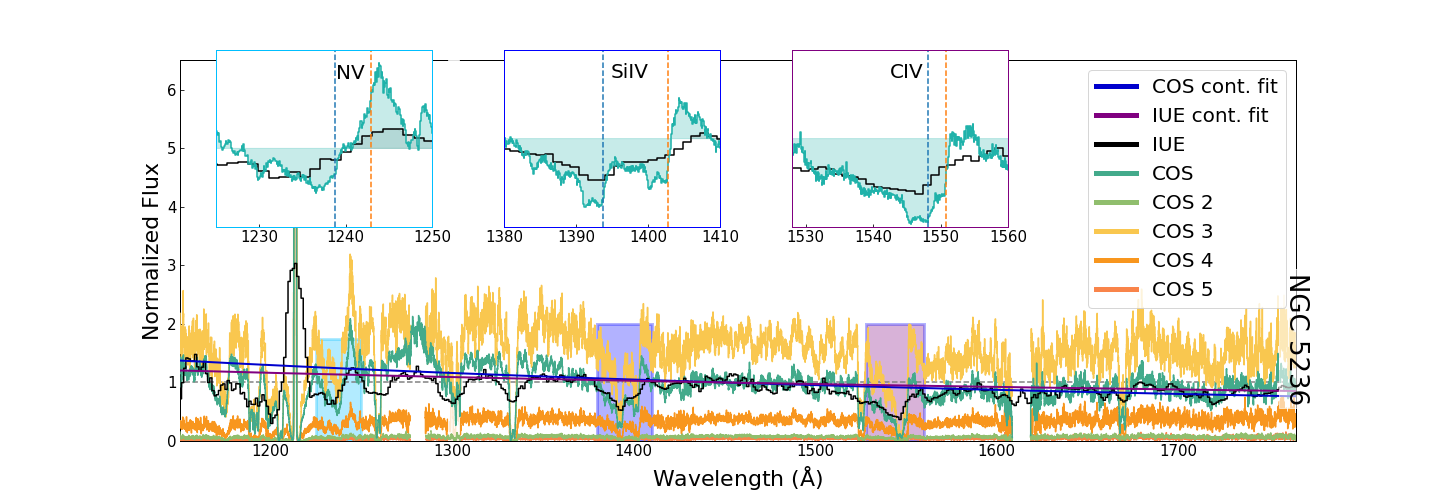} 
 \end{center}
\end{figure*}

\begin{figure*}
\begin{center}
    \includegraphics[width=0.75\textwidth,trim=20mm 0mm 20mm 17mm,clip]{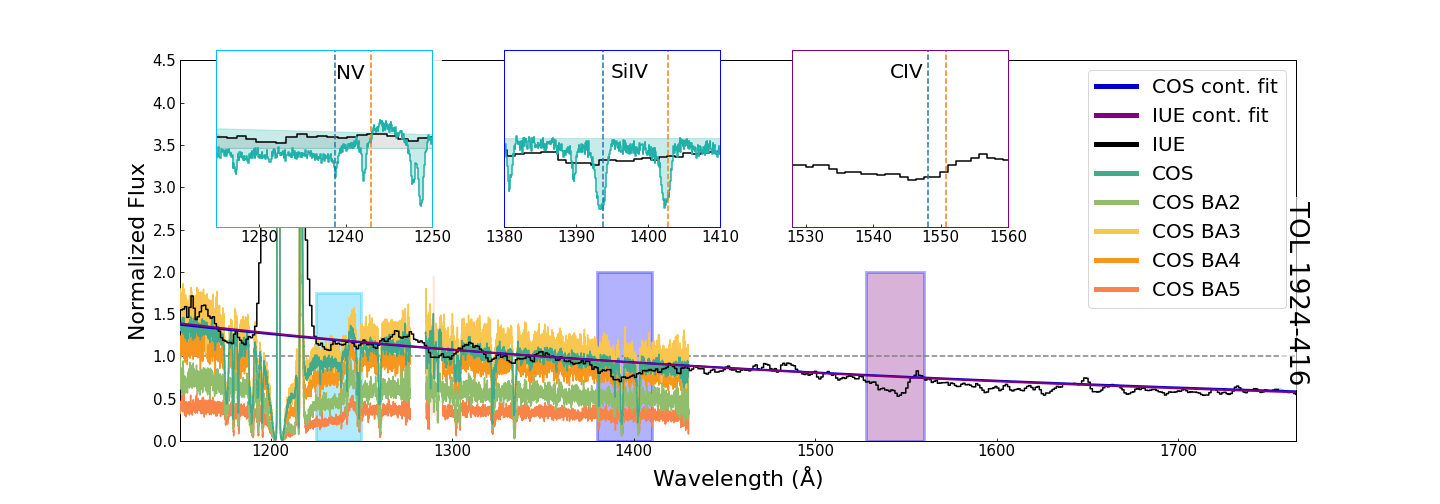} \\
    \includegraphics[width=0.75\textwidth,trim=20mm 0mm 20mm 17mm,clip]{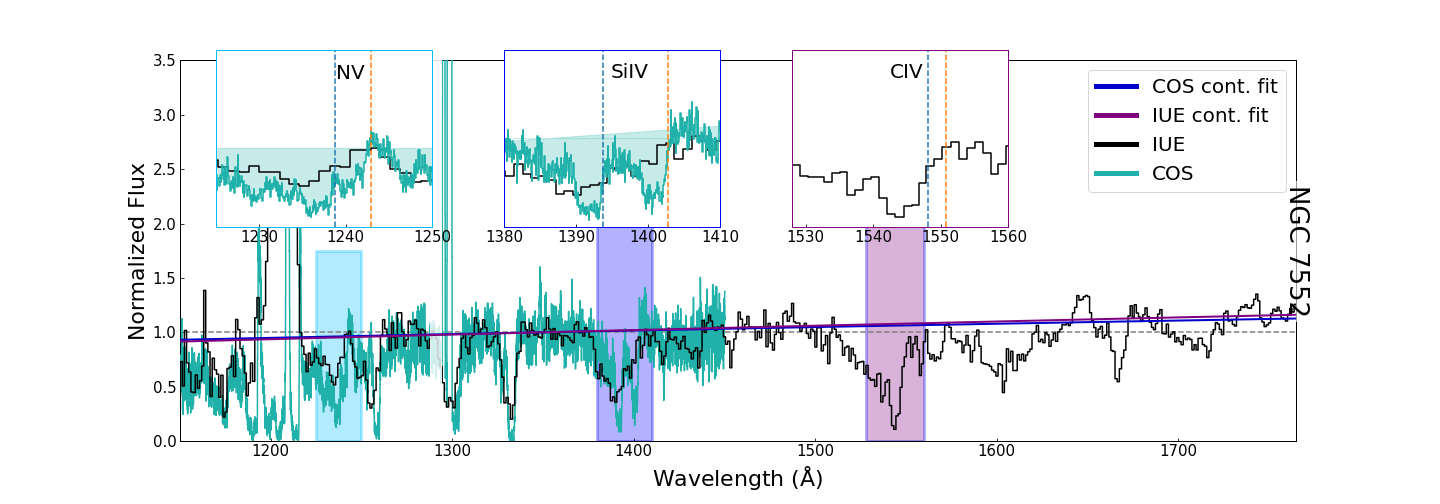} \\
    \includegraphics[width=0.75\textwidth,trim=20mm 0mm 20mm 17mm,clip]{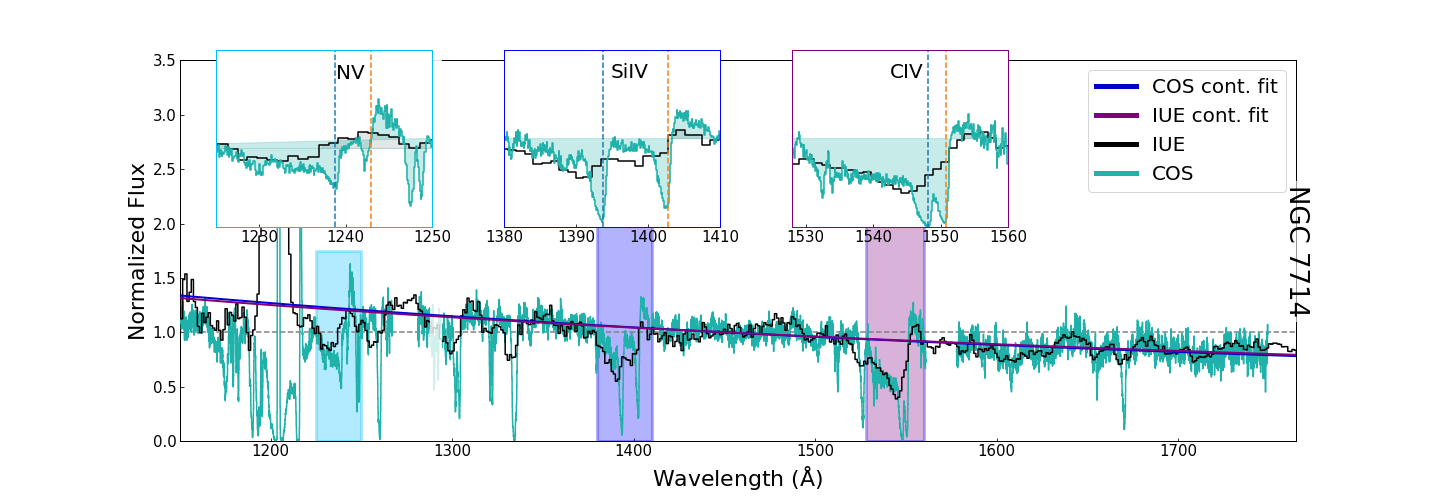}  
\caption{Comparison of IUE and coadded COS rest-frame FUV spectra for the galaxies in our sample.
The IUE spectra are relatively low resolution (\W/$\Delta$\W$\sim300$)
and consists of the integrated light within a large aperture (10\arcsec$\times$20\arcsec).
In contrast, the COS spectra have significantly higher spectral resolution 
(\W/$\Delta$\W$=15,000$) and higher S/N,
but only within the much smaller 2\farcs5 COS aperture.
All IUE and coadded COS spectra that extend past 1450 \AA\ are normalized at 1450 \AA,
while coadded COS spectra with shorter wavelength coverage are normalized at 1350 \AA.
For galaxies with multiple COS pointing, each spectrum is normalized by the coadded
COS spectrum normalization, allowing relative differences in shape and absolute flux
to be compared.
The $\beta$-slope fits to the IUE and coadded COS spectra are overplotted as 
purple and blue lines respectively.}
 \end{center}
\end{figure*}

\end{document}